\documentclass[11pt]{article}
\usepackage{framed,multirow}

\usepackage{amssymb}
\usepackage{latexsym}

\usepackage{url}

\usepackage{xcolor}
\definecolor{newcolor}{rgb}{.8,.349,.1}

\usepackage[mathscr]{eucal}
\usepackage[all]{xy} 
\usepackage{graphicx}
\usepackage{caption}
\usepackage{subcaption}
\usepackage{multicol}
\usepackage{amssymb,amsfonts,amsthm}    
\usepackage{color}                              
\usepackage[centertags]{amsmath}
\usepackage{amsfonts,amssymb,amsthm,newlfont}
\usepackage{geometry}
\usepackage{algorithmic}
\usepackage{algorithm}
\usepackage{hyperref}
\usepackage{mathrsfs}
\usepackage{enumerate}

\usepackage{amsfonts}
\usepackage{algorithm}

\usepackage{nomencl}

\usepackage{epsfig, bm, array}
\usepackage{ulem}
\usepackage{multirow}

\theoremstyle{plain}
\newtheorem{thm}{Theorem}[section]
\newtheorem{prop}[thm]{Proposition}
\newtheorem{lemma}[thm]{Lemma}

\newtheorem{remark}[thm]{Remark}

\newenvironment{pf}{{\noindent \it \bf Proof.}}{{\hfill$\Box$}\\}

\newcommand{\ceil}[1]{\left\lceil #1 \right\rceil}

\numberwithin{equation}{section}
\numberwithin{algorithm}{section}

\newcolumntype{L}[1]{>{\raggedright\let\newline\\\arraybackslash\hspace{0pt}}m{#1}}
\newcolumntype{C}[1]{>{\centering\let\newline\\\arraybackslash\hspace{0pt}}m{#1}}
\newcolumntype{R}[1]{>{\raggedleft\let\newline\\\arraybackslash\hspace{0pt}}m{#1}}

\begin{document}


\title{Numerical algorithms and simulations of boundary dynamic control for optimal mixing in unsteady Stokes flows}
\author{   
Xiaoming Zheng\thanks{Department of Mathematics, Central Michigan University, Mount Pleasant, MI 48859, USA  (e-mail:  zheng1x@cmich.edu)}
\and
Weiwei Hu\thanks{Department of Mathematics, University of Georgia, Athens, GA 30602, USA  (e-mail: Weiwei.Hu@uga.edu)}
\and
Jiahong Wu  \thanks{Department of Mathematics, Oklahoma State University, OK 74078 USA  (e-mail: jiahong.wu@okstate.edu)}
}

\date{}
\maketitle
\tableofcontents

\begin{abstract}
This work develops an efficient and accurate optimization algorithm to study the optimal mixing problem driven by boundary control  of  unsteady Stokes flows, based on the theoretical foundation laid by Hu and Wu in a series of work. The scalar being mixed is purely advected by the flow and the control is a force exerted  tangentially on the domain boundary  through the Navier slip conditions. The control design has potential  applications in many industrial processes such as rotating wall driven mixing, mircomixers with acoustic waves, and artificial cilia mixing. 

The numerical algorithms have high complexity, high accuracy demand, and high computing expense, due to the multiscale nature of the mixing problem and the optimization requirements.
A crucial problem is the computation of the Gâteaux  derivative of the cost functional. To this end,  a hybrid approach based on variational formula and finite difference is built with high accuracy and efficiency to treat various types of control input functions. 
We have experimented with various optimization schemes including the steepest descent algorithm, the conjugate gradient method and two line search options (backtracking and exact line search). We are able to identify and implement the best combinations.

The numerical simulations show that the mixing efficacy is limited when only one single type of control is applied, but can be enhanced when more diverse control types and more time segmentation are utilized. The mix-norm in the optimal mixings decays exponentially. 
The numerical study in this work demonstrates that boundary control alone could be an effective  strategy for mixing in incompressible flows.
\end{abstract}





\section{Introduction}\label{intro}
Transport and mixing in fluids are of fundamental importance in many processes in nature and industry. A long-lasting and central problem is to design an optimal  control that  enhances  transport and mixing or steers a scalar field to a desired distribution, which has drawn great attention to researchers  in many fields.

\subsection{Motivations and applications}
\label{sec_examples}
\begin{figure}[!htbp]
\begin{center} 
\includegraphics[scale=0.18]{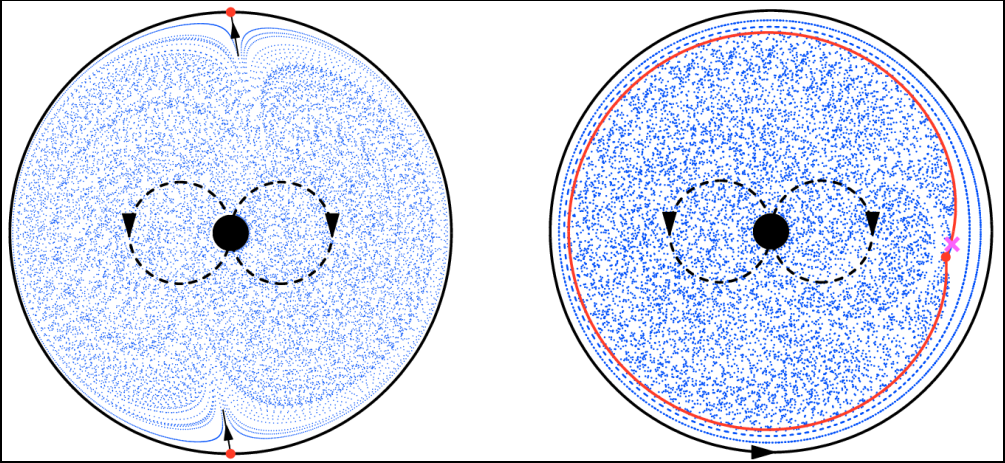}[a]
\includegraphics[scale=0.2]{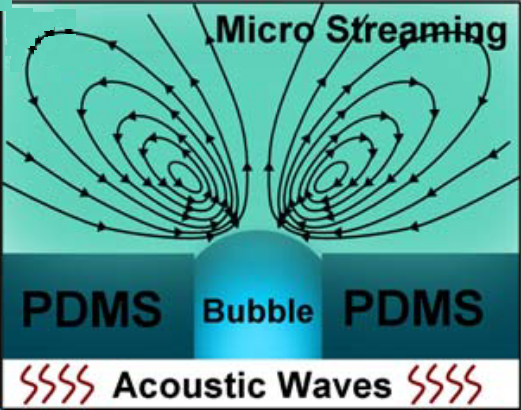}[b]\\
\includegraphics[scale=0.13]{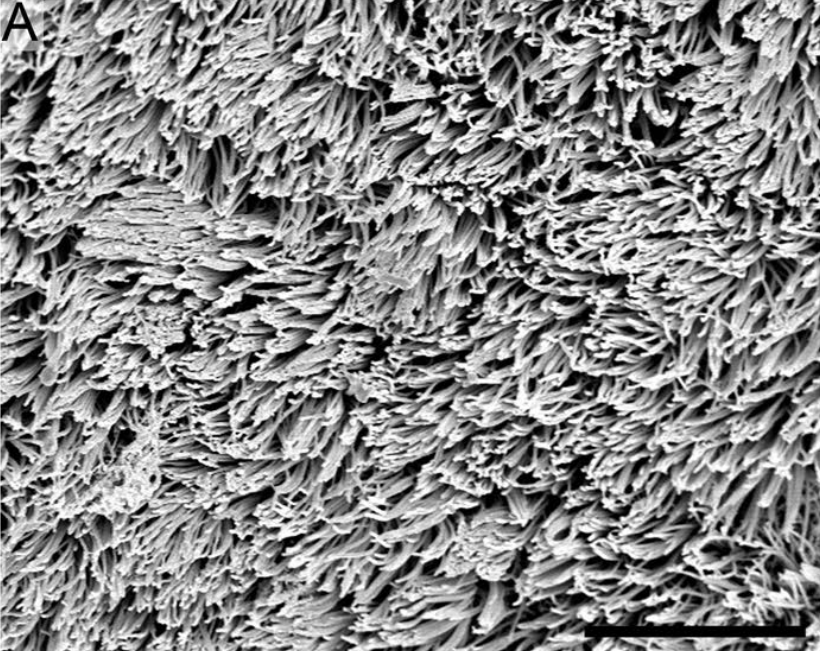}[c]
\includegraphics[scale=0.13]{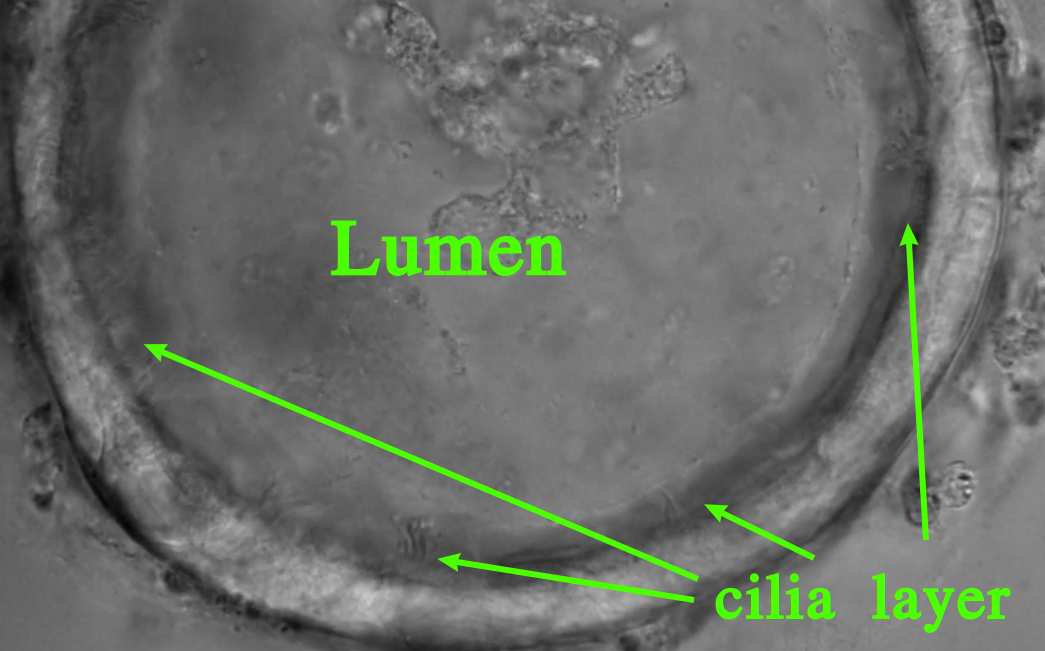}[d]
\caption{\scriptsize
[a] Left: numerical solution of figure-eight internal stirring with a fixed wall where the arrows point to wall separatrices. Right: rotating wall breaks the separatrices. Taken from \cite{thiffeault2011moving} with permission.
[b] Schematic of the velocity field generated by a micro air bubble activated by acoustic waves, which is  embedded in the polydimethylsiloxane (PDMS)  sidewall. Taken from \cite{Ahmed2009} with permission. 
[c] Scanning electron microscopy  image of  human tracheal epithelial cilia. Taken from \cite{Nakamura2020} with permission. 
[d] The mixing of mucus driven by cilia beating within the lumen of the airway organoids derived from human lung stem cells. The cilia layer is located on the boundary of the lumen. Taken from the video  \url{https://www.youtube.com/watch?v=1Q8RL1g9txk} related to the paper\cite{Hui2018}.
\label{example_boundary_mixing}
}
\end{center}
\end{figure}

Boundary control, by implementing energy sources through the boundary of the mixer,  has been observed or used individually  or synergistically with other approaches for transport and mixing in many scenarios. 
One straightforward boundary control protocol is moving or rotating the container walls to facilitate mixing. In the mixing of two immiscible viscous fluids under low Reynolds numbers in a rectangular cavity \cite{chakravarthy1996mixing, vikhansky2002enhancement}, the top and bottom walls are moved where the moving velocity is employed as the control input to steer mixing, measured by the area or length of the  fluid interface. 
In a series of studies \cite{ gouillart2008slow, gouillart2007walls,gouillart2010rotation, thiffeault2011moving}, it is discovered  that the fixed wall with no-slip boundary condition can slow down the internal mixing from exponential decay into power decay due to the separatrices near the wall; however, rotating walls with a constant angular velocity can recover the exponential decay by removing the separatrices (see Figure\,\ref{example_boundary_mixing}[a]). These studies use theoretical analysis and/or scientific computing instead of real physical devices. 

Instead of moving an entire piece of a sidewall, some boundary control strategies apply controls on individual spots  of the fixed sidewall. For example, some micromixers use acoustic waves to perturb  mircobubbles embedded in the sidewall of the mixer, whose oscillation can create high pressure and velocity in the bulk liquid in the mixer \cite{Ahmed2009, D0LC01173H} (see Figure\,\ref{example_boundary_mixing}[b]). This mixing method is considered simple and effective  to overcome the low Reynolds numbers in microfluids due to high viscosity and long microchannel.

Another example of the boundary control is the cilia induced mixing \cite{Islam2022}. 
Cilia are microscopic hair-like structures extensively present in vertebrates and  they are located on the epithelial surfaces of internal organs such as the respiratory tract (see Figure\,\ref{example_boundary_mixing}[c]). 
The cilia beating generates metachronal waves, which is an effective way to transport fluid and perform mixing \cite{Supatto2008, Nawroth2017}.
Attracted by the functions of biological cilia, researchers have created artificial cilia, driven by magnetic or electric field, or pneumatics, to generate microfluidic flow, with possible practices in microfluidic devices like lab-on-chip \cite{Islam2022}.
There exist some numerical studies of cilia mixing such as 
\cite{Lukens2010, Ding2014, Chateau2018, Guo2020}, where all of these work consider the direct interaction between fluid and the cilium structure and the mixing is measured by the mixing number according to redistribution of tracer particles advected by the flow. 
When the cilium length is significantly smaller than the size of the mixer (see Figure\,\ref{example_boundary_mixing}[d]), the cilia beating can be approximated as boundary conditions applied on the mixer.

\subsection{Objectives and challenges}
\label{sec_objectives}
Despite the motivations and applications mentioned above, boundary control for  transport and mixing is still a new field with very few studies. 
Recently, Hu and Wu in \cite{ hu2020approximating, hu2018boundarycontrol, hu2018boundary, hu2019approximating} have established a theoretical framework of boundary control for optimal mixing via the incompressible  flows, where the boundary control is the tangential force exerted on the mixer boundary \eqref{Stokes3}. In addition,  the scalar or density being mixed is assumed to be driven by advection only and the diffusion is neglected,  which corresponds to the case of large P\'{e}clet number (the ratio of the rate of advection to the rate of diffusion).

The objective of this work is to develop efficient numerical algorithms for the optimization problem proposed by Hu and Wu and then use them to investigate the efficacy of boundary control for fluid mixing. 
This work, to the authors' best knowledge,  is the first numerical study of  optimal mixing via  boundary control of the unsteady Stokes flow. Indeed, there are barely any numerical algorithms  developed  for solving the optimal control for mixing governed by the coupled  flow-transport system in a general open bounded domain.
Although the optimal mixing and stirring of passive scalars via pure advection has been extensively discussed
by means of theoretical analysis and numerical simulations in recent years (cf.~\cite{ chakravarthy1996mixing, alberti2019exponential,elgindi2019universal, lin2011optimal, lunasin2012optimal, thiffeault2012using,   yao2014mixing, crippa2019polynomial, iyer2014lower,  gubanov2010towards,  mathew2007optimal,vikhansky2002enhancement, liu2008mixing, seis2013maximal}), 
all these studies focus on  prescribed velocity fields and none of them consider the real-time control of the unsteady flow dynamics driven by control forces.

This work features  features high complexity, high accuracy demand, and high computing expense.
The first complexity is a cascade of four events from the  control to the objective cost function as shown in \eqref{cascade_4relations},  in contrast to 3 steps from flow to cost in the existing work mentioned above. 
\begin{equation}
\text{ control }
\stackrel{\text{unsteady Stokes}} {\longrightarrow} 
\text{ flow }
\stackrel{\text{advection}} {\longrightarrow}
\text{ mixed scalar } 
\longrightarrow 
\text{ cost functional}.
\label{cascade_4relations}
\end{equation}
The entire cascade will be called repeatedly in optimization algorithms, which would entail a high computing expense. However, this can  be partially relieved by utilizing a finite basis of the control space and the linear   relation between flow velocity and the control (given zero  initial velocity field. From the viewpoint of real world applications, a finite number of control inputs is a more realistic assumption since it is not practical to create arbitrarily distributed force fields for stirring.
Through this approach, only the velocity fields corresponding to the control basis are needed and stored before the optimization process.  Indeed, a control input is a linear combination of the control basis functions (see Equation\,\ref{def_g}) and the associated velocity field is a linear combination of the velocity basis with the same  coefficients (see Equation\,\ref{LC_Lg}). 

The mixing problem is intrinsically multiscale, where the optimal mixed scalar has delicate structures of thin filaments everywhere in the domain.
This complexity requires high accuracy in  the flow and advection solvers. 
In the flow solver, one complexity is how to enforce the divergence free condition in the numerical methods of the unsteady Stokes equations, which is important in  computing the transport equations and  the gradient of the cost functional (see Equation \eqref{eqn_gradsystemVF}). An iterative projection method for solving the Navier-Stokes equations \cite{zheng2023iterative} is applied in this work, which obtains the weakly divergence free velocity with the Taylor-Hood finite element method. 
In the evolution of the mixed scalar, the high order approximation is desirable due to its better ability to capture the microscale structures. However, high order approximations would slow down the evolution and thus the entire optimization process. Thus, a compromise between approximation order and evolution speed has to be made. 
Furthermore, a better mixing quality is often related to a larger control input and thus a larger flow velocity magnitude  (see details in Section\,\ref{sec_mixingfeatures}), which would induce small time steps in the advection solvers for stability reasons. If the velocity basis in all the time steps is stored in hard drive, it will result in a large amount of data storage, where a care is needed to balance the data storage quota and accuracy demand.

The development of optimization algorithms also has remarkable complexities. For instance, the accuracy of the gradient of the cost functional is crucial to the convergence of the optimization algorithms. The finite difference method is accurate but has high computing expense when the dimension of the control space  is large. The variational formula is much more efficient but may give disastrous results for a certain type of control functions. A hybrid approach will be proposed to combine the advantages of these two methods based on extensive experiments. 
Another complexity is the choice of the line search methods and optimization schemes. The back tracking method is fast but may not provide a local minimizer. In the work \cite{mathew2007optimal}, the exact line search is used with conjugate gradient method to solve an optimal mixing problem. The exact line search is computationally expensive because it needs many iterations of the cascade of \eqref{cascade_4relations} but it provides a local minimizer.
In the optimal control problem of an advection-reaction-diffusion system, a linearization line search method is proposed in \cite{Glowinski2022}, which will be examined in this work (see details in Section\,\ref{sec_linearization}). Both the steepest descent and the conjugate gradient optimization schemes, along with these line search choices,  will be tested for convergence, efficiency, and robustness.

The rest of this paper is outlined as follows. Section\,\ref{section2} presents the optimization problem of boundary control design for optimal  mixing in unsteady Stokes flows, along with the derivation of the Gâteaux derivative of the cost functional and   the first-order necessary optimality conditions for solving the optimal control.  Section\,\ref{comp} introduces the optimization algorithms, including the choice of the control input basis, the computation of the velocity basis,  the transport equations,  the cost functional and its Gâteaux  derivative, the line search methods, and the optimization schemes. Section\,\ref{sec_simulations}  first reports some basic properties of the control functions used in this work, such as flow patterns and mixing characteristics, and then applies the optimization algorithms to investigate the efficacy of boundary control in mixing optimization.  The conclusions are presented in Section \ref{conclusions}. 

\section{Boundary control design for optimal mixing\label{section2}}
Here, we briefly introduce the mathematical model and the first-order optimality conditions established in  \cite{hu2020approximating}.

\subsection{Optimization problem}
\label{sec_optimization_problem}
Consider a passive scalar field advected by an unsteady Stokes flow in an open bounded and connected  domain $\Omega\subset \mathbb{R}^{d}$, $d=2$, with a sufficiently smooth boundary $\Gamma$. The governing equations for the scalar density $\theta$, velocity $v$, and pressure $p$ are described  by  
  \begin{align}
      & \frac{\partial \theta}{\partial t}+ v\cdot \nabla \theta=0,  \label{EQ01}\\
&\frac{\partial v}{\partial t} - \Delta v+ \nabla p=0,   \label{Stokes1}\\
&\nabla \cdot v = 0,   \label{Stokes2}
  \end{align}
with the Navier slip boundary conditions  (cf.~\cite{kelliher2006navier, navier1823memoire}),
      \begin{align}
   v\cdot n|_{\Gamma}=0 \quad   \text{and}   \quad (2 n\cdot \mathbb{D}(v)\cdot \tau+k v\cdot \tau)|_{\Gamma}=g,\label{Stokes3}
    \end{align}
    and the initial condition
\begin{align}
(\theta(0), v(0))=(\theta_{0}, v_{0}). \label{ini}
\end{align} 
Here, $\mathbb{D}(v)=(1/2)(\nabla v+(\nabla v)^{T})$ is the strain rate tensor, and $n$ and $\tau$ are the outward unit normal and tangential vectors to the domain boundary $\Gamma$. 
The Navier slip boundary conditions allow the fluid to slip along the boundary  with resistance  under the tangential force and the friction between the fluid and the wall is proportional to $-v$ with the positive coefficient of proportionality  $k$. 
In this model, the boundary control input $g$ is specialized in the tangential direction, that is, $g\tau$ is the force exerted only in the tangential direction. Physically, this boundary condition can be regarded as a model in the tangential direction of the cilia beating  in the inner membrane of vertebrate organs, as described at the end of Section\,\ref{sec_examples}.

The notation $L^2(G)$ is used to denote the Lebesgue space of square integrable functions over a set $G$, and $H^s(G), s\geq 0,$ the subset of $L^2(G)$ of functions whose weak derivatives  up to order $s$ are also square integrable. Note $H^0(G)=L^2(G)$. Let 
$$V^{s}_{n}(\Omega)=\{ v \in H^{s}(\Omega)\colon \text{div}\ v=0,\  v\cdot n|_{\Gamma}=0\}, \quad \text{for} \ s\ge 0.$$
Throughout  this paper, we use $(\cdot, \cdot)$  and $\langle \cdot, \cdot \rangle_\Gamma$ for the $L^{2}$-inner products in 
the interior of the domain $\Omega$ and on the boundary $\Gamma$, respectively.

The objective in this work is to seek  a control input $g\in U_{ad}$ that minimizes the following cost functional at a given final time $T>0$:
\begin{align*}
J(g)=\frac{1}{2} \|\theta(T)\|^{2}_{(H^{1}(\Omega))'}
+\frac{\gamma}{2}  \| g\|^{2}_{U_{\text{ad}}}, 
\end{align*}
 subject to the PDE constraints \eqref{EQ01}--\eqref{ini}, where $\gamma>0$ is the control weight parameter and  $U_{ad}=L^2(0, T; L^2(\Gamma))$ 
 is the set of admissible controls equipped   with the norm $\|\cdot\|_{U_{ad}}$ given by 
\begin{equation}
||g||_{U_{ad}} 
=\left (\int_0^T \int_\Gamma   |g(x,t)|^2 dx dt\right)^{1/2}, \quad
\forall g\in U_{ad}.
\end{equation}
The choice of $U_{ad}$  is often determined based on the physical properties as well as the need   to guarantee the existence of an optimal solution. The detailed explanation   can be found in   \cite{hu2020approximating}.
In this work, we adopt  the dual norm $\|\cdot\|_{(H^{1}(\Omega))'}$ that quantifies  the weak convergence   as the mix-norm to quantify  mixing \cite{mathew2005multiscale, lin2011optimal, thiffeault2012using},   where   $(H^{1}(\Omega))'$ is the dual space of $H^{1}( \Omega)$. To make it explicit,  we define $f$ as the solution of 
\begin{align}
(-\Delta +I)f= \theta \quad \mbox{in } \Omega, \qquad & \frac{\partial f}{\partial n}= 0 \quad \mbox{on } \Gamma.  \label{eqn_adjointT}
\end{align}
Let 
\begin{equation*}
\Lambda=(-\Delta+I)^{1/2}. 
\end{equation*}
 Then  $\Lambda$ is a self-adjoint and positive operator. Thus  $f =\Lambda^{-2}\theta$ and 
 \begin{equation}
 \|\theta\|_{(H^{1}(\Omega))'}
= (\Lambda^{-1} \theta, \Lambda^{-1} \theta)^{1/2}
= (\Lambda^{-2} \theta, \theta)^{1/2}
=(f, \theta)^{1/2}. \label{mx_norm}
 \end{equation}
 %
We impose $\theta_0$ to be a spatially mean-zero function, that is,
$\bar\theta_0\triangleq \frac{1}{|\Omega|}\int_\Omega\theta_0(x)dx=0$.  Then when perfect mixing is achieved, the mix-norm is zero. 
This is the same treatment as in \cite{lin2011optimal}.
It is straightforward to show that  the spatial mean value of $\theta$ is time-invariant, i.e., $\bar{\theta}(t)=\bar{\theta}_0, \forall t>0$.
 
With the help of \eqref{eqn_adjointT}--\eqref{mx_norm}, $J$ can be rewritten as
\begin{align}
J(g)
=\frac{1}{2} (\Lambda^{-2}\theta(T), \theta(T))
+\frac{\gamma}{2}  \int^T_{0}\langle g, g\rangle_{\Gamma}\,dt
.
 \label{cost}
\end{align}
Note that the boundary control of the velocity field  gives rise to  a nonlinear control problem of the scalar equation, due to  the one-way coupling through the  advective  term $v\cdot \nabla \theta$, and therefore, the problem \eqref{cost} is non-convex.  The existence of an optimal solution  $g\in U_{ad}$   is proven  in  \cite{hu2020approximating}. Moreover, when $d=2$  and $\gamma$ is sufficiently large, the optimal solution is unique.

In this work, we set $v_0=0$ for simplicity. Since  the state variables $v$ and $\theta$ 
depend on  $g$,  we  use the notations 
\begin{equation}
v= v(g) \quad \text{and} \quad \theta=\theta(g). 
\end{equation}
Furthermore, we define the control-to-state operator  
\begin{equation}
L\colon g\in U_{ad}\mapsto v(g) \in L^2(0,T;V^0_n(\Omega) ),
\label{def_L}
\end{equation}
where $v(g)$ is solution of \eqref{Stokes1}--\eqref{ini} with inhomogeneous  boundary input $g$.
With the zero initial velocity condition,  it is easy to see that $L$ is a linear operator,  that is, 
\begin{equation}
L(\alpha_1 g_1+ \alpha_2 g_2) = \alpha_1 L(g_1) + \alpha_2 L(g_2), \quad
\forall \alpha_1, \alpha_2\in\mathbb{R}, \,
\forall g_1, g_2\in U_{ad}.
\label{L_linear}
\end{equation}
The detailed properties of $L$ are introduced in \cite{hu2018boundary, hu2020approximating}.

\subsection{ First-order necessary optimality conditions}

To solve the optimal control problem \eqref{cost},  we apply   a variational inequality  \cite{lions1971optimal},  that is,
if $g$ is an optimal solution, then
\begin{align}
DJ(g;\varphi)\geq 0, \quad  \forall \varphi\in U_{ad}, \label{var_ineq}
\end{align}
where  $DJ(g;\varphi) $ stands for the Gâteaux derivative of  $J$ with respect to $g$   in the direction $\varphi\in U_{ad}$. 
A rigorous  definition is given by
\begin{equation*}
DJ(g;\varphi)=\lim_{\delta\to 0}\frac{J(g+\delta \varphi)-J(g)}{\delta}
=\frac{dJ(g+\delta \varphi)}{d\delta}|_{\delta=0}, \quad   \forall \varphi\in U_{ad}.
\end{equation*}
If the limit exists for all $\varphi\in U_{ad}$, then $J$ is called Gâteaux differentiable at $g$.
The Riesz representation of the Gâteaux  derivative in  $U_{ad}$,  denoted as $DJ(g)$, which is the gradient of $J$ at $g$ \cite{hinze2008optimization}, satisfies 
\begin{align}
( DJ(g), \varphi)_{U_{ad}}
\triangleq 
DJ(g;\varphi)
 = \int_0^T  \gamma  \langle g, \varphi\rangle_{\Gamma} + (\theta(g) \nabla\rho(g), L\varphi) \,dt, 
 \quad \forall \varphi\in U_{ad},
\label{eqn_gradientJ}
\end{align}
where  $\rho(g)$ is the adjoint state satisfying 
 \begin{align}
 & \frac{ \partial}{\partial t} \rho 
  + v(g)\cdot \nabla \rho=0,  \label{adj}\\
  &\rho(T)=\Lambda^{-2}\theta(g)(T).
    \label{adj_final}
  \end{align}
Derivation of \eqref{eqn_gradientJ}--\eqref{adj_final} is briefly stated in Appendix\,\ref{appendix_derivation_Gateaux}.  Since  there are no local constraints on 
$U_{ad}$ \cite{lions1971optimal},   the first-order necessary optimality condition for $g$ to be a local minimizer is given by
\begin{align}
DJ(g) =0 \text{ in } U_{ad}.
\label{Ori_pt_cond}
 \end{align}

In addition, the following  relation  between $\theta$ and $\rho$ holds, which is proven in  Appendix\,\ref{proof_prop}
and is used to verify the numerical code as shown in Appendix\,\ref{sec_simple_check}.
\begin{prop}\label{lemma_invariance_innerprod}
For a fixed final time $T>0$ and $g\in U_{ad}$, let $\rho$ be the solution to the adjoint system  \eqref{adj}--\eqref{adj_final}.
Then  the quantity $\int_\Omega \rho(x,t) \theta(x,t)dx$ is invariant with respect to $t\in [0,T]$.
\end{prop}

In this work, the domain is a two dimensional unit disk, i.e.,  $\Omega=\{(x, y)\colon x^2+y^2<1\}$,
the terminal time is $T=1$,  the friction coefficient is  $k=0.5$, and the control weight is $\gamma=$1e-6. 
We adopt a scientific notation with `e' in many programming languages to denote a very large or small floating point number, such as  6.23e-5 for $6.23\times 10^{-5}$. 
The initial value of $\theta$ is $\theta_0=\sin(2\pi y)$ (Figure\,\ref{fig_alltypes} at $t=0$), the same as in \cite{mathew2007optimal}. The choice of control functions is discussed in Section\,\ref{sec_controlbasis}.


\section{Optimization algorithms}\label{comp}

\subsection{General optimization algorithm}
\label{sec_general_optimization_algo}
The gradient decent based optimization strategies  such as steepest descent method and  conjugate gradient method will  be used in solving the optimality conditions.  The fundamental idea used in this work is generating a sequence $g^n$, $n=0, 1, \cdots$ with a recursive relation 
\begin{equation}
g^{n+1} = g^n + \eta^n d^n,
\label{eqn_fundamental}
\end{equation}
where $d^n$ is a descent search direction of $J$ at $g^n$ (i.e.,  $DJ(g^n;d^n)<0$)  and $\eta^n\ge 0$ is a step length. 
The entire optimization process is outlined in Algorithm\,\ref{alg_overview}.
\begin{algorithm}
  \caption{General Optimization Algorithm for the Mixing Problem}
  \label{alg_overview}
  \begin{itemize}
    \item Input: mesh of size $h$,  initial guess $g^0$, control basis.
    \item Output: solution $g$.
    \item [1.] Compute and store velocity basis for the control basis (see Section\,\ref{sec_basis}).
    \item [2.] Optimization.
    For $n=0, 1, \cdots$, 
    \begin{enumerate}
    \item[(1)] If $g^n$ is a local minimizer, stop and output it as the solution.
    \item[(2)] Compute a descent search direction $d^n$ of $J$ at $g^n$ (see Section\,\ref{sec_descentdirection}).
    \item[(3)] Compute a step length $\eta^n$ in the direction $d^n$ (see Section\,\ref{sec_linesearch}), then $g^{n+1} = g^n + \eta^n d^n$.
    \end{enumerate}

  \end{itemize}
\end{algorithm}

A relay approach through a sequence of refined meshes is used to improve computational efficiency. That is, the optimization problem is first solved on a coarse mesh, whose solution is passed as the initial guess for the optimization process on a finer mesh. The scheme is described in Algorithm\,\ref{alg_relay}. In this work, we use three meshes with resolution $h=0.1, 0.05, 0.025$ in a unit disk domain.
\begin{algorithm}
  \caption{Relay Algorithm for the Mixing Problem}
  \label{alg_relay}
  \begin{itemize}
    \item [1.] Create a sequence of meshes of mesh size $h_1>h_2>\cdots$.
    \item [2.] Apply Algorithm\,\ref{alg_overview} on mesh $h_1$ with initial guess $g^0$ and denote the solution as $g_{h_1}$.
    \item [3.] Apply Algorithm\,\ref{alg_overview} on mesh $h_2$ with initial guess $g_{h_1}$ and denote the solution as $g_{h_2}$.
    \item [4.] Relay from mesh $h_2$ to mesh $h_3$, $\cdots$

  \end{itemize}
\end{algorithm}

\subsection{Control basis, velocity basis, and advection evolutions} 
\label{sec_basis}
\subsubsection{Finite dimensional control basis}
\label{sec_controlbasis}
We focus on a finite dimensional control space $U^M_{ad}\triangleq \text{span}\{ g^b_j\}^M_{j=1}$, where  $\{ g^b_1, \cdots, g^b_{M} \}\subseteq U_{ad}$ are linearly independent.  Therefore, any control $g\in U^M_{ad}$ can be written as 
\begin{equation}
g = \sum_{j=1}^M \alpha_j g^b_j. 
\label{def_g}
\end{equation}  

In this work, the control basis functions $g^b_j$ are built by time segmenting the elementary functions $1$,  $\cos(k\omega)$, $\sin(k\omega)$, where $k=1,2$ and  $\omega$ is the polar angle of the point $(x, y)$ on the unit circle.
The time segmentation is defined as follows. 
Let $N$ be the number of time segments, and 
$\Delta s=\frac{1}{N}$ is the uniform segment size. 
Define the time segmentation function $\chi^N_i(t)$ as
\begin{equation}
\chi^N_i(t)=
\left\{ 
\begin{array}{lc}
    1, \quad \text{if} \quad t\in ((i-1)\Delta s,\,i\Delta s), \\
0,\quad  \text{otherwise},
  \end{array}   
\right\},
\quad 
i=1, \cdots, N.
\end{equation}
A control basis function $g^b$ is one of above elementary functions multiplying a time segmentation function, that is, 
\begin{equation}
g^b(x,t) =  \chi^N_i(t) \cdot \text{ one of }
\{1, \cos(\omega), \sin(\omega), \cos(2\omega), \sin(2\omega) \}.
\end{equation}
The control basis functions generated by the same elementary function are called of the same Type. For example, Type 1 is the set of functions generated by multiplying 1 with time segmentation functions, Type $\cos(\omega)$ is generated by multiplying $\cos(\omega)$ with time segmentation functions, etc.
\subsubsection{Velocity basis: generation and storage}
\label{sec_velocitybasis}
Due to the linearity of the operator $L$ in \eqref{L_linear}, the velocity field generated by $g$ in \eqref{def_g} can be written as 
\begin{equation}
v(g) = L(g) = \sum^{M}_{j=1}\alpha_{j} L(g^b_{j}).
\label{LC_Lg}
\end{equation}
This linear relation produces a big advantage in computations: we only need to compute the velocity basis 
\begin{equation}
v^b_j = L(g^b_j), \quad j=1,\cdots, M,
\end{equation}
before the optimization process and store it in the computer hard drive. Whenever there is a need to compute $L(g)$, the formula \eqref{LC_Lg} will be used to compose the velocity for $g$ from the stored velocity basis. 
An iterative projection method with Taylor-Hood finite elements is employed to solve the unsteady  Stokes equations
 \eqref{Stokes1}--\eqref{ini}, where the details are given in Appendix\,\ref{section_stokes_solver}.

The linearity of the operator $L$ holds only when the initial velocity $v_0=0$. If $v_0\ne 0$,  we  denote the  velocity generated by $v_0$ and $g=0$ as $v^b_{v_0, g=0}$. Then the full solution $v$ can be written as $v=\sum_{j=1}^{M} \alpha_j v^b_j + v^b_{v_0, g=0}$. However,  in our numerical experiments, we restrict our discussion to  the cases with $v_0=0$. 

Limited by storage, every basis velocity is saved with a  not-too-small time step $\Delta t_{V}$, which is typically several folds of the time step used in the Stokes solver.
Denote $N_V=\frac{T}{\Delta t_V}$. Thus, there are $N_V+1$ moments of velocity storage in the time window $[0,1]$.  In other words, for each basis velocity $v^b_j$, $j=1, \cdots, M$, its values at time $t^i_V=i \Delta t_{V}$, $i=0, 1, \cdots, N_V$, are saved into files. In practice, we use $T=1$ and $\Delta t_V=0.01$, so $N_V+1=101$.

If the Navier-Stokes equations with the nonlinear convection are considered, then the relation between $v$ and $g$ is no longer linear even when $v_0$ is zero, where a solver for the Navier-Stokes equations has to be called to obtain $v(g)$ whenever $g$ changes. Therefore, the lineararity of unsteady Stokes equation saves a lot of the computation time.

\subsubsection{Evolution of advection equations with sparsely stored velocity data}
\label{sec_evolution_transport}
A discontinuous Galerkin (DG) method is employed to solve the advection equations for the density $\theta$ and its adjoint state $\rho$, where a brief introduction is given in Appendix\,\ref{sec_advection}. Due to the CFL condition \eqref{CFL_condition}, the time step of the DG method, $\Delta t_{DG}$, is generally far smaller than the velocity storage time step $\Delta t_V$, where $\Delta t_V$ is often 20 to 40 folds larger than $\Delta t_{DG}$. Therefore, the stored velocity data is sparse relative to the requirement of the DG evolution method. 
We use the embedding and interpolation scheme in Algorithm\,\ref{alg_evolution} to evolve $\theta$, where the one for $\rho$ is the similar.
\begin{algorithm}
\caption{Evolution of transport equation for $\theta$ (or $\rho$) with sparsely stored velocity data}
\label{alg_evolution}
\begin{itemize}
\item Input: control $g=\sum_{j=1}^M \alpha_j g^b_j$, initial value $\theta_0$, basis velocity data $v^b_j(t^i_V)$, $j=1,\cdots, M$ at time $t^i_V$, $i=0, 1, \cdots, N_V$.
Note: $N_V=\frac{T}{\Delta t_V}$.

\item Output: $\theta$ at time $t^i_V$, $i=0, 1, \cdots, N_V$.
\item Evolution: at time $t^i_V$, $i=0, 1, \cdots, N_V-1$, 
\begin{itemize}
\item [(1)] Compose velocity $v(g)$ at $t^i_V$ and $t^{i+1}_V$: $v(g)(t^i_V)=\sum_{j=1}^M \alpha_j v^b_j(t^i_V)$,  $v(g)(t^{i+1}_V)=\sum_{j=1}^M \alpha_j v^b_j(t^{i+1}_V)$.

\item [(2)] Compute $V_{\max} = \max( ||v(t^{i}_V)||_{max}, ||v(t^{i+1}_V)||_{max})$.

\item [(3)] Use the CFL condition  \eqref{CFL_condition} to compute a tentative DG time step $\widetilde{\Delta t}^i_{DG}=\frac{h\cdot \text{CFL}_{L^2}}{V_{\max}}$. To get an integer number of steps of evolution in the time interval $[t^i_V, t^{i+1}_V]$, we let $N_i=\ceil{\frac{\Delta t_V}{\widetilde{\Delta t}^i_{DG}}}$, the ceiling function of the time steps ratio. Afterwards, define $\Delta t^i_{DG}=\frac{\Delta t_V}{N_i}$.

\item [(4)]  Interpolate the velocity at any time $t\in [t^i_V, t^{i+1}_V]$, $v_I(t)$,  required by the DG method by  $v_I(t) = \frac{t^{i+1}_V - t }{\Delta t_V}\cdot v(g)(t^i_V) +  \frac{t- t^i_V}{\Delta t_V}\cdot v(g)(t^{i+1}_V)$.

\item [(5)] Use the DG method to evolve $\theta$ from $t^i_V$ to $t^{i+1}_V$ with time step size $\Delta t^i_{DG}$ and the interpolated velocity $v_I(t)$.
\end{itemize}
\end{itemize}
\end{algorithm}

The backward evolution of $\rho(t)$ from $t=T$ to $t=0$ through the advection equation \eqref{adj} can be reformulated to a forward evolution process by the following transformation. Let $s=T-t$  and   $\tilde\rho(s)=\rho(t)$ and $\tilde{v}(s)=-v(t)$. Then  $\tilde\rho$ satisfies 
\begin{eqnarray}
\frac{\partial\tilde\rho(s)}{\partial s} + \tilde{v}(s) \cdot \nabla \tilde\rho(s)=0, 
\quad  \tilde\rho(0)=\rho(T).
\label{eqn_rho_forward}
\end{eqnarray}

To evaluate the second integral in \eqref{eqn_gradsystemVF},
both $\theta$ and $\rho$ are stored at the same time moments as the velocity basis, that is, time $t^i_V$, $i=0, 1,\cdots, N_V$ as mentioned in Section\,\ref{sec_velocitybasis}.
It turns out the majority time of the entire optimization process is spent on the simulation of $\theta$ and $\rho$, because whenever there is a need to compute the cost functional or its gradient, the evolution of $\theta$ and/or $\rho$ will be computed. To balance the efficiency and accuracy, we choose to use a second order Runge-Kutta scheme in time for the transport equations and a second degree polynomial approximation for $\theta$ and $\rho$ in space.

\subsection{Computation of  the cost functional $J(g)$}
\label{sec_costJ}
The cascade \eqref{cascade_4relations} or the computation  from a control inpout to the cost functional is computed through Algorithm\,\ref{alg_cost}. 
\begin{algorithm}
\caption{Computation of cost $J(g)$}
\label{alg_cost}
\begin{itemize}
\item Input: control $g$, initial value $\theta_0$, basis velocity data $v^b_j$, $j=1, \cdots, M$.
\item Output: cost $J(g)$.
\item Steps:
\begin{itemize}
\item [(1)] Evolve $\theta$ with $g$, $\theta_0$, and the basis  velocity data by Algorithm\,\ref{alg_evolution} to obtain $\theta(T)$.
\item [(2)] Compute the adjoint state $\rho(T)$ from the Neumann elliptic problem \eqref{eqn_adjointT}. We use a continuous piecewise quadratic finite elementh method to solve this problem.
\item [(3)] Compute the cost $J(g)$ by computing the integrals in the first formula of \eqref{cost}.
\end{itemize}
\end{itemize}
\end{algorithm}

\subsection{Computation of the gradient $DJ(g)$}
With the choice of a finite control basis, the gradient $DJ(g)$ in $U^M_{ad}$ is also a linear combination of $g^b_j$'s,  i.e., 
\begin{equation}
DJ(g) = \sum_{j=1}^{M} DJ(g)_j \cdot g^b_j,
\quad DJ(g)_j\in\mathbb{R}.
\label{def_Jgradient}
\end{equation} 
Letting $\varphi=g^b_i$, $i=1, \cdots, M,$ in \eqref{eqn_gradientJ} (using the first equality),  we get the following linear system
\begin{align}
 \sum_{j=1}^{M} (g^b_i,g^b_j)_{U_{ad}}\cdot DJ(g)_j
 = DJ(g; g^b_i), 
 \quad i=1,\cdots,M.
 \label{eqn_gradsystem}
\end{align}
Let $G$ be the matrix $G_{ij}\triangleq (g^b_i,g^b_j)_{U_{ad}}$, $i,j=1,\cdots, M$,  and the vector 
$b=(DJ(g)_1, \cdots, DJ(g)_M)^T$.
Thereafter,  the norm $||DJ(g)||_{U_{ad}}$  is given by 
\begin{equation}
||DJ(g)||_{U_{ad}}
 = \sqrt{\left( \sum_{i=1}^{M} DJ(g)_i g^b_i,
 \sum_{j=1}^{M} D J(g)_j g^b_j \right)}
 = \sqrt{b^T G b}. \label{J_sqrt}
\end{equation}

\subsubsection{Finite Difference (FD) method}
A simple method of computing the directional derivative is a Finite Difference (FD) approximation \cite{Griva2009}:
\begin{equation}
DJ(g; \varphi) \approx
\frac{J(g+\delta\cdot \varphi) - J(g)}{\delta}, 
\end{equation}
where $\delta$ is a small scalar. 
In our numerical implementations, a typical value of $\delta$ is 1e-5 or 1e-4.
With this approach, the right side of linear system \eqref{eqn_gradsystem} is replaced by
\begin{align}
DJ(g;g^b_i) = 
\frac{J(g+\delta \cdot g^b_i) - J(g)}{\delta}, 
 \quad i=1,\cdots,M.
 \label{eqn_gradsystemFD}
\end{align}

\subsubsection{Variational Formula (VF) with adjoint system} 
\label{sec_VF}
This method uses the Variational Formula (VF) \eqref{eqn_gradientJ} (the second equality), where the right side of \eqref{eqn_gradsystem} becomes 
\begin{align}
DJ(g;g^b_i)
 = \gamma \int_0^T \langle g, g^b_i\rangle_{\Gamma}\,dt 
 + \int_0^T \left(  \theta(g) \nabla\rho(g), L(g^b_i)\right) dt, 
 \quad i=1,\cdots,M.
 \label{eqn_gradsystemVF}
\end{align}
The second integral in \eqref{eqn_gradsystemVF} is evaluated with the trapezoidal rule in each interval $[t^i_V, t^{i+1}_V]$ for $i=0, 1, \cdots, N_V-1$ by using the data of $\theta$, $\rho$, and $L(g^b_i)$. 
The entire VF scheme is stated in Algorithm\,\ref{alg_VF}.
\begin{algorithm}
\caption{VF (Variational Formula) method of computing $DJ(g)$}
\label{alg_VF}
\begin{itemize}
\item Input: control $g$, initial value $\theta_0$, basis velocity data $v^b_j$, $j=1, \cdots, M$.
\item Output: $DJ(g)$.
\item Steps:
\begin{itemize}
\item [(1)] Evolve $\theta$ from $t=0$ to $t=T$ with Algorithm\,\ref{alg_evolution}.
\item [(2)] Compute the adjoint state $\rho(T)$ from 
\eqref{adj_final}, that is, the Neumann elliptic problem \eqref{eqn_adjointT} with $\theta=\theta(T)$. 
\item [(3)] Evolve $\rho$  with Algorithm\,\ref{alg_evolution} by solving the system \eqref{eqn_rho_forward}.
\item [(4)] Compute $DJ(g)$ with equations \eqref{eqn_gradsystem} and \eqref{eqn_gradsystemVF}.
\end{itemize}
\end{itemize}
\end{algorithm}

\subsubsection{Comparison of VF and FD methods in 1-D control spaces}
\label{compare_VF_FD_1D}
The finite difference method requires to compute a forward evolution process for each basis function $g^b_i$, $i=1,\cdots,M$, in order to compute $J(g+\delta g^b_i)$. Plus another forward evolution of $\theta$ in $J(g)$, the FD scheme requires $M+1$ forward evolutions to compute $DJ(g)$.  In contrast, using the variational formula takes only two evolutions: one forward for $\theta$ and one backward for $\rho$. 
 In this sense, the VF method is more appealing when  $M$ is large. However, the VF method has much higher complexity: one elliptic solver for $\rho(T)$ and the integration of 
 $\int_0^T \left(  \theta(g)  \nabla\rho(g), L(g^b_i)\right) dt$.
Especially, the calculation of $\theta\nabla\rho$ involves the spatial  derivative of $\rho$, which has one less order accuracy than $\rho$ itself. In certain cases, it may result in too large errors.

\begin{figure}[htbp]
\includegraphics[scale=0.09]{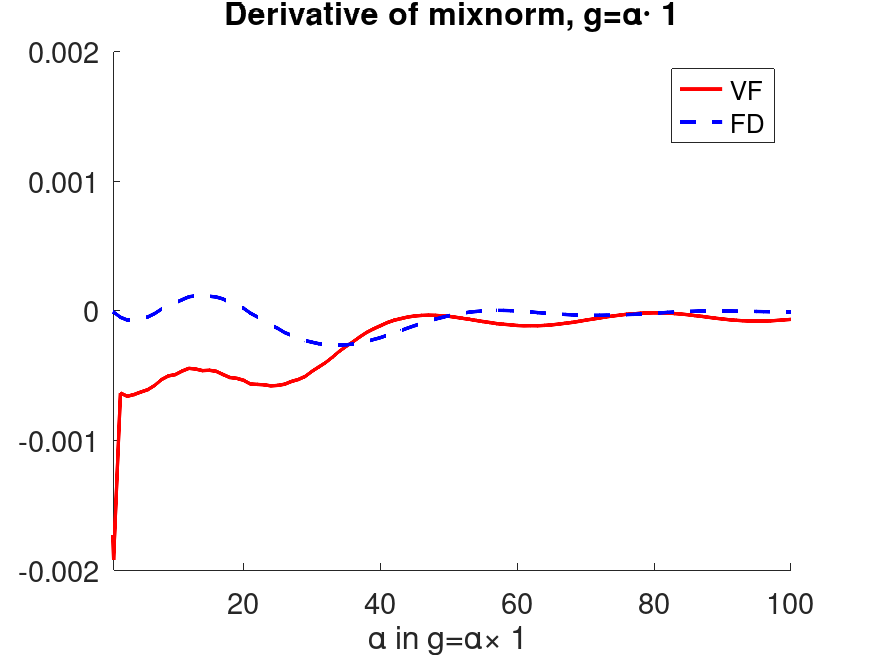}[a]
\includegraphics[scale=0.09]{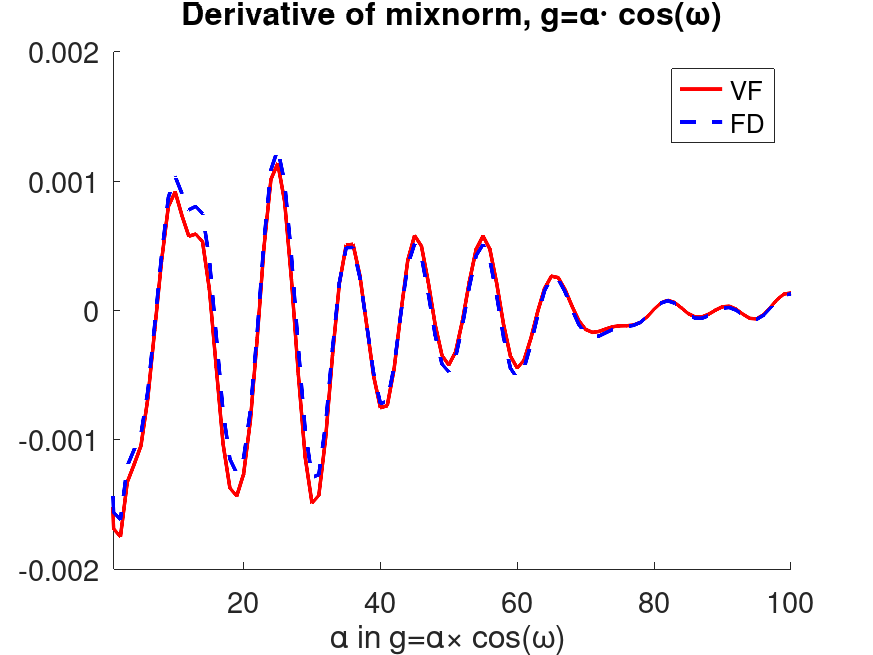}[b]
\includegraphics[scale=0.09]{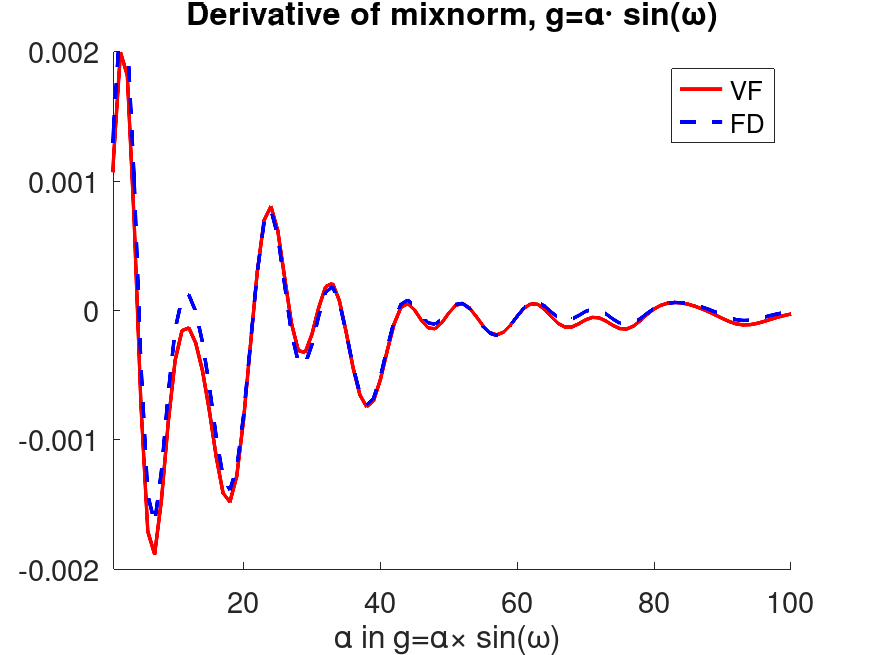}[c]
\includegraphics[scale=0.09]{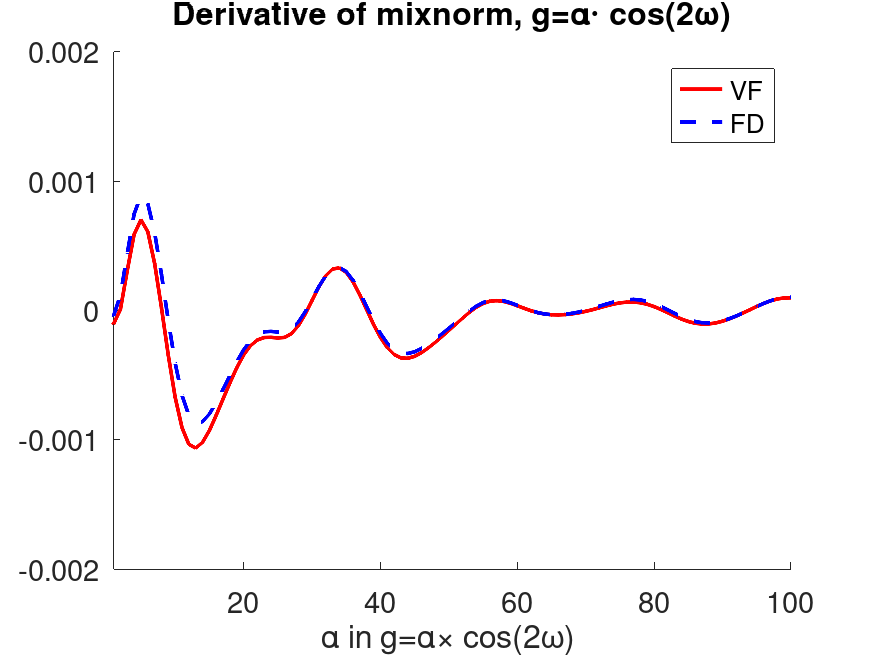}[d]
\includegraphics[scale=0.09]{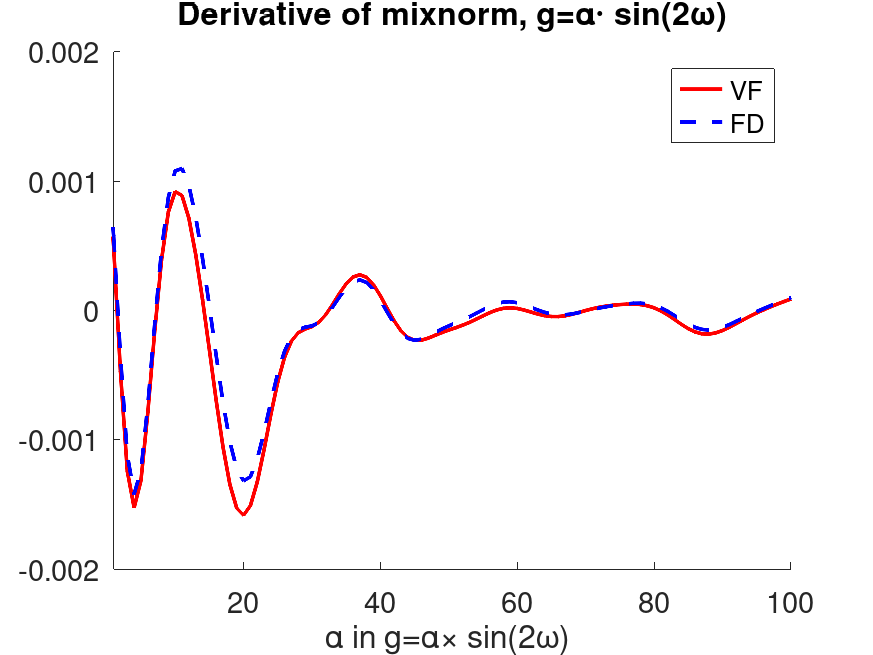}[e]
\caption{Computed derivatives of the mix-norm  $J_\theta(g)=\frac{1}{2} ||\theta(g)||^2_{(H^1(\Omega))'}$ by the VF method (red solid line) and the FD method (blue dashline) on mesh $h=0.1$  at $t=1$. 
The control $g=\alpha, \alpha \cos(\omega), \alpha \sin(\omega), \alpha \cos(2\omega), \alpha \sin(2\omega)$, $\alpha\in[0,100]$ from left to right.
\label{cat0981} 
}
\end{figure}

To compare the performance of the VF and FD methods, we give one experiment on the five elementary control functions used in this work: $g^b=1$, $\cos(\omega)$, $\sin(\omega)$, $\cos(2\omega)$, $\sin(2\omega)$. 
Because the mix-norm in the cost functional, $J_\theta(g)\triangleq \frac{1}{2} ||\theta(g)||^2_{(H^2(\Omega))'}$,  
is the only challenging part and the major source of error in the entire gradient  calculation, this experiment just focuses on this term. The derivatives of this term computed by these two methods are shown in Figure\,\ref{cat0981}, where the computations are taken for integer values of $\alpha\in [0,100]$ in $g=\alpha g^b$. Overall, both methods agree far better for the cosine and sine functions than the function 1. 
We denote $D_{VF} J_\theta(g)$ and $D_{FD} J_\theta(g)$ as the gradient  of $J_\theta(g)$ with VF and FD methods, respectively. Let the average absolulte error be 
$\text{AAE}(g^b)=\frac{1}{101}\sum_{\alpha=0}^{100}|D_{VF} J_\theta (\alpha g^b) - D_{FD} J_\theta(\alpha g^b)|$
and the average relative error be 
$\text{ARE}(g^b)=\frac{1}{101}\sum_{\alpha=0}^{100}|D_{VF} J_\theta (\alpha g^b) - D_{FD} J_\theta(\alpha g^b)| /|D_{FD} J_\theta(\alpha g^b)|$.
These two errors for these control basis functions are shown in Table\,\ref{Table0cat_0023}. We observe the  first-order  convergence of the average absolute errors when the mesh is refined, with the error of the control 1 is at least twice of the errors of other control basis functions. 
The average relative error is not a well-defined metric since it is not symmetric, so we cannot expect any convergence. However, it shows that the average relative error of the control 1 is far larger than those of other controls (at least 20 folds larger). 

\begin{table}[htbp]
\begin{center}
\caption{Errors of derivatives of the mix-norm $J_\theta(g)=\frac{1}{2} ||\theta(g)||^2_{(H^1(\Omega))'}$ by  VF and FD methods. AAE=Average Absolute Error, ARE=Average Relative Error.}
\label{Table0cat_0023}
\scriptsize
\begin{tabular}{|c|c|c|c|c|c||c|c|c|c|c|}
\hline
 & AAE & AAE & AAE & AAE & AAE & ARE & ARE & ARE & ARE & ARE \\
$h$ & 1  & $\cos(\omega)$  &  $\sin(\omega)$  &  $\cos(2\omega)$ 
 &  $\sin(2\omega)$ 
 & 1  & $\cos(\omega)$  &  $\sin(\omega)$  &  $\cos(2\omega)$ 
 &  $\sin(2\omega)$   \\
\hline
0.1  & 2.06e-4 & 6.47e-5 & 6.41e-5 & 4.36e-5 & 6.37e-5
     & 3.65e+1 & 2.83e-1 & 1.25e0  & 5.45e-1 & 6.88e-1 \\
0.05 & 8.01e-5 & 3.67e-5 & 2.58e-5 & 1.55e-5 & 2.81e-5 
     & 2.38e+2 & 2.05-e1 & 1.33e0  & 2.77e-1 & 6.71e-1  \\
0.025& 3.59e-5 & 1.69e-5 & 1.19e-5 & 7.77e-6 & 1.11e-5 
     & 2.33e+1 & 1.16e-1 & 7.05e-1 & 2.19e-1 & 6.74e-1 \\
\hline
\end{tabular}
\end{center}
\end{table}

\subsubsection{Comparison of VF and FD methods in 2-D control spaces}
We further explore the different performance between VF and FD methods in two tests where in each test, the control space is spanned by two time-segmented basis functions. We denote  $g=\alpha_1 g^b_1+\alpha_2 g^b_2$ and  $\alpha=(\alpha_1, \alpha_2)$. Here, we test on the whole gradient  where $\gamma=$1e-6.

\begin{figure}[htbp]
\begin{center}
\includegraphics[scale=0.3]{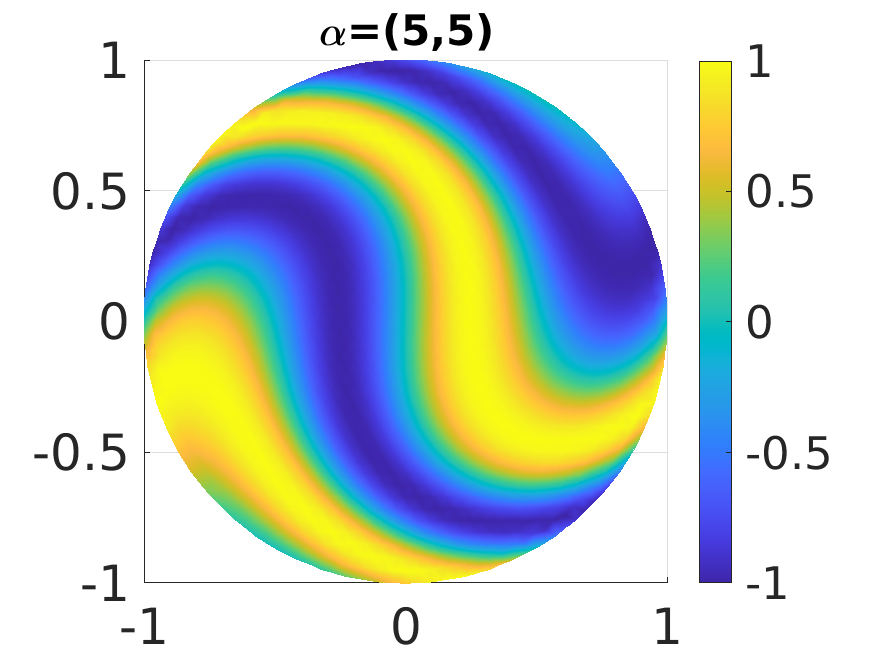} [a]
\includegraphics[scale=0.3]{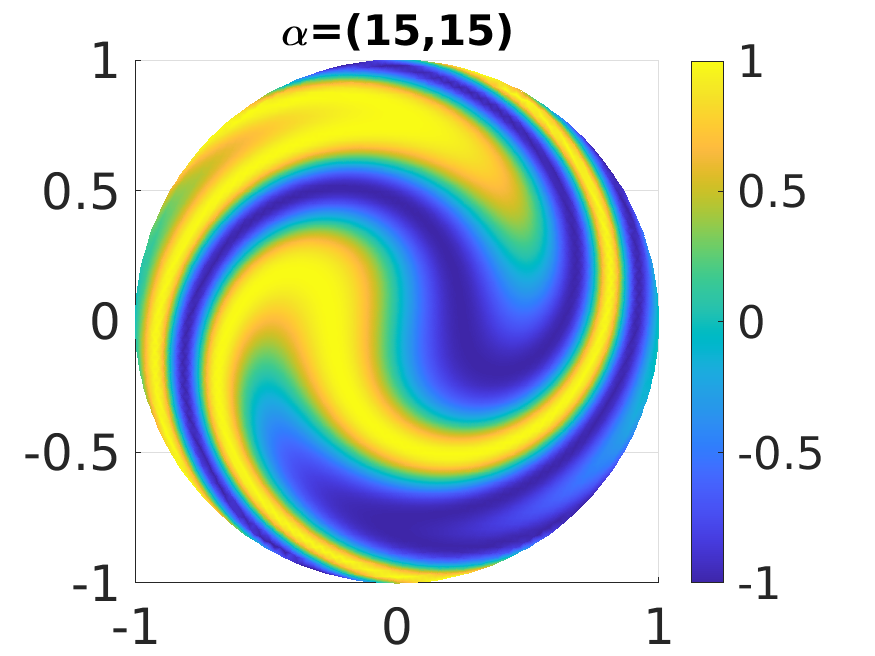} [b]
\includegraphics[scale=0.3]{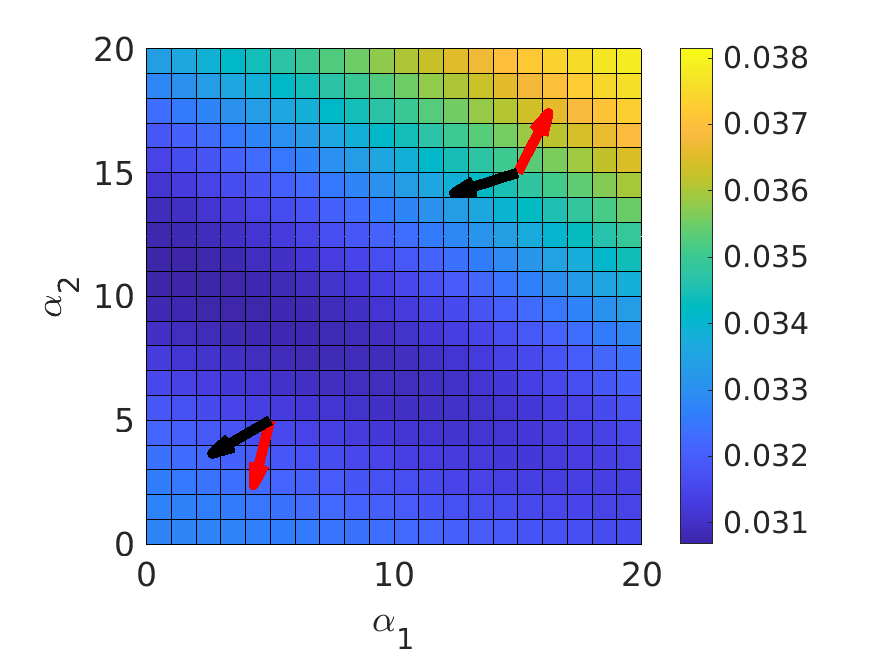}[c]
\caption{
[a]: $\theta(T)$ when $\alpha=(5,5)$. 
[b]: $\theta(T)$ when $\alpha=(15,15)$. 
[c]: cost map $J(g)$ when the mesh size $h=0.1$  and $\gamma=$1e-6. The red vectors are the derivatives from the FD method, and the black vectors are the derivatives from the VF method. The  vectors are scaled to have the same length.}
\label{fig_comp_VA_FD_cat001}

\end{center}
\end{figure}

In the first test,  $g^b_1=1_{[0,0.5]}$ (1 when $t\in [0,0.5]$ and 0 when $t\in (0.5, 1]$) and $g^b_2=1_{[0.5, 1]}$. 
The results are shown in Table\,\ref{Table01_comp_VF_FD_k1e-6_2D}. In this table, the FD method gives consistent approximations when the mesh is refined. The FD results are also consistent when some different $\delta=$1e-5, 1e-4, 1e-3 values are used in \eqref{eqn_gradsystemFD} (data not shown). This suggests the FD results are more reliable when the exact derivative is unknown. 
The VF results have huge relative errors compared with those of the FD method and they even have opposite directions when $\alpha=(15,15)$ and $h=0.1$ (see Figure \,\ref{fig_comp_VA_FD_cat001},  the VF derivative at $(15,15)$). The correctness of the  directional derivative from the FD method can be verified in Figure\,\ref{fig_comp_VA_FD_cat001} by checking with the cost map.  
The cost map is the colored plot of the costs computed on integer points of $\alpha=(\alpha_1, \alpha_2)$. Therefore, the VA result in this case does not give a descent direction.  The wrong  directional derivative  is catastrophic in the optimization method used in this work because the line search fails with a non-descent search direction.
\begin{table}[htbp]
\begin{center}
\caption{Gradient approximated by VF and FD  methods.
The control $g=\alpha_1 1_{[0,0.5]}+ \alpha_2 1_{[0.5,1]}$.  }
\label{Table01_comp_VF_FD_k1e-6_2D}
\scriptsize
\begin{tabular}{|c|c|c|c||c|c|c|}
\hline
 & $\alpha=(15,15)$ &  $\alpha=(15,15)$ &  relative
 & $\alpha=(5,5)$ &  $\alpha=(5,5)$ & relative \\
$h$ & VF & FD & error & VF & FD & error\\
\hline
0.1   & (-6.67e-4, -2.13e-4)  
      & (8.49e-5, 1.64e-4)
      & 455\%
      & (-7.68e-4, -4.45e-4) 
      & (-1.98e-5, -7.79e-5)
      & 1036\%
       \\
\hline
0.05  & (-2.37e-4, 1.62e-6)  
      & (8.96e-5, 1.65e-4)
      & 194\%
      & (-3.46e-4, -2.40e-4)
      & (-2.76e-5, -8.18e-5)
      & 412\%
     \\
\hline
0.025 & (-6.09e-5, 9.09e-5) 
      &(8.94e-5, 1.65e-4)
      & 90\%
      &(-1.74e-4, -1.55e-4)  
      & (-2.77e-5, -8.17e-5)  
      & 190\%
       \\
\hline
\end{tabular}
\end{center}
\end{table}

\begin{figure}[htbp]
\begin{center}
\includegraphics[scale=0.3]{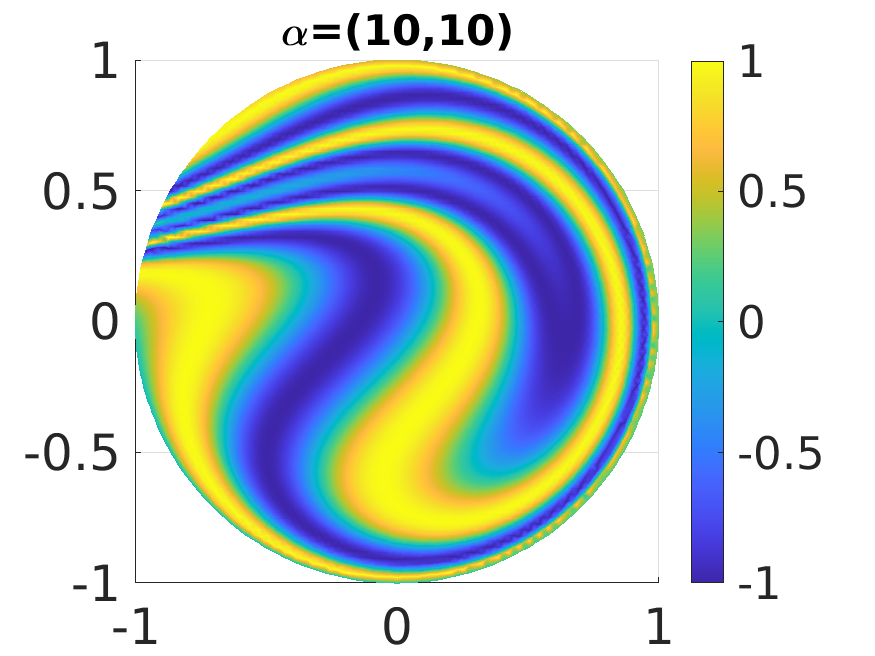} [a]
\includegraphics[scale=0.3]{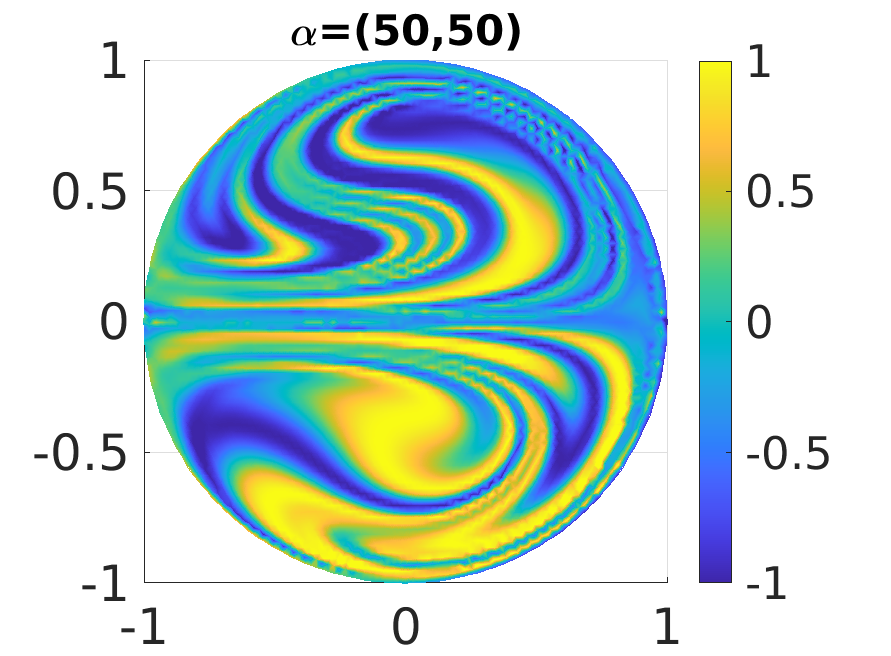} [b]
\includegraphics[scale=0.3]{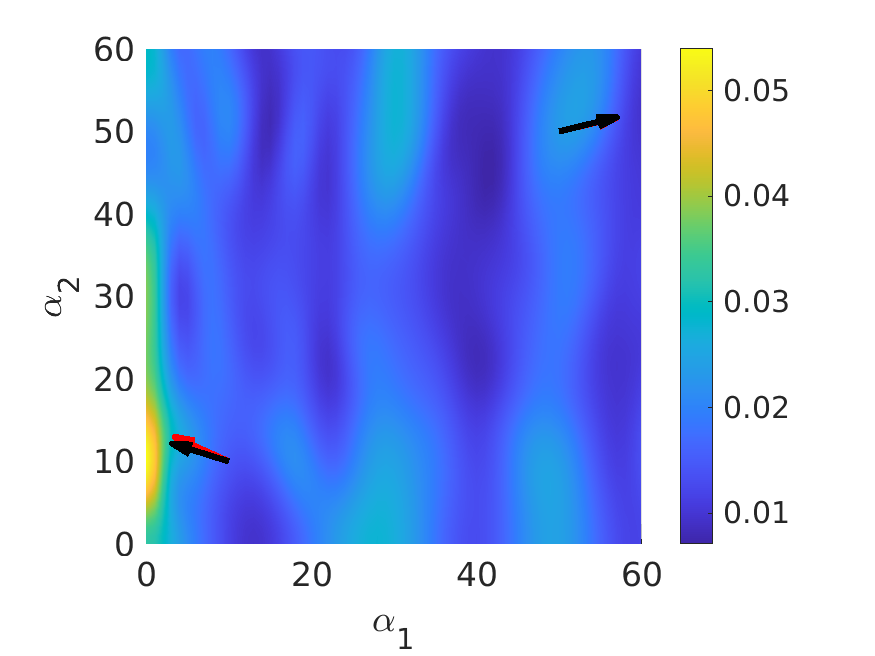} [c]
\caption{
[a]: $\theta(T)$ when $\alpha=(10,10)$.
[b]: $\theta(T)$ when $\alpha=(50,50)$.
[c]: cost map when the mesh size $h=0.1$  and $\gamma=$1e-6 and the computed derivative $DJ(g)$ at $\alpha=(10, 10)$ and $(50,50)$. 
The red vectors are from the FD method and the black vectors from the VF method. At $\alpha=(50,50)$, the vectors from the two methods are indistinguishable by eyes. The vectors are scaled to have the same length. 
}
\label{fig_comp_VA_FD_cat002}
\end{center}
\end{figure}

The second test is given to $g^b_1=\cos(\omega)\cdot 1_{[0,0.5]}$ and 
$g^b_2=\sin(\omega) \cdot 1_{[0.5,1]}$. From the results shown in Table\,\ref{Table03_comp_VF_FD_k1e-6}, the VF and FD methods are very close. The morphologies of $\theta$ at $t=T$ corresponding to two different $\alpha$ values are shown in Figure\,\ref{fig_comp_VA_FD_cat002}. 
Similar observatons are obtained when the control bases are 
$\cos(2\omega)$ and $\sin(2\omega)$ and their time segmentations (results not shown).
\begin{table}[htbp]
\begin{center}
\caption{Gradient approximated by VF  and FD  methods.
The control $g=\alpha_1 \cos(\omega)\cdot 1_{[0,0.5]} + \alpha_2 \sin(\omega)\cdot 1_{[0.5,1]}$. 
}
\label{Table03_comp_VF_FD_k1e-6}
\scriptsize
\begin{tabular}{|c|c|c|c||c|c|c|}
\hline
 & $\alpha=(50,50)$ &  $\alpha=(50,50)$ &  relative
 & $\alpha=(10,10)$ &  $\alpha=(10,10)$ & relative \\
$h$ & VF & FD & error & VF & FD & error\\
\hline
0.1   & (3.79e-4, 9.31e-5)    
      & (5.11e-4, 1.22e-4)     
      & 26\%
      & (-6.99e-4, 2.20e-4)     
      & (-6.27e-4, 2.84e-4)  
      & 14\%
       \\
\hline
0.05  & (6.31e-4, 1.39e-4)      
	  & (6.55e-4, 1.41e-4)      
	  & 3.59\%
      &  (-6.51e-4, 2.44e-4)     
      & (-6.20e-4, 2.87e-4)  
      & 7.9\%
     \\
\hline
0.025 &(6.42e-4, 1.48e-4)      
      & (6.27e-4, 1.42e-4)      
      & 2.51\%
      & (-6.33e-4, 2.73e-4)    
      & (-6.20e-4, 2.86e-4) 
      & 2.69\%
       \\
\hline
\end{tabular}
\end{center}
\end{table}

\subsubsection{A hybrid approach}
\label{sec_hybrid}
Because of the dramatically different performance of the VF method on Type 1 controls and other types of controls ($\cos(\omega)$, $\sin(\omega)$, $\cos(2\omega)$, $\sin(2\omega)$), we adopt an ad hoc hybrid approach: using the FD method to compute directional derivatives for Type 1 controls and the VA method for other types.  
That is, in \eqref{eqn_gradsystem}, 
\begin{eqnarray}
DJ(g;g^b_i) = 
 \left\{
 \begin{array}{cc}
\frac{J(g+\delta \cdot g^b_i) - J(g)}{\delta}, 
& g^b_i \in \text{  Type 1 }; \\
\gamma \int_0^T \langle g, g^b_i\rangle_{\Gamma}\,dt 
 + \int_0^T \left(  \theta(g) \nabla\rho(g), L(g^b_i)\right) dt, 
& g^b_i \in \text{ other types},
 \end{array}
 \right.
 \label{eqn_gradsystem_hybrid}
\end{eqnarray}

\subsubsection{Summary of numerical methods for computing $DJ(g)$ }
A summary of these three computation methods for the Gâteaux   derivative is given in Table\,\ref{table_summary_DJ}. Note that the derivative of the mix-norm is the only computationally demanding part and the main source of error.
\begin{table}[htbp]
\scriptsize
\begin{center}
\caption{Comparison of different methods for computing $DJ(g)$. $M$ is the dimension of control basis and $M_{Type 1}$ is the number of Type 1 control basis functions. 
\label{table_summary_DJ}}
\begin{tabular}{|L{3cm}|L{4cm}|L{3cm}|L{5cm}|}
\hline
Method & VF (Variational Formula) & FD (Finite Difference) 
& Hybrid (FD for Type 1 and VF for Type 2)\\
\hline
Evolution of $\theta$ and/or $\rho$ & 2  & $M+1$  & $3+M_{Type 1}$ \\
\hline
Other computations &  1 elliptic solver, $M$ integrals of $\int_0^T \left(  \theta(g)  \nabla\rho(g), L(g^b_i)\right) dt$ 
& none &  1 elliptic solver, $M-M_{Type 1}$ integrals of $\int_0^T \left(  \theta(g)  \nabla\rho(g), L(g^b_i)\right) dt$ \\
\hline
Accuracy in mix-norm derivative & poor in Type 1, accurate in other types
& accurate in all types & accurate in all types\\
\hline
\end{tabular}
\end{center}
\end{table}

\subsection{Line search methods: computation of step size $\eta^n$}
\label{sec_linesearch}
Here we discuss some line search methods in finding the step size $\eta^n$ in the update formula
\eqref{eqn_fundamental}.

\subsubsection{Backtracking method and Armijo condition}
The backtracking technique (e.g., \cite{Griva2009}) is finding $\eta^n$ such that it is the first value in the sequence 
\begin{equation}
\{ \eta^n_i=\frac{\epsilon_b}{2^i}:  i=0, 1, \cdots \}
\label{beta_epsilon}
\end{equation}
satisfying the following sufficient descent condition (also called Armijo condition), 
\begin{equation*}
J(g^n+\eta^n d^n) \le  J(g^n) + \eta^n\cdot  \mu\cdot  
DJ(g^n;d^n), 
\end{equation*}
where $\epsilon^n_b, \mu$ are positive constants. This method only guarantees the sufficient descent, not a local minimizer. Thus, it does not produce an exact line search.  The value of the first value $\epsilon^n_b$ is empirically determined and in our work, the values between 1 and 8 are good candidates when $d$ is a unit vector in the $U_{ad}-$norm. The parameter $\mu\in (0,1)$ according to \cite{Griva2009} and we use $\mu=0.3$.
The backtracking scheme is summarized in Algorithm\,\ref{alg_backtracking}.

\begin{algorithm}
\caption{Backtracking line search with Armijo condition}
\label{alg_backtracking}
\begin{itemize}
\item Input: control $g^n$, search direction $d^n$, $J(g^n)$,  parameters $\mu$, $\epsilon_b$, 
$\text{ back\_MAXITER}$.
\item Output: $\eta^n$, $g^{n+1}$, $J(g^{n+1})$.
\item Steps
\begin{itemize}
\item  [(1)] If $d^n$ is not a descent direction (that is, $DJ(g^n;d^n)\ge 0$), then stop and report problem.

\item [(2)] Backtracking iteration. 
For $i=1, 2, \cdots, \text{ back\_MAXITER}$, 
\begin{enumerate}
\item $\eta^n_i =\epsilon_b/2^{i-1}$.
\item $g^{n+1}_i=g^n + \eta^n_i d^n$.
\item Compute the cost $J(g^{n+1}_i)$ by using Algorithm\,\ref{alg_cost} with input $g^{n+1}_i$.
\item If $J(g^{n+1}_i) \le  J(g^n) + \eta^n_i \mu  DJ(g^n;d^n)$, stop and return $\eta^n=\eta^n_i$, 
$g^{n+1}=g^{n+1}_i$, and $J(g^{n+1})$.
\end{enumerate}
\end{itemize}
\end{itemize}
\end{algorithm}

In this work, the backtracking method is typically combined with the steepest descent method, where $d^n=-\nabla J(g^n)$. Therefore, the Armijo formula for the steepest descent method becomes 
\begin{equation}
J(g^n - \eta^n \nabla J(g^n)) \le 
 J(g^n) - \eta^n \mu ||D J(g^n)||^2_{U_{ad}}. \label{sufficientdescent}
\end{equation}

\subsubsection{Exact line search:  a coupled bisection-secant method}
\label{sec_exactlinesearch}
In some optimization methods, an exact line search is needed, such as in the conjugate gradient method,  to  guarantee the new search direction $d^{n+1}$ is a descent direction (see Section\,\ref{sec_CG} Equation \eqref{cat0020}). 
That is, $\eta^n$ is a minimizer of 
\begin{equation}
\min_{\eta\ge 0} f(\eta) \triangleq J(g^n+ \eta d^n).
\label{line_search_problem}
\end{equation}
Note $f'(\eta) = DJ(g^n+\eta d^n; d^n)$.

To get an exact solution $\eta$ of \eqref{line_search_problem}, we use a coupled bisection and secant method. The strategy is first finding an interval $[0, \eta_1]$, as small as possible, where $f'(0)<0$ and $f'(\eta_1)>0$, and then searching for a root of $f'(\eta)=0$ in this interval. The condition $f'(0)<0$ is equivalent to that $d^n$ is a descent direction of $J$ at $g^n$. Because $\lim_{||g||\to \infty} J(g)=\infty$,  a value $\eta^n_1$ satisfying $f'(\eta_1)>0$ must exist. To find $\eta_1$, we adopt a forward tracking process as shown in Algorithm\,\ref{alg_exactlinesearch} Step 2. When $f'(0)<0$ and $f'(\eta_1)>0$, there exists a root of $f'(\eta)=0$ in $(0,\eta_1)$ with the continuity assumption of $f'$. To find a root,  we first use several steps of bisection method in order to reduce the search interval size, defined by the distance between the last two  bisection solutions ($|\eta^n_{BIS-1} - \eta^n_{BIS}|$), sufficiently small. This is important to the secant method that has faster convergence but requires that the initial guess values are sufficiently close to the exact root. 
The details of the bisection and secant methods of finding a root of a nonlinear function can be found, e.g, in \cite{Burden2016}. The whole exact line search scheme is briefly described in Algorithm\,\ref{alg_exactlinesearch}.
\begin{algorithm}
\caption{Exact line search with bisection-secant method}
\label{alg_exactlinesearch}
\begin{itemize}
\item Aim: finding a root of $f'(\eta)=0$ in an interval $[0, \eta_1]$ where $f'(0)<0$ and $f'(\eta_1)>0$. The value of $\eta_1$ will be found in this algorithm.
\item Input: control $g^n$, search direction $d^n$, $J(g^n)$,  parameter $\epsilon_{bisection}$, 
$\epsilon_{secant}$.
\item Output: $\eta^n$, $g^{n+1}$, $J(g^{n+1})$.
\item Steps
\begin{itemize}
\item [(1)] If $d^n$ is not a descent direction (that is, $DJ(g^n;d^n)\ge 0$), then stop and report problem.

\item [(2)] Find an $\eta^n_1>0$ such that $f'(\eta^n_1)>0$. This is done by a forward tracking process: $\eta^n_1$ is the first value of the sequence $\eta=\{1, 2, 2^2, \cdots \}$ that satisfies $f'(\eta)>0$. 

\item [(3)] Apply the bisection method of finding a root of $f'(\eta)=0$ in $[0, \eta^n_1]$ and stop 
when $|\eta^n_{BIS-1} - \eta^n_{BIS}|\le \epsilon_{bisection}$. Here, $\eta^n_{BIS-1}, \eta^n_{BIS}$ are the last two values of bisection solution. In practice, we use $\epsilon_{bisection}=1$.

\item [(4)] Apply the secant method of finding a root of $f'(\eta)=0$ with the initial values as $\eta^n_{BIS-1},  \eta^n_{BIS}$. Stop when $|DJ(g^n+\eta d^n; d^n)|<\epsilon_{secant}$ and return $\eta^n=\eta$, 
$g^{n+1}=g^n+\eta d^n$, and $J(g^{n+1})$.
In practice, we choose $\epsilon_{secant}=$1e-10.

\end{itemize}
\end{itemize}
\end{algorithm}

The forward tracking of finding $\eta_1$, bisection, and secant are all iterative and in each iteration,  the directional derivative $DJ(g^n+\eta^n_i d^n; d^n)$ is computed for an iterative index $i$.  Because this derivative is only in one direction $d^n$,  we adopt the FD method which uses two  evolutions of $\theta$, one for $J(g^n+\eta^n_i d^n)$, one for $J(g^n+ (\eta^n_i+\delta) d^n)$. This is simpler than the VF method (see comparisons in Table\,\ref{table_summary_DJ} when $M=1$). If $N_{exact}$ steps are used in the whole algorithm, then there are $2N_{exact}$ evolutions. From our experience, this whole process of the exact line search takes about 8 iterations on average, which is about 16 evolutions of $\theta$. This is far more expensive than the backtracking method which uses only 2 evolutions on average in each line search. 

To guarantee the step size $\eta^n$ is a local minimizer instead of a local maximizer or saddle point, the interval $[0, \eta_1]$ should be small enough such that it does not contain any other roots of $f'$. But it is difficult to actualize it because it is too time consuming to find all the roots  in this interval. Fortunately,  among over  thousands of exact line searches in this work, we only find only one case where the step size increases the cost value. Therefore, we claim this method ``almost guarantees descent''.

\subsubsection{Linearization method}
\label{sec_linearization}
A linearization process has been proposed in  \cite{Glowinski2022} to approximate the step size in the line search in an optimal control problem subject to an reaction-advection-diffusion system. This motivates us to develop a similar approach. 
We first linearize the relation between $\theta$ and $g^n+\eta d^n$ as 
\begin{equation}
\theta(g^n+\eta d^n) \approx \theta(g^n) + \eta\cdot
D\theta(g^n; d^n)
\end{equation}
and denote $z\triangleq D\theta(g^n; d^n)$.
Then the objective function $J(g^n+\eta d^n)$ is replaced by
the linearized version
\begin{equation}
J_{L}(g^n+\eta d^n) =  \frac{1}{2} (\Lambda^{-2} 
(\theta(g^n)+\eta z), \theta(g^n)+\eta z) (T)
 + \frac{\gamma}{2} \int_0^T 
 \langle g^n+\eta d^n, g^n+\eta d^n \rangle_\Gamma dt.
\end{equation}
Its derivative on $\eta$ is
\begin{eqnarray}
DJ_L(g^n+\eta d^n; d^n)
= ( \Lambda^{-2} (\theta(g^n) + \eta z) , z )(T)
+ \gamma \int_0^T \langle g^n+\eta d^n, d^n \rangle_\Gamma dt.
\end{eqnarray}
Letting it be zero, we get the critical value 
\begin{eqnarray}
\eta^n = - \frac{(\Lambda^{-2} \theta(g^n), z)(T) + 
\gamma \int_0^T \langle g^n, d^n\rangle _\Gamma dt}
{(\Lambda^{-2}z,z)(T) + \gamma \int_0^T\langle d^n,d^n\rangle_\Gamma dt} 
= - \frac{(DJ(g^n), d^n)_{U_{ad}}}
{(\Lambda^{-2}z,z)(T) + \gamma \int_0^T\langle d^n,d^n\rangle_\Gamma dt}.
\label{eta_eqn}
\end{eqnarray}
To determine $z$, we take the Gâteaux  derivative on the equation \eqref{EQ01} and the initial value \eqref{ini}  and obtain 
\begin{eqnarray}
\frac{\partial z}{\partial t}
+ v(g^n)\cdot \nabla  z
+ v(d^n)\cdot \nabla \theta(g^n) &=&0, \label{z_eqn}\\
z(t=0)&=&0. \label{z_init}
\end{eqnarray}
Note $v(g^n)=L(g^n)$ and $v(d^n)=L(d^n)$.
Thus, to evaluate $\eta^n$, we first evolve  $z$ with \eqref{z_eqn}--\eqref{z_init} and then compute it from \eqref{eta_eqn}.

In this method, the product $\eta^n d^n$ is scale invariant, i.e., if $d^n$ multiplies a positive number $a$, then $\eta^n$ value will be decreased by $a$. Indeed, if $d^n$ is increased by $a$ folds, then $z$ will be also increased by $a$ folds (because $z$ is linear on $d^n$ in \eqref{z_eqn}), and then $\eta^n$ in \eqref{eta_eqn} will be decreased by $a$.

There are two issues with this linearization methods based on our numerical tests. First, the step sizes obtained by this method are often ten to a few hundred times smaller than those  computed by the backtracking and  exact line search methods, which makes this method very inefficient. 
Second, when this method is combined with  the conjugate gradient method, the cost value often increases. 
This is because the combined method cannot guarantee that the new search direction is descent, that is, $DJ(g^{n+1};d^{n+1})<0$.  Indeed,  in the calculation in \eqref{cat0019} and \eqref{cat0020}, $DJ(g^{n+1};d^n)$ is not guaranteed to be zero. Instead,  $DJ_L(g^{n+1};d^n)$ is zero in this linearization method due to the choice of $\eta^n$ in \eqref{eta_eqn}.  
That is, $\eta^n$ is a local minimizer of $J_L(g^n+\eta d^n)$, instead of $J(g^n+\eta d^n)$.
Due to the nature of linearization, this method should provide a good approximation of the exact line search only when the exact step size is sufficiently close to zero, which is not often the case. Therefore, this method is not used in our work.

\subsubsection{Summary of line search methods}
Table\,\ref{table_summary_LineSearch} summarizes the performance of these line search methods based on the simulations of this work. The linearization method is not used extensively in this work due to its low efficiency. We mainly use the backtracking and exact line search methods. 
\begin{table}[htbp]
\scriptsize
\begin{center}
\caption{Comparison of line search methods of computing the step size $\eta$.
\label{table_summary_LineSearch}}
\begin{tabular}{|L{2cm}|L{4cm}|L{4cm}|L{4cm}|}
\hline
Method & Backtracking& Exact line search
& Linearization\\
\hline
Evolutions of transport equations & 2 to 3 on average & 15 on average & 1 \\
\hline
Guarantee descent? & yes & almost yes & no \\
\hline
Exact local minimizer? & no & yes & no \\
\hline
Comments & mainly used with steepest descent method 
& mainly used with conjugate gradient method 
& low efficiency: solution $\eta$ is often too small. Not used in this work. \\
\hline
\end{tabular}
\end{center}
\end{table}

\subsection{Optimization methods: choices of descent direction $d^n$}
This section describes the implementation of the General Optimization Algorithm \ref{alg_overview} with specific choice of the descent direction $d^n$, one being the negative derivative and one being the conjugate gradient direction. 
\label{sec_descentdirection}

\subsubsection{Steepest descent (SD) method}
The steepest descent method uses the negative Gâteaux  derivative as the descent search direction, i.e., 
$d^n=-DJ(g^n)$.  This method is described in Algorithm\,\ref{alg_steepestdescent}. 
\begin{algorithm}
\caption{Steepest descent method}
\label{alg_steepestdescent}
\begin{itemize}
\item Input: initial control $g^0$, maximum iteration number $\text{MAXITER}$, stopping criterion $\epsilon$, 
\item Output: a local minimizer of $J$.
\item Before iteration: compute $J(g^0)$.
\item For $n=0, 1, \cdots, \text{MAXITER}$, 
\begin{itemize}
\item [(1)] Compute $DJ(g^n)$ with FD or VF or Hybrid method.
\item [(2)] If $||DJ(g^n)||_{U_{ad}}/(1+J(g^n))<\epsilon$, stop and output $g^n$ as a local minimizer.
\item [(3)] Let $d^n=-DJ(g^n)$.
\item [(4)] Use a line search method with $g^n$ and $d^n$ to compute $\eta^n$ and then 
obtain $g^{n+1}=g^n+\eta^n d^n$ and $J(g^{n+1})$. 
\end{itemize}
\end{itemize}
\end{algorithm}
Through trials, we find the exact line search applied to the SD method not only requires many evolutions in each line search, but also takes many steepest descent steps to converge. Therefore, we will only use backtracking with steepest descent method.
In this work, we set $\text{MAXITER}=10000$ and $\epsilon=$1e-5 for both steepest descent and conjugate gradient methods in most cases.

\subsubsection{Conjugate gradient (CG) method}
\label{sec_CG}
The conjugate gradient method (e.g. \cite[section 13.4]{Griva2009}) is widely used in optimization and its  application in this work is given in Algorithm\,\ref{alg_CG}. 
\begin{algorithm}
\caption{Conjugate gradient method}
\label{alg_CG}
\begin{itemize}
\item Input: initial control $g^0$, maximum iteration number $\text{MAXITER}$, stopping criterion $\epsilon$, 
\item Output: a local minimizer of $J$.
\item Before iteration: compute $J(g^0)$, $DJ(g^0)$ and let $d^n=-DJ(g^0)$.
\item For $n=0, 1, \cdots, \text{MAXITER}$, 
\begin{itemize}
\item [(1)] If $||DJ(g^n)||_{U_{ad}}/(1+J(g^n))<\epsilon$, stop and output $g^n$ as a local minimizer.
\item [(2)] Use the exact line search Algorithm\,\ref{alg_exactlinesearch} with $g^n$ and $d^n$ to compute $\eta^n$ and then 
obtain $g^{n+1}=g^n+\eta^n d^n$ and $J(g^{n+1})$. 
\item [(3)] Compute $DJ(g^{n+1})$ with FD or VF or Hybrid method.
\item [(4)] Compute the parameter $\beta^n=\frac{||D J(g^{n+1})||^2_{U_{ad}}}{||D J(g^n)||^2_{U_{ad}}}$.
\item [(5)] Compute the new search direction $d^{n+1} = -D J(g^{n+1}) + \beta^n d^n$.
\end{itemize}
\end{itemize}
\end{algorithm}
Note the exact line search is used with the conjugate gradient method to ensure that $d^{n+1}$ is a descent direction. Indeed, 
\begin{eqnarray}
DJ(g^{n+1};d^{n+1}) &=& 
DJ(g^{n+1}; -DJ(g^{n+1})+\beta^n d^{n})
\label{cat0019}
\\
&=& - ||DJ(g^{n+1})||^2_{U_{ad}} + \beta^n DJ(g^{n+1}; d^n).
\label{cat0020}
\end{eqnarray}
To guarantee the negativity of $DJ(g^{n+1};d^{n+1})$ when $||DJ(g^{n+1})||^2_{U_{ad}}$ approaches the tolerance $\epsilon$, the value $|\beta^n DJ(g^{n+1}; d^n)|$ should be  smaller than $\epsilon^2$. In this work, we use $\epsilon=$1e-5 and the tolerance in the exact line search as 1e-10, that is, $g^{n+1}$ is accepted when $|DJ(g^{n+1}; d^n)|<$1e-10 in the exact line search.

\subsection{A convergence test of the optimization algorithms}  
\label{sec_2Dtest}
\begin{figure}[htbp]
\begin{center}
\includegraphics[scale=0.3]{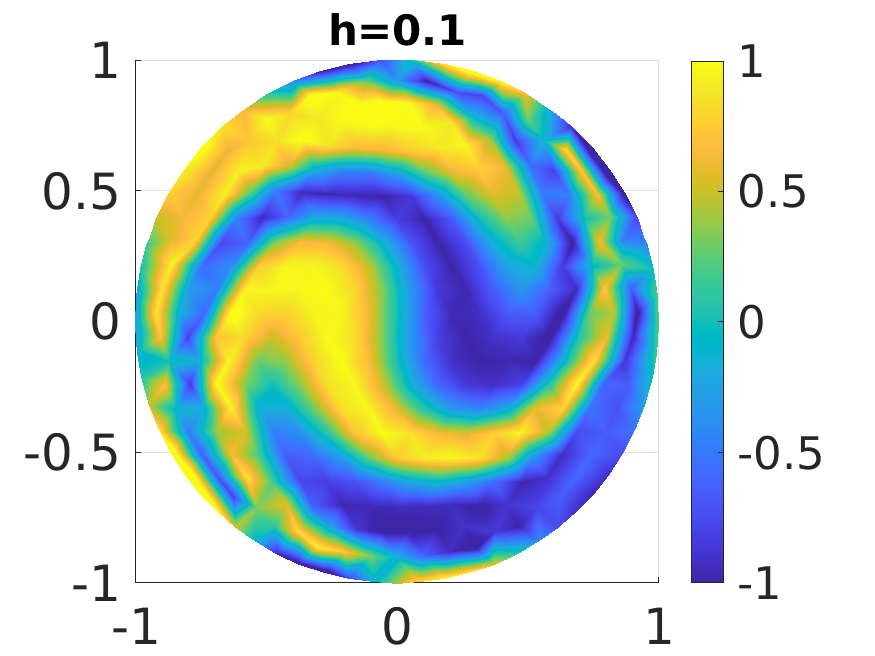}[a]
\includegraphics[scale=0.3]{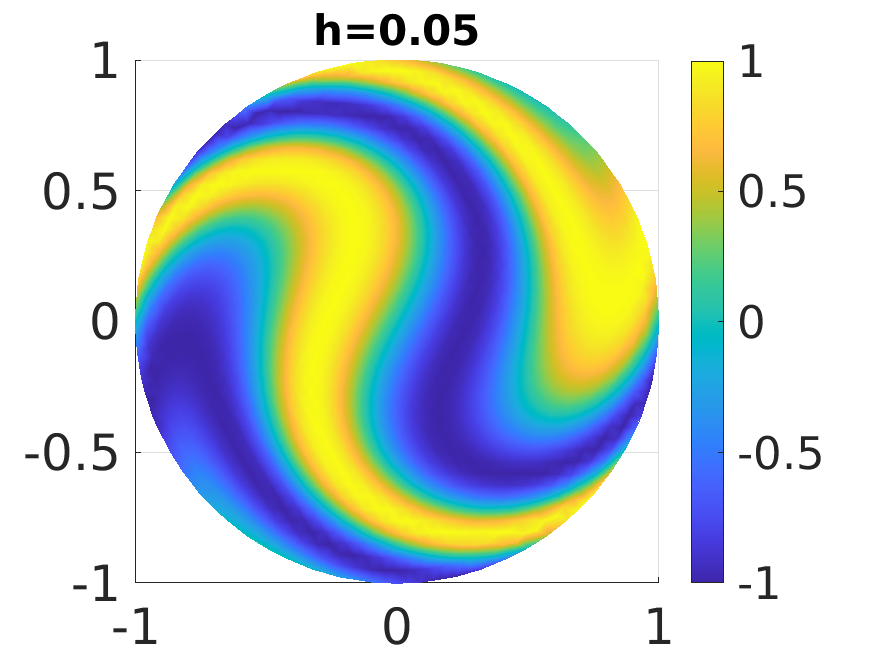}[b]
\includegraphics[scale=0.3]{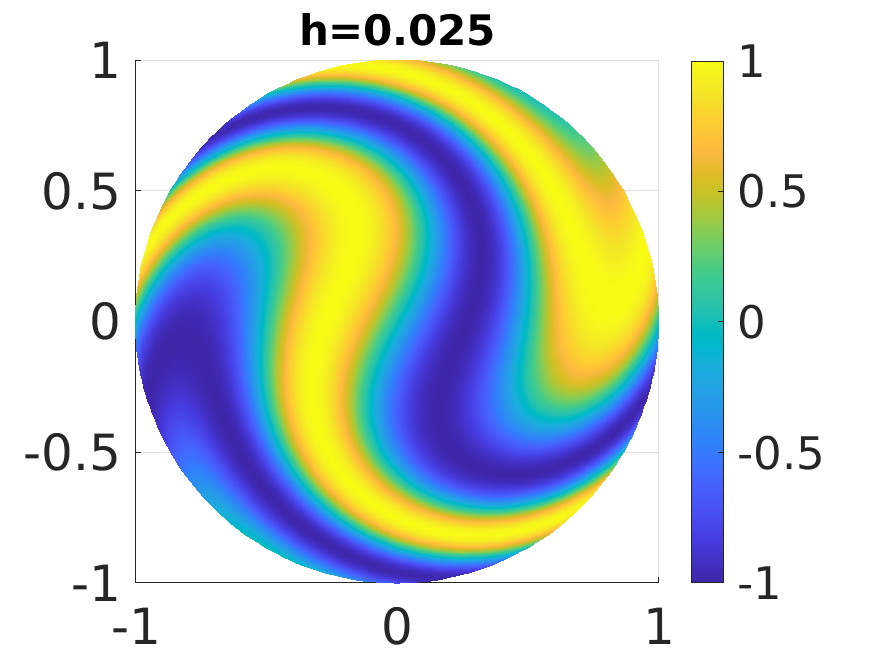}[c]
\includegraphics[scale=0.35]{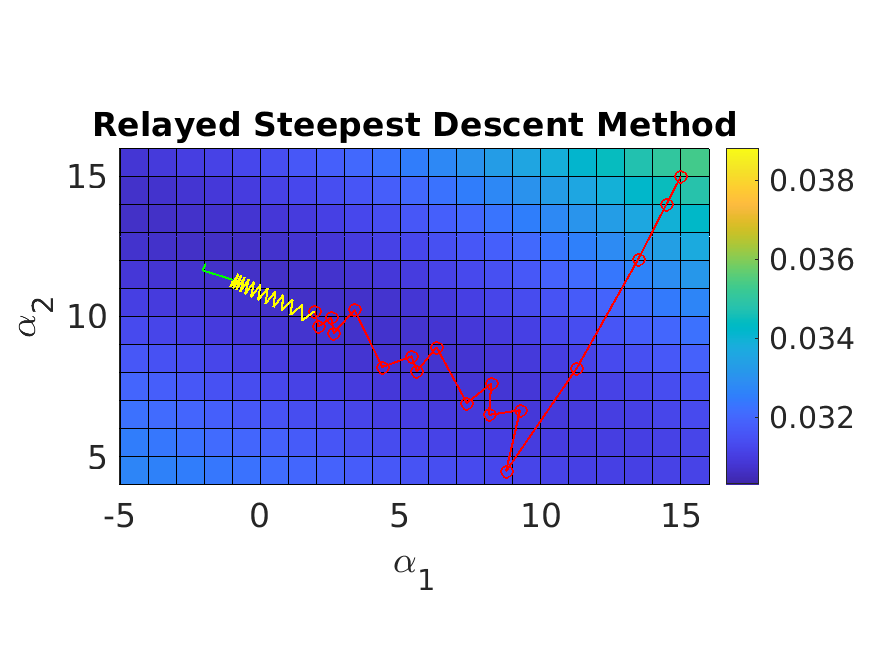}[d]
\includegraphics[scale=0.35]{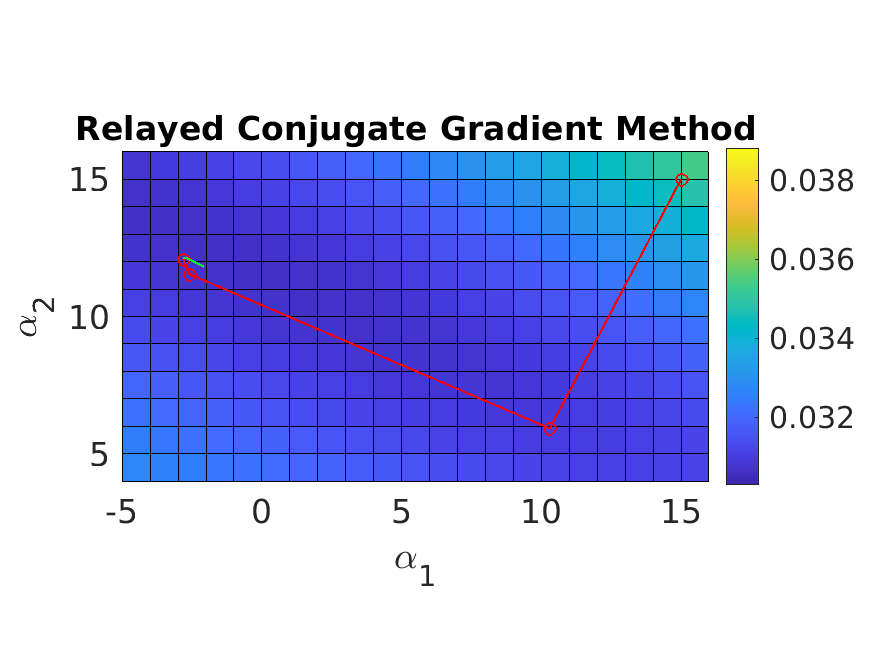}[e]
\caption{
[a,b,c]: Plots of $\theta$ at $t=1$ in the conjugate gradient algorithm.
[d,e]: paths of iterative solutions $\alpha^n=(\alpha^n_1, \alpha^n_2)$ of the relayed methods on the cost map.  The red marker and line refer to the solutions on the mesh with $h=0.1$, the yellow ones with $h=0.05$, and the green ones with $h=0.025$. The iteration information is given in Table\,\ref{1func2D_new}.
\label{fig_SD2D}
}
\end{center}
\end{figure}

In this convergence study, we compare the steepest descent method with the backtracking line search and the conjugate gradient method with the exact line search. 
The control function space is chosen as $U^2_{ad} =\text{span}\{ g^b_1=1_{[0,0.5]}, g^b_2=1_{[0.5,1]} \}$.
In the stopping criterion, $||DJ(g^n)||_{U_{ad}}/(1+J(g^n))<\epsilon$ of Algorithm\,\ref{alg_steepestdescent} and \ref{alg_CG}, $\epsilon$ is set as 1e-5, 5e-6, 1e-6 when $h=0.1, 0.05, 0.025$ respectively. The initial guess is $\alpha^0=(15,15)$. Both the steepest descent and conjugate gradient solutions converge approximately to the same local minimizer $\alpha=(-2.06, 11.81)$ (see Table\,\ref{1func2D_new}), where the corresponding $\theta$ at $t=1$ is plotted in Figure\,\ref{fig_SD2D}[c]. The steepest descent method shows the typical zigzag path of solutions  (Figure\,\ref{fig_SD2D}[b]), as seen in \cite[page 408]{Griva2009}. Although the relayed conjugate gradient method uses far less iteration steps towards the minimizer than the relayed steepest descent method (5 steps vs 44 steps), it indeed takes roughly the same amount of CPU time (between 7 and 9 hours). This is because the conjugate gradient method uses many more evolutions in each line search, which results in roughly the same number of total evolutions (roughly 90). Most importantly, the relay schemes significantly save the computational time: they converge within 9 hours but the non-relayed schemes take one or two days on the finest mesh used in the relayed schemes. The solution paths (Figure\,\ref{fig_SD2D}[de]) manifest the search on the coarsest mesh gets close to the final solution, which makes the remaining search on the finer meshes much easier. Although the mixed scalar looks rough on the coarsest mesh (Figure\,\ref{fig_SD2D}[a], it does not prevent the relay algorithm to converge to a local minimizer. This is a hallmark of all the relay simulations in this work.  

\begin{table}[htbp]
\scriptsize
\begin{center}
\caption{Convergence information of SD with backtracking line search and CG with exact line search. 
Abbreviations: SD=steepest descent, CG=conjugate gradient, 
2 hr 30 min =2 hours 30 minutes,  steps = SD or CG iteration steps, 
93 eolvs in LS = 93 evolutions of $\theta$ in line search.
\label{1func2D_new}
}
\begin{tabular}{|L{3cm}|L{3cm}|L{3cm}|L{3cm}|L{2.5cm}|}
\hline
 & non-relay SD  & relay SD  & non-relay CG & relay CG \\
\hline
h=0.1
CPU time:
& 15 min
& $\Leftarrow$ same as left 
& 15 min
&$\Leftarrow$ same as left \\
steps: 
& 17 steps 
&
& 3 steps
&  \\
minimizer:
& (1.94,  10.19) 
& 
& (-2.81,  12.09)
&  \\
cost:
& 3.07013e-02 
& 
& 3.06672e-02
&\\
\hline 
h=0.05, CPU time:
& 5 hr
& 1 hr 49 min
& 2 hr 2 min
& 6 min \\
steps: 
& 39 steps 
& 24 steps 
& 2 steps 
& 0 steps \\
minimizer:
& (-1.01,  11.33)
& (-0.97,  11.31)
& (-2.11, 11.81)
& (-2.81,  12.09) \\
cost:
& 3.10956e-02
& 3.10957e-02
& 3.10929e-02
& 3.10941e-02\\
\hline
h=0.025, 
CPU time:
&  48 hr 41 min 
&  5 hr 22 min
&  24 hr
&  8 hr 24 min \\
steps: 
& 40 steps 
& 3 steps
& 3 steps
& 2 steps \\
minimizer:
& (-1.66, 11.61)
& (-2.00, 11.77)
& (-2.06, 11.81) 
& (-2.06,  11.81)
\\
cost:
& 3.119671e-2
& 3.119639e-2
& 3.119638e-2
& 3.119638e-2
\\
\hline
relay total CPU time, steps,  evols in LS:
&  
& 7 hr 26 min, 44 steps, 93 evols
&  
& 8 hr 45 min, 5 steps, 88 evols\\
\hline
\end{tabular}
\end{center}
\end{table}

\section{Optimization simulations}
\label{sec_simulations}
This section applies the optimization algorithms developed in this work to study the boundary controlled mixing  with the control basis functions mentioned in Section\,\ref{sec_controlbasis} through extensive numerical experiments. We first describe the flow patterns and mixing features of each control basis function, and then combine them together to study the optimal mixing. All the numerical simulations are performed in Michigan State University’s High Performance Computing Center (HPCC).

The physical setup and the initial values are introduced at the end of Section\,\ref{sec_optimization_problem}. When the  optimization algorithms are called for a set of control basis $\{g^b_i: i=1, \cdots, M\}$, we apply the relay Algorithm\,\ref{alg_relay} with three meshes of $h=0.1, 0.05, 0.025$. To handle multiple local minimizers,  5 different initial guesses of $\alpha=(\alpha_1, \cdots, \alpha_M)$ are tested on the coarsest mesh for each set of control basis. These initial vectors $\alpha$ are randomly  chosen where each component $\alpha_i, i=1, \cdots, M$, is uniformly distributed from -100 to 100. For each initial guess of $\alpha$, we apply both the steepest descent and conjugate gradient methods to find the optimal solutions, where these two solutions are generally different.
Afterwards, the one with the smallest cost is relayed to the intermediate mesh and finally is sent to the finest mesh.   The computation of the cost gradient is by the hybrid method in Section\,\ref{sec_hybrid}. The line search for the steepest descent method is the backtracking Algorithm\,\ref{alg_backtracking} and the one for the conjugate gradient method is the exact line search Algorithm\,\ref{alg_exactlinesearch}. The parameter $\epsilon$ in the stopping criterion of SD and CG methods is $\epsilon=$1e-5 for all the simulations in this section. 

\subsection{Flow patterns of control basis functions}
\label{sec_flowpatterns}
\begin{figure}[!htbp]
\begin{center}
\includegraphics[scale=0.17]{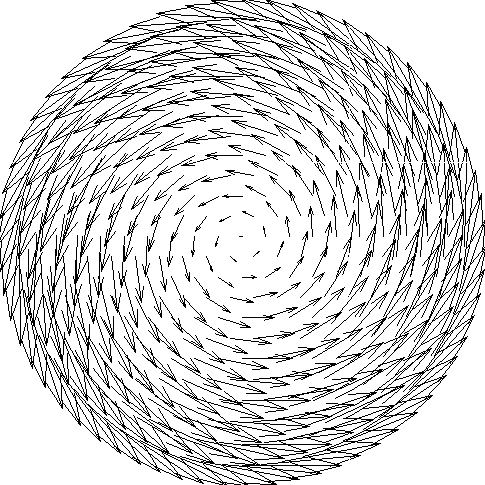}\,
\includegraphics[scale=0.17]{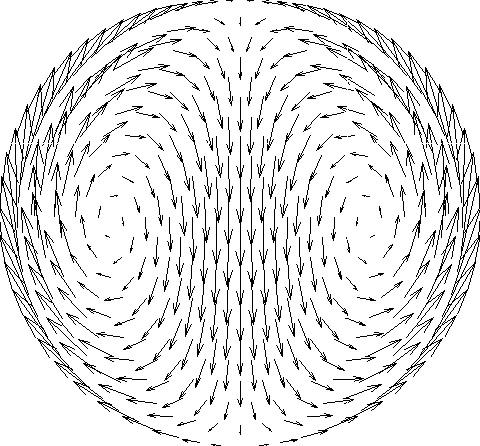}\,
\includegraphics[scale=0.17]{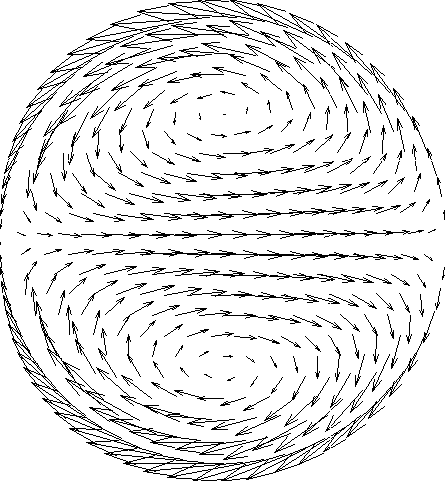}\,
\includegraphics[scale=0.17]{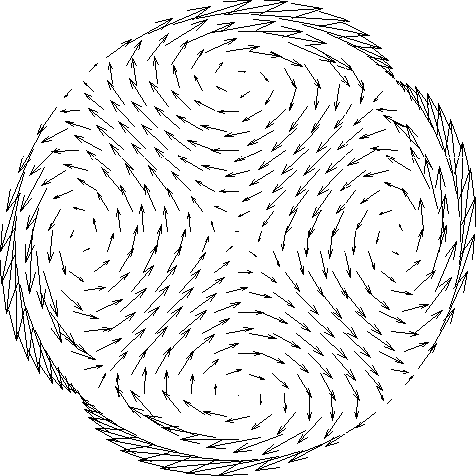}\,
\includegraphics[scale=0.17]{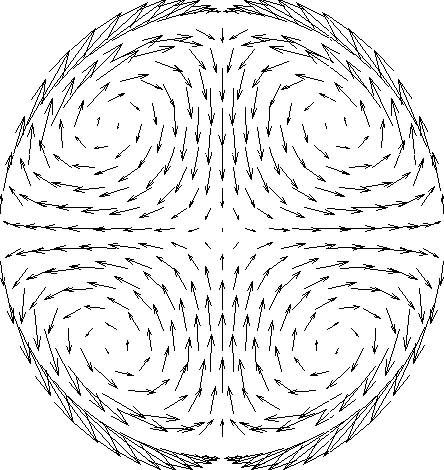}
\caption{Flow patterns for $g=1, \cos(\omega), \sin(\omega), \cos(2\omega), \sin(2\omega)$
 at time $t=1$ (from left to right). 
\label{fig_flowpatterns}
}
\end{center}
\end{figure}

In this work, the controls are divided into five types based on five elementary functions: $1$, $\cos(\omega)$, $\sin(\omega)$,  $\cos(2\omega)$,  $\sin(2\omega)$, where $\omega\in [0,2\pi)$ (see details in Section\,\ref{sec_controlbasis}). The $U_{ad}$-norm is $\sqrt{2\pi}$ for $g=1$ and  $\sqrt{\pi}$ for other functions.
Their flow patterns at time $t=1$ are shown in Figure\,\ref{fig_flowpatterns}. There exist one vortex for $g=1$, two vortices for $\cos(\omega)$ and $\sin(\omega)$, and four vortices for $\cos(2\omega)$ and $\sin(2\omega)$.

\begin{figure}[!htbp]
\begin{center}
\includegraphics[scale=0.33]{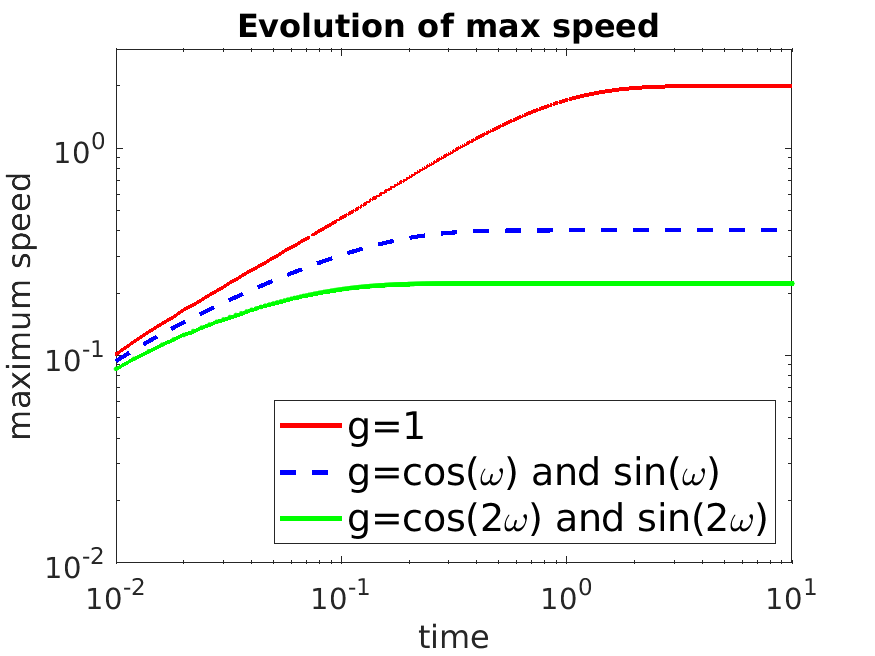}[a]
\includegraphics[scale=0.33]{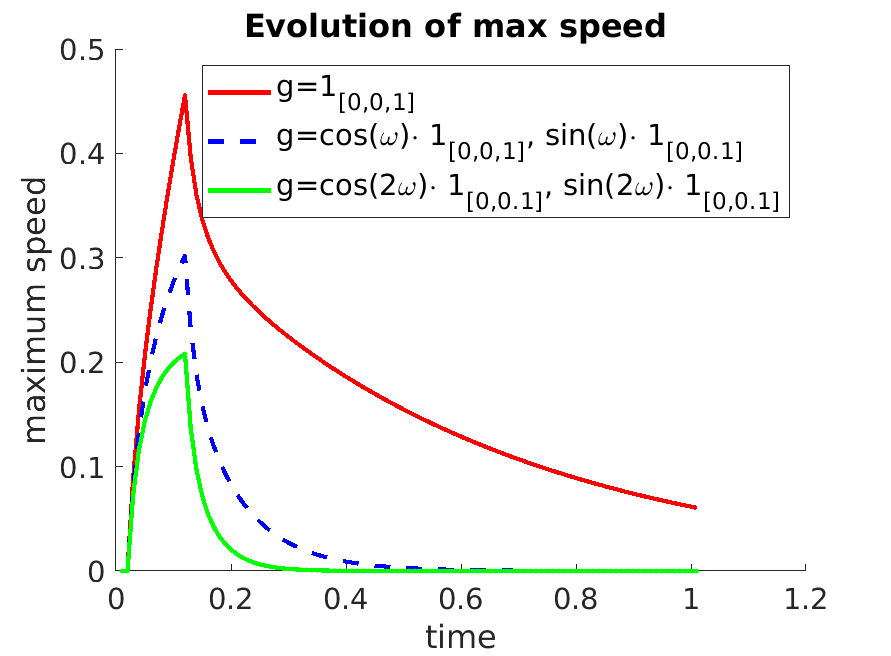}[b]
\caption{
Evolution of the maximum speed of velocity for five elementary control functions.
[a]: the controls are applied in the entire time interval $[0,1]$.
[b]: the controls are only applied in the time interval $[0,0.1]$.
The fluid-wall friction parameter $k=0.5$.
\label{fig_umax}
}
\end{center}
\end{figure}

When $g=1$, a radially symmetric analysis (Appendix\,\ref{sec_radialsymmetry}) shows the velocity field has a unique steady state with radial component $v_r=0$ and angular component $v_\varphi=g/k r$, along with a zero pressure field.  
Apparently this steady state velocity does not induce any mixing because it is simply a rigid rotation. Therefore,  the mixing for $g=1$ occurs only when the velocity is unsteady. 
The evolution of the maximum speeds in the domain of these elementary control functions are shown in Figure\,\ref{fig_umax}[a].
The flow of $g=1$ reaches the steady state around $t=3.8$, while the flows of $\cos(\omega)$ and $\sin(\omega)$ reach the steady states with maximum speed $0.4$ around $t=0.6$ . 
The flows of $\cos(2\omega)$ and $\sin(2\omega)$ reach the steady states  with maximum speed $0.22$ around $t=0.3$. 
Note when the initial velocity is zero, the time scale for the flow induced by $\alpha g$, $\alpha\ne 0$, to reach the steady state is independent of $\alpha$,  due to the linearity of the flow to the control.

When the control is only applied in a time segmentation interval, the flow velocity will decay to zero over time after the control is turned off due to viscous dissipation and boundary wall friction. Figure\,\ref{fig_umax}[b] shows the evolution of the 
maximum speed where the five elementary functions are applied only in the time interval $[0, 0.1]$. When the time segmentation interval is $[0.1n, 0.1(n+1)]$, $n=1, \cdots, 9$, the corresponding flow can be obtained by simply shifting by $0.1n$ units to the positive time direction the flow generated by the same elementary  function applied on $[0,0.1]$. It is noticed that the flow decays to zero far faster when it is generated by a cosine or sine function than by the function 1. This is produced by the extra dissipation between multiple vortices from a cosine or sine control function, in contrast to only one vortex from the control 1 (see Figure\,\ref{fig_flowpatterns}).

\subsection{Optimization by each single control type}
\label{sec_mixingfeatures}
This part is devoted to the mixing properties of each of the five types of control basis functions. First, we compute the mix-norms and costs at $t=1$ with Algorithm\,\ref{alg_cost} for the controls $g=\alpha g^b$, where $g^b$ is one of the five elementary functions and $\alpha$ takes integer values in $[0,100]$. This corresponds to the time segmentation number $N=1$. Afterwards, we use the optimization algorithms to compute the optimal solution when  $N=10$ for each type of control basis functions. 

\begin{figure}[htbp]
\begin{center}
\includegraphics[scale=0.33]{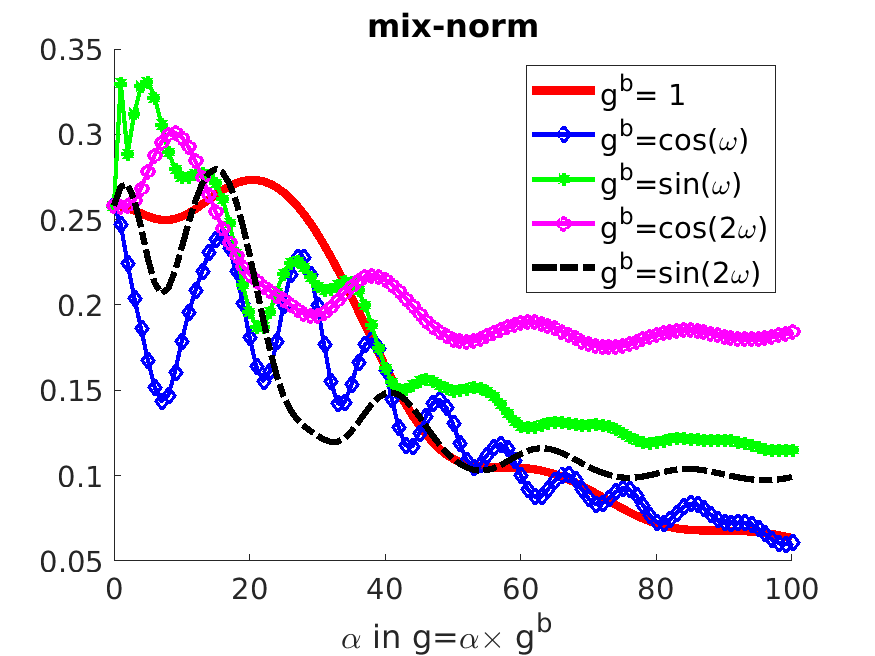}[a]
\includegraphics[scale=0.33]{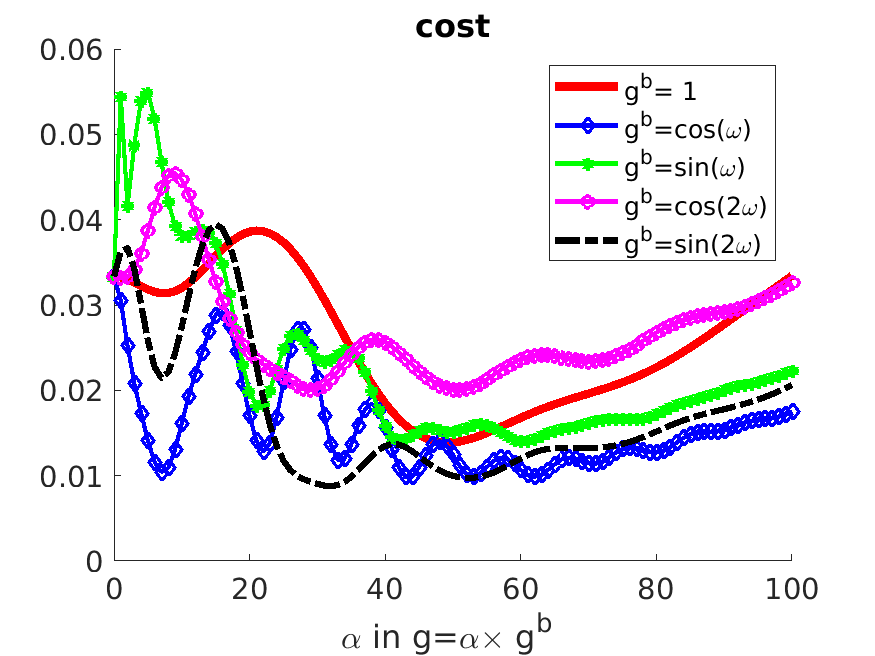}[b]
\end{center}
\caption{Mix-norms and costs of five elementary functions when $\alpha\in[0,100]$.
\label{cat0982} 
}
\end{figure}

The most striking property is the existence of multiple local minimizers of the mix-norm for most control basis functions when the coefficient $\alpha$ varies, according to Figure\,\ref{cat0982}[a]. When $\gamma=$ 1e-6, the  cost also has multiple local minimizers (Figure\,\ref{cat0982}[b]). Because one initial guess only leads to one local minimizer in an optimization algorithm, multiple initial guesses are needed in order to achieve the global minimizer.

\begin{figure}[htbp]
\begin{center}
\includegraphics[scale=0.19]{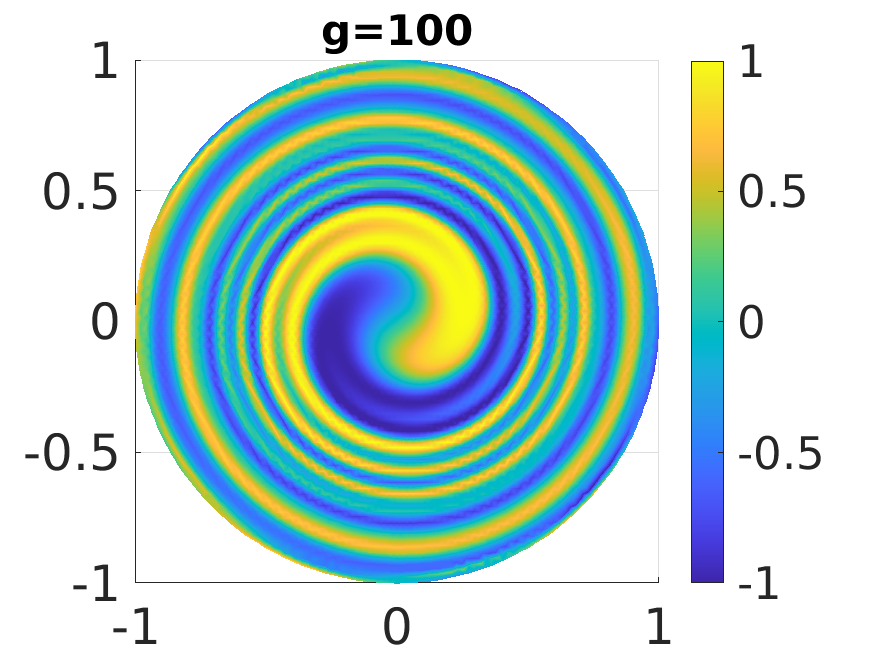}[a]
\includegraphics[scale=0.19]{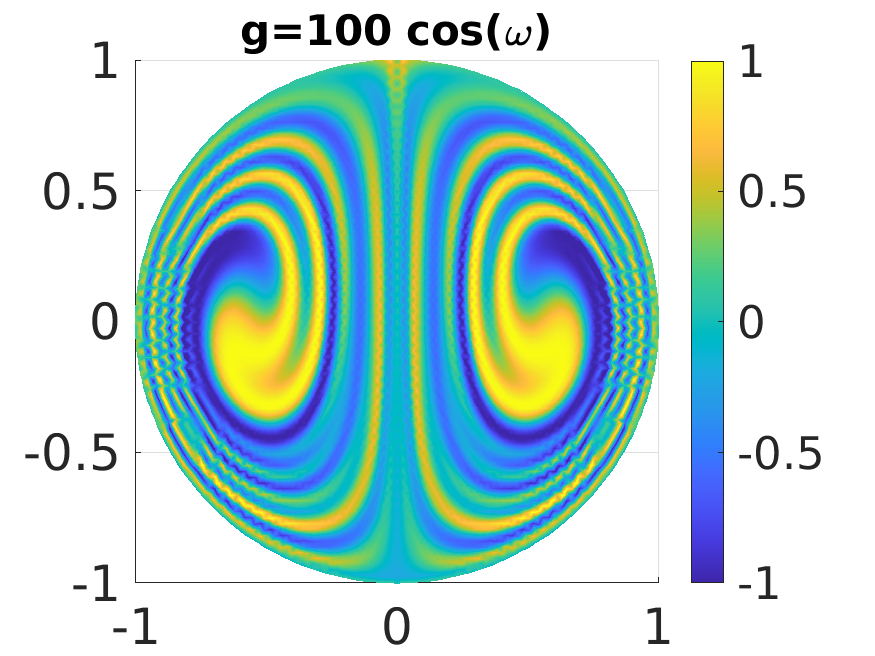}[b]
\includegraphics[scale=0.19]{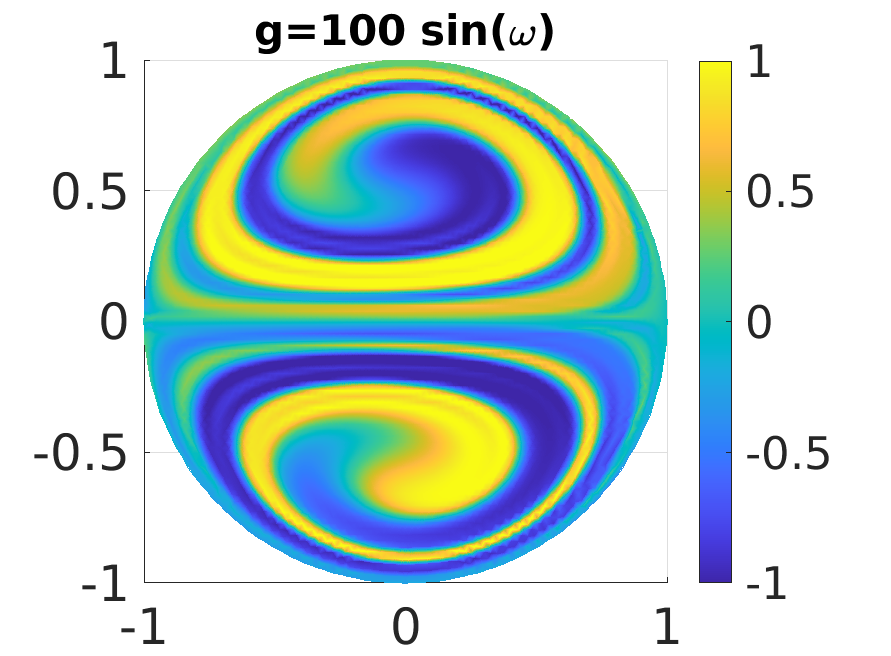}[c]
\includegraphics[scale=0.19]{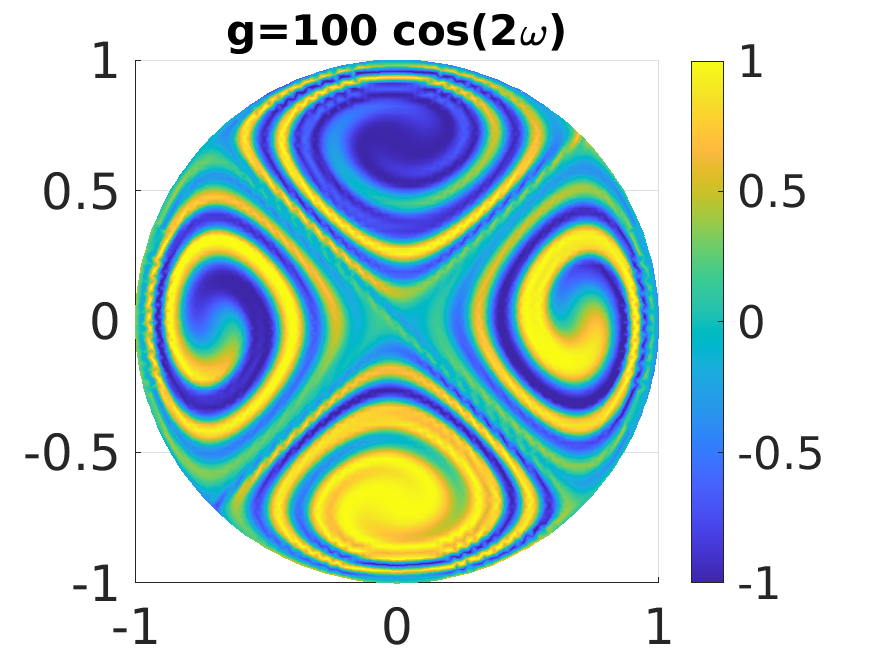}[d]
\includegraphics[scale=0.19]{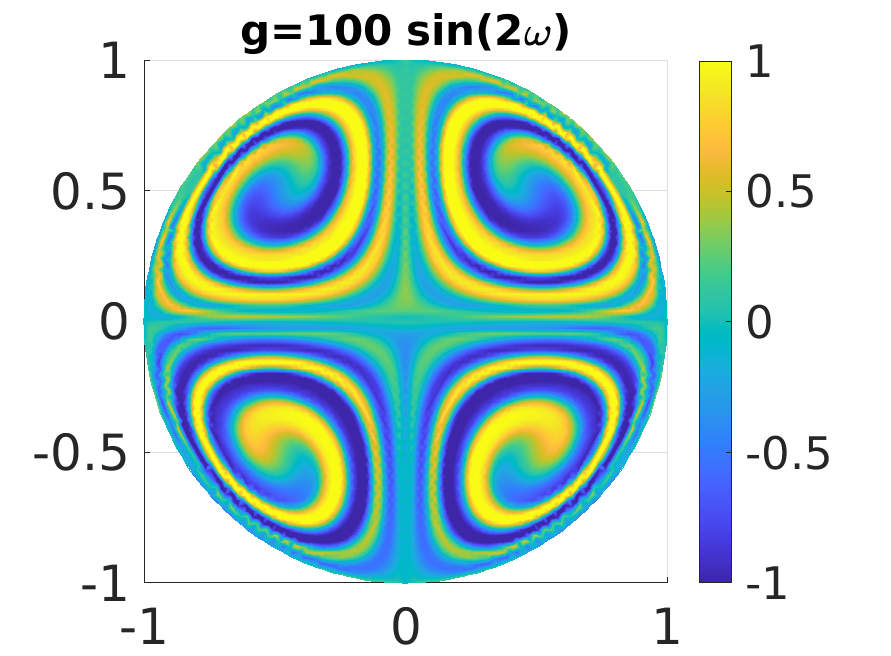}[e]
\includegraphics[scale=0.19]{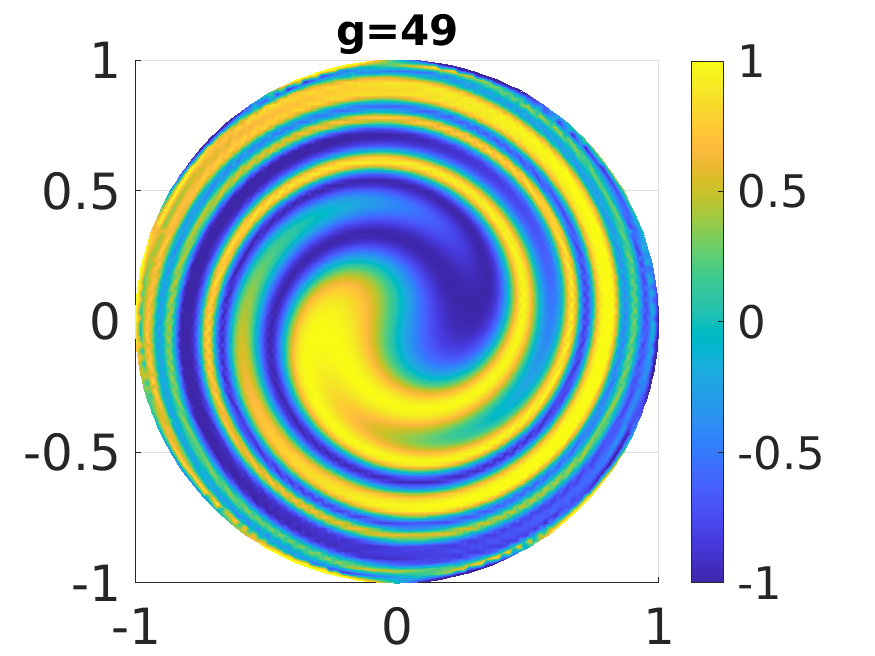}[f]
\includegraphics[scale=0.19]{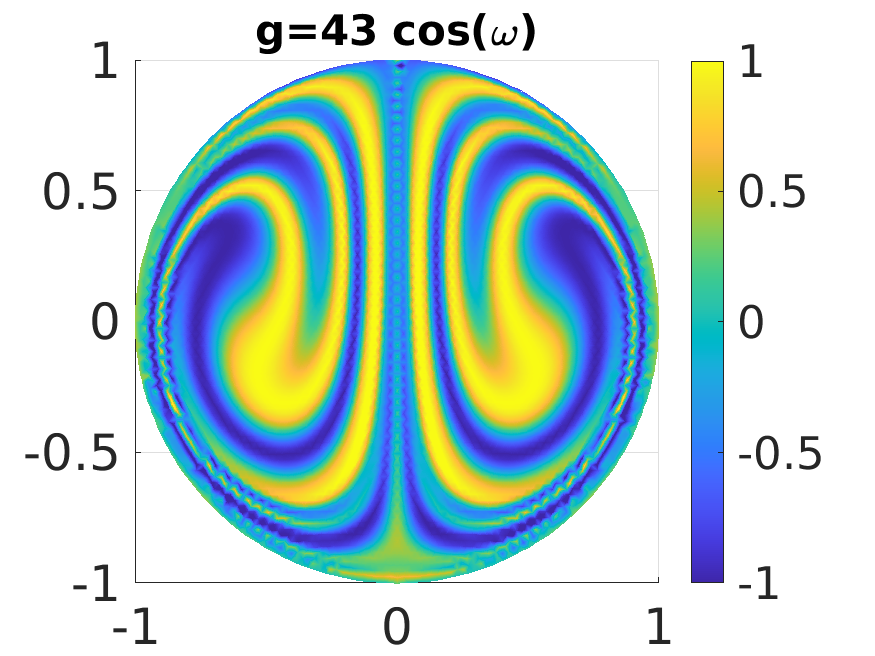}[g]
\includegraphics[scale=0.19]{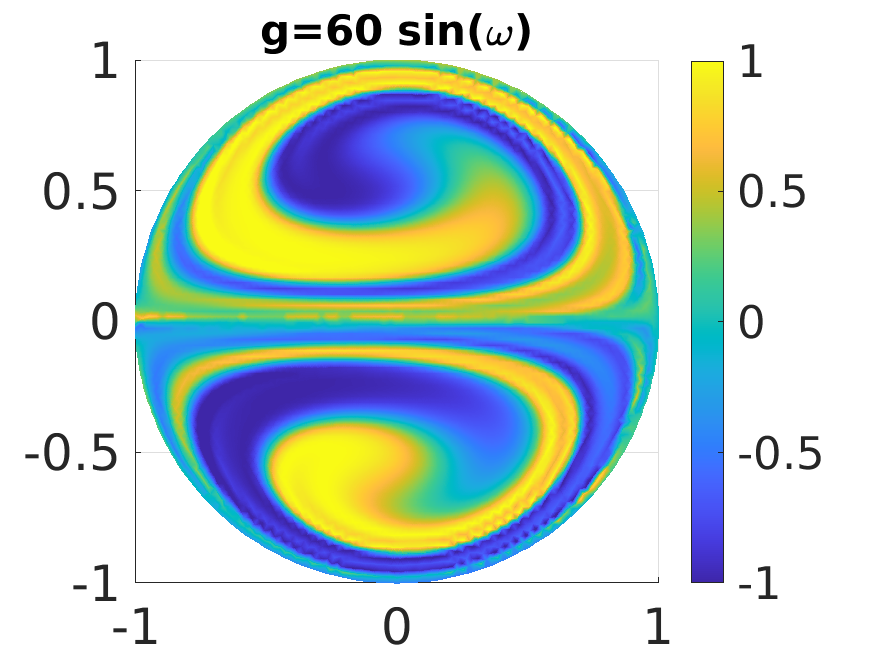}[h]
\includegraphics[scale=0.19]{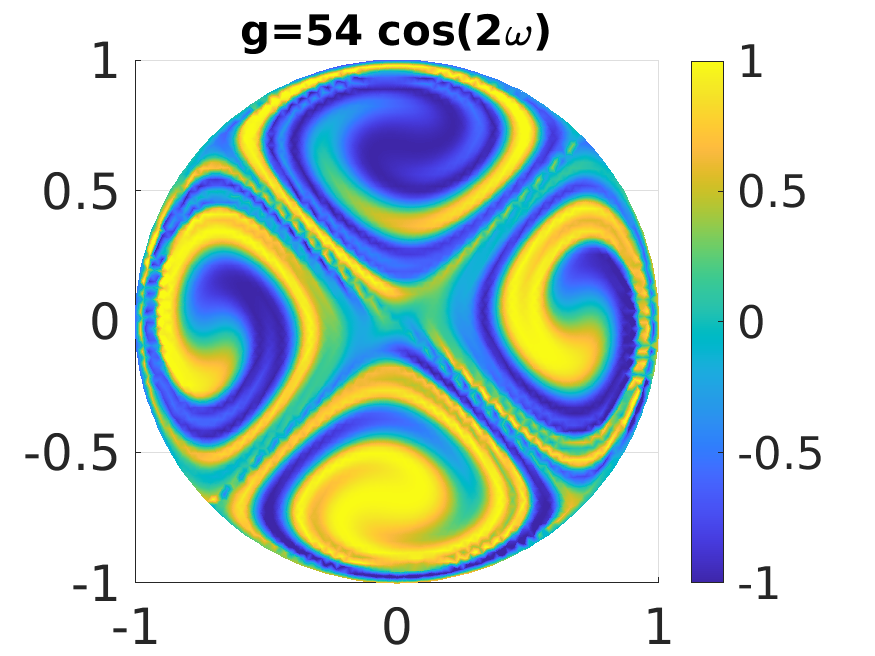}[i]
\includegraphics[scale=0.19]{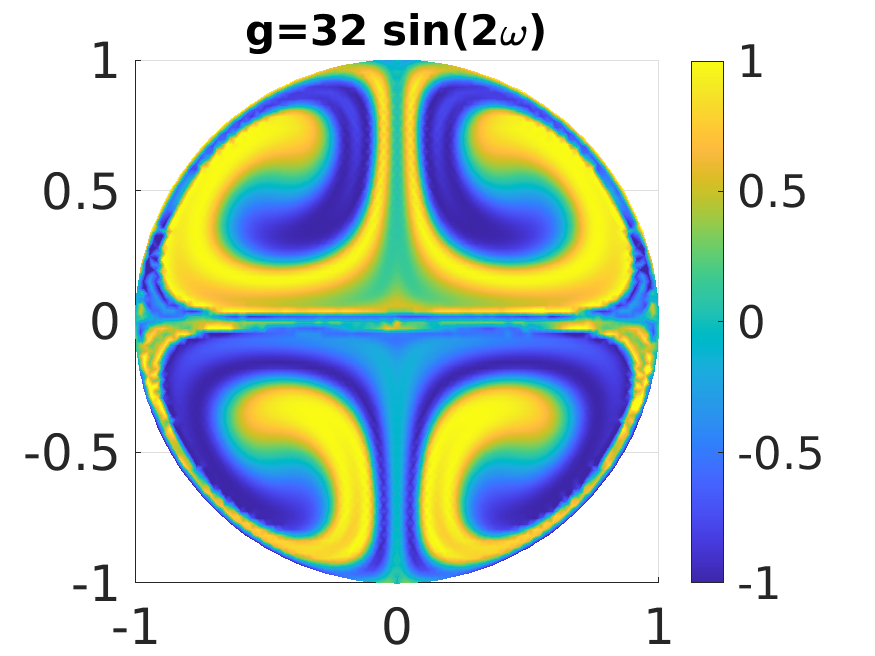}[j]\\
\includegraphics[scale=0.19]{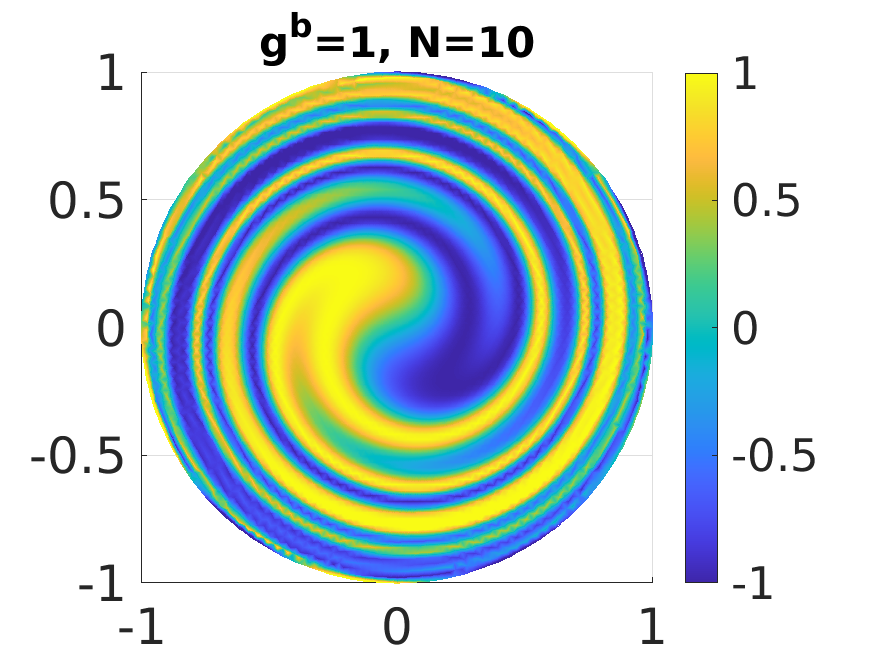}[k]
\includegraphics[scale=0.19]{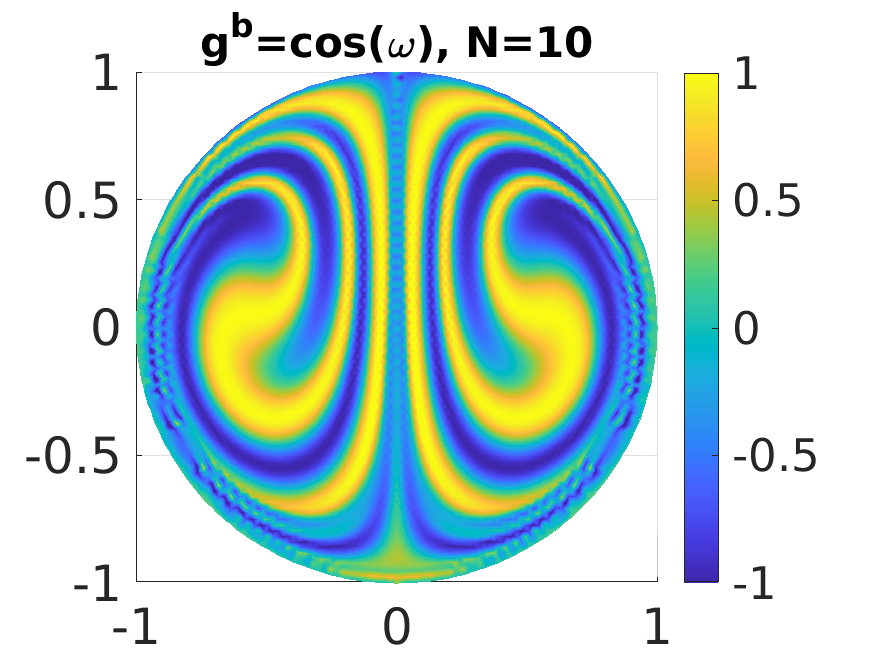}[l]
\includegraphics[scale=0.19]{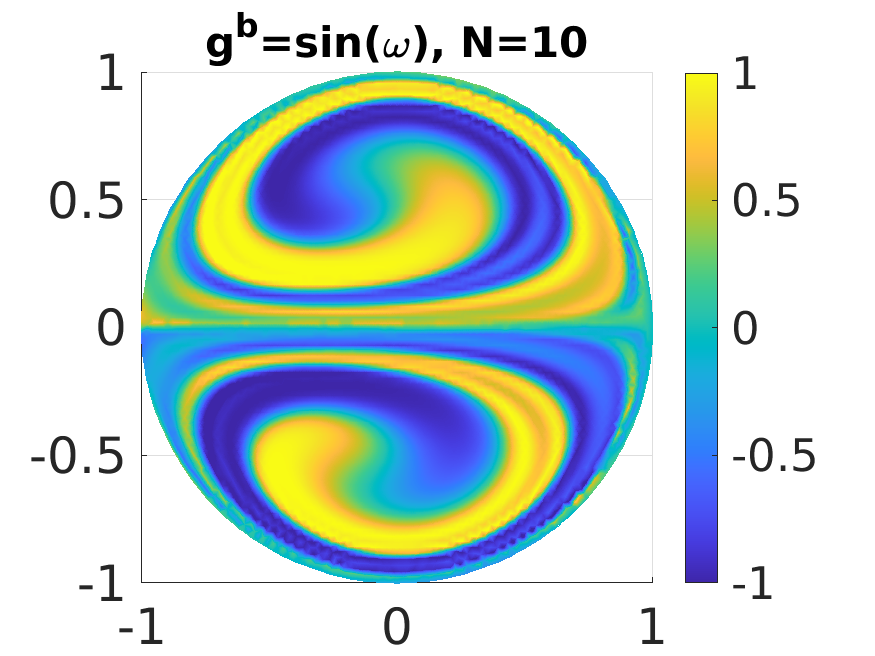}[m]
\includegraphics[scale=0.19]{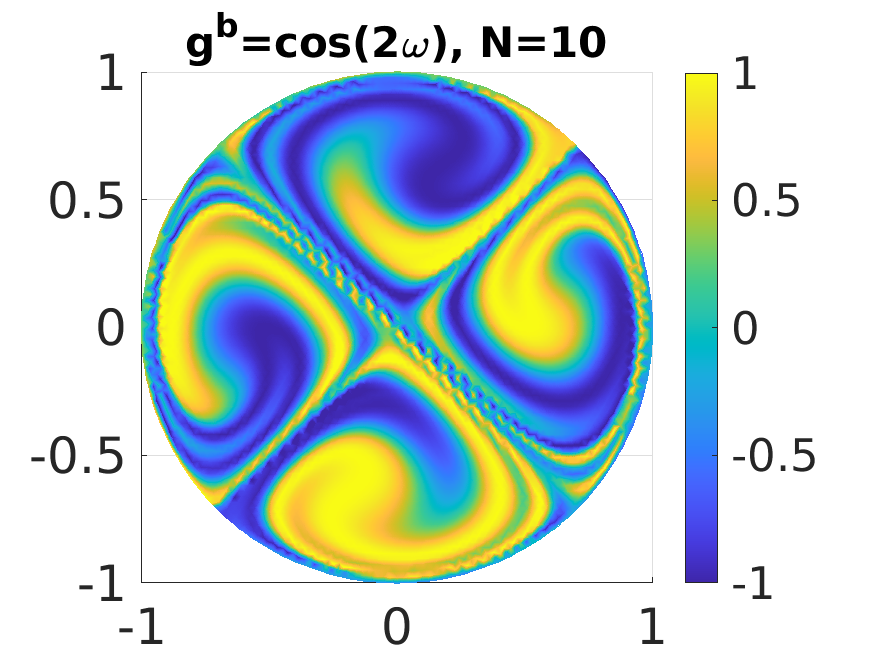}[n]
\includegraphics[scale=0.19]{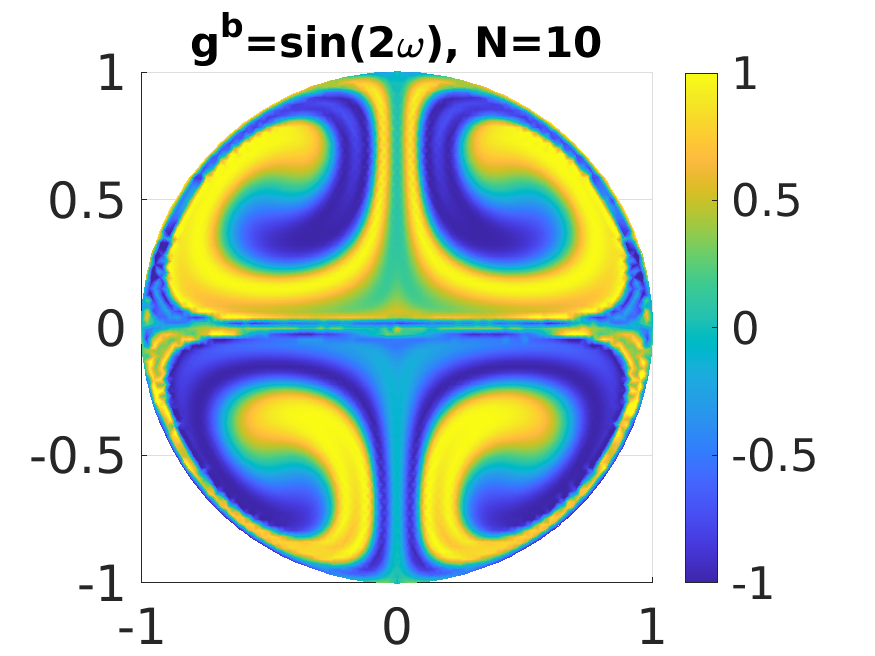}[o]
\end{center}

\caption{Plotings of $\theta$ at $t=1$ on the mesh of $h=0.025$. 
The first row is when $\alpha=100$ in $g=\alpha g^b$,  the second row is for the optimal solutions when $N=1$, and the third row is for the optimal solutions when  $N=10$.
\label{cat2202} 
}
\end{figure}

The second property is that the better mixing quality, identified with the lower mix-norm, corresponds to the larger control strength in general (Figure\,\ref{cat0982}[a]), and thus the larger velocity magnitude because the flow velocity is linearly dependent on the control. The mixed scalar fields at $t=1$ when $\alpha=100$,  the largest control strength considered,  are shown in Figure\,\ref{cat2202}[a-e], each of which has almost the smallest mix-norm in the same control type. 
On the other hand, the scalar fields at $t=1$ with the smallest costs in the same control type when $N=1$ are shown in Figure\,\ref{cat2202}[f-j]. The data of the mix-norms, g-norms, and costs of these  simulations are displayed in Table\,\ref{Table0cat_1098}.  From the relation between the scalar field renderings and their mix-norms, it appears that a better mixed field is characterized by thinner and longer filaments.

\begin{table}[htbp]
\begin{center}
\caption{Mixing information of single type of control basis functions.
CRP=Cost Reduction Percentage from N=1 case to N=10 case.
}
\label{Table0cat_1098}
\scriptsize
\begin{tabular}{|c||ccc||ccc||ccc|c|}
    \hline
    & \multicolumn{3}{c}{ $\alpha=100$ and $N=1$} \vline \vline
    & \multicolumn{3}{c}{Optimal solution when $N=1$} \vline \vline
    & \multicolumn{3}{c}{Optimal solution when $N=10$} \vline &CRP\\
    \hline
     & mix-norm & g-norm & cost & mix-norm & g-norm & cost & mix-norm & g-norm & cost &\\
    \hline
    1             & 6.34e-2 & 2.51e+2 & 3.34e-2  
                  & 1.12e-1 & 1.23e+2 & 1.39e-2 
                  & 1.10e-1 & 1.17e+2 & 1.30e-2 & 6\%\\
    \hline
   $\cos(\omega)$ & 6.07e-2 & 1.77e+2 & 1.76e-2 
                  & 1.18e-1 & 7.62e+1 & 9.84e-3 
                  & 9.32e-2 & 9.44e+1 & 8.79e-3 & 11\%  \\
    \hline
    $\sin(\omega)$ & 1.15e-1 & 1.77e+2 & 2.24e-2 
                   & 1.30e-1 & 1.06e+2 & 1.41e-2  
                   & 1.11e-1 & 1.10e+2 & 1.22e-2 & 13\%\\
    \hline
    $\cos(2\omega)$ & 1.84e-1 & 1.77e+2 & 3.27e-2
                    & 1.80e-1 & 9.57e+1 & 2.08e-2 
                    & 1.82e-1 & 6.87e+1 & 1.89e-2 & 9\%\\
    \hline
    $\sin(2\omega)$ & 9.93e-2 & 1.77e+2 & 2.06e-2
                    & 1.20e-1 & 5.67e+1 & 8.78e-3 
                    & 1.13e-1 & 6.46e+1 & 8.42e-3 & 4\%\\
    \hline                        
\end{tabular}
\end{center}
\end{table}

\begin{figure}[htbp]
\begin{center}
\includegraphics[scale=0.19]{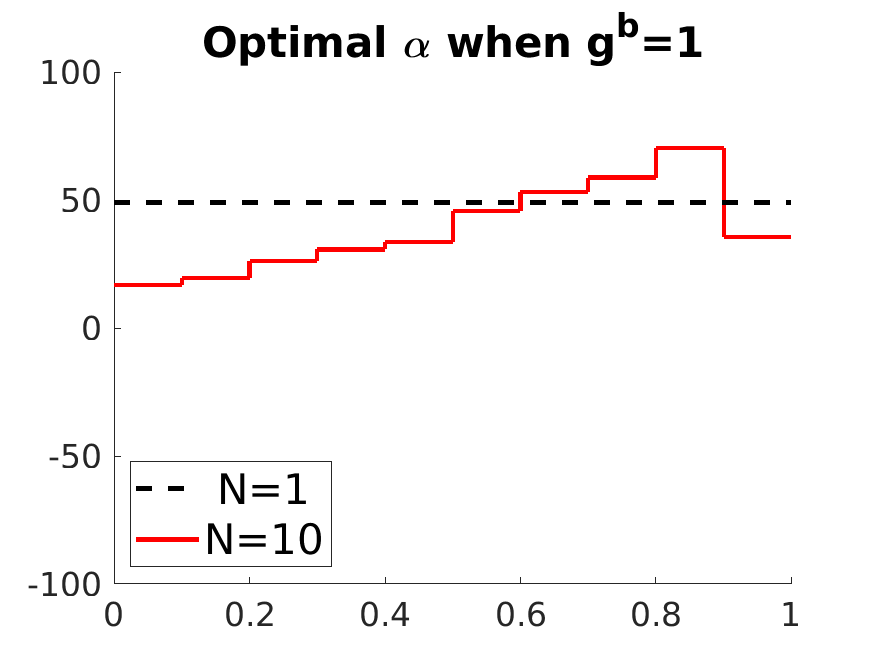}[a]
\includegraphics[scale=0.19]{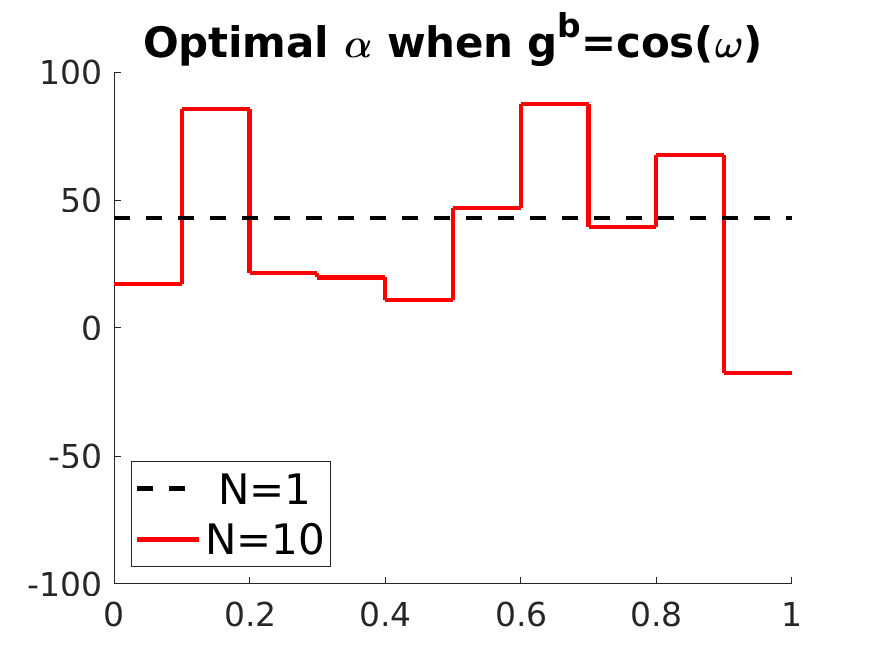}[b]
\includegraphics[scale=0.19]{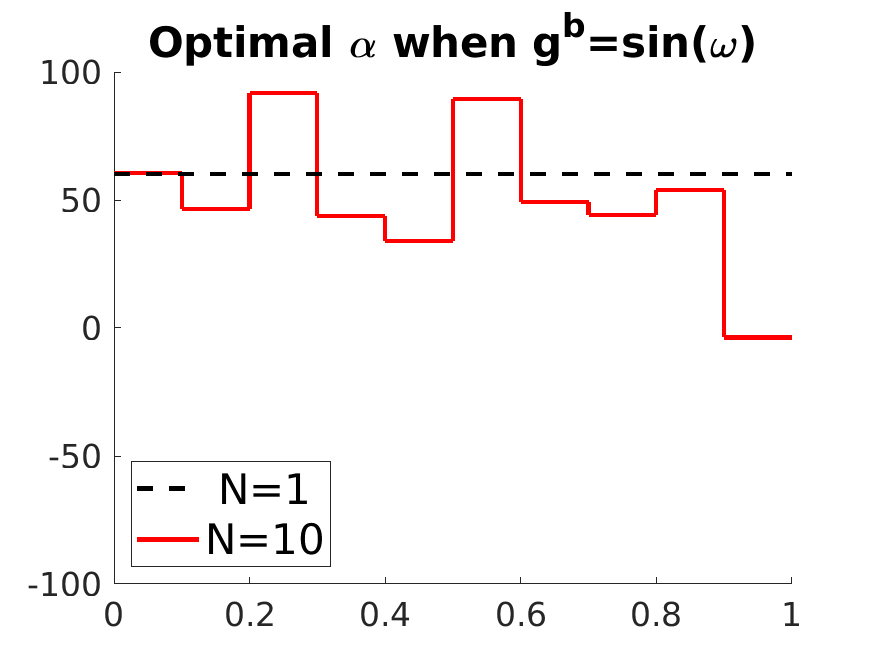}[c]
\includegraphics[scale=0.19]{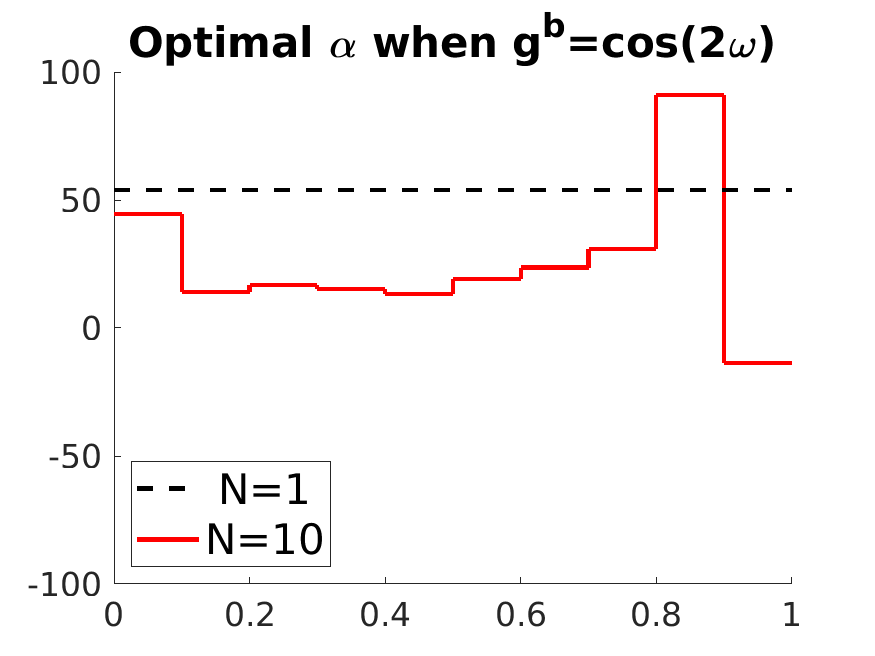}[d]
\includegraphics[scale=0.19]{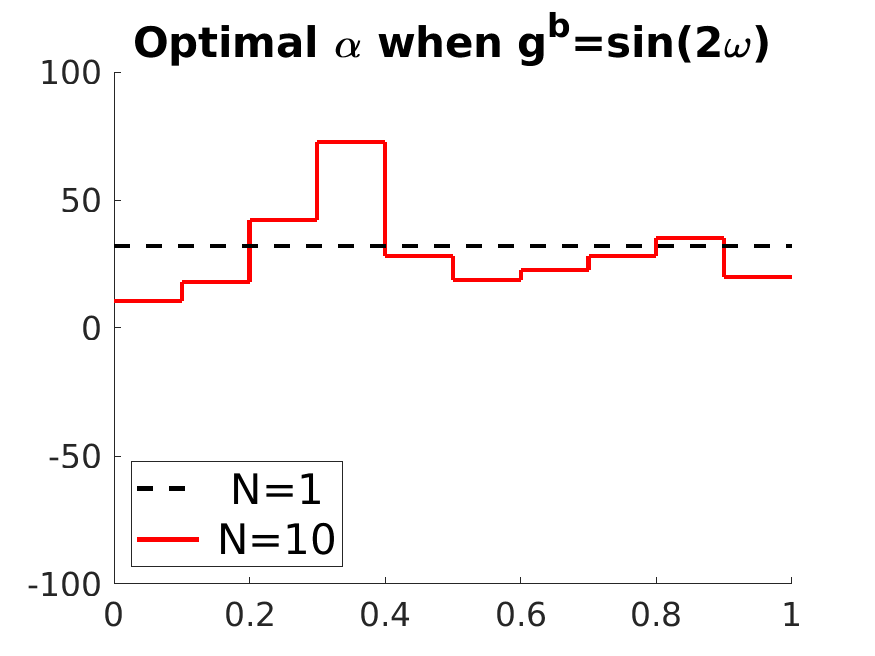}[e]
\end{center}
\caption{Optimal solutions $\alpha$ for $N=1$ and $N=10$ for different control basis functions. The black dashlines are for $N=1$ and the red solid lines are for $N=10$. When $N=1$, $\alpha=49, 43, 60, 54, 32$ from left to right.
\label{cat2201} 
}
\end{figure}


The third property is that the mixing quality of one control type is limited by its specific flow pattern. By comparing Figure\,\ref{fig_flowpatterns} and Figure\,\ref{cat2202}, we can tell Type 1 takes the entire domain as a single mixing zone,   Type $\cos(\omega)$ and Type $\sin(\omega)$ divide the domain into two separate mixing zones, and Type $\cos(2\omega)$ and Type $\sin(2\omega)$ divide the domain into four isolated mixing zones. In each mixing zone, the mixing is performed by rotating the scalar around the center. If a mixing zone is predominantly occupied by one value or one color, then the mixing would not be effective due to the lack of mass exchange between different zones. For example, in the four mixing zones of the control $\cos(2\omega)$, the color of $\theta$ is predominantly blue in the upper zone and predominantly yellow in the lower zone all the time during the mixing process no matter how $\alpha$ changes (see Figure\,\ref{fig_alltypes} at $t=0$ and Figure\,\ref{cat2202}[d,i,n]). This is why the mix-norm refuses to decrease when $\alpha$ exceeds 50 for $\cos(2\omega)$ (see Figure\,\ref{cat0982}[a]).

The purpose of time segmenting is to provide control flexibility in time to reduce cost. This is modestly successful because the cost reduction rates from $N=1$ to $N=10$ are only between 4\% and 13\%, as seen in Table\,\ref{Table0cat_1098}. The mix-norms when $N=10$ are also smaller than those when $N=1$ in the same type of control except for Type $\cos(2\omega)$. 
To easily plot the control solution, we express the control as 
$g=\alpha g^b$ where $\alpha = \sum_{i=1}^N \alpha_i \chi_i^N(t)$ and  $g^b$ is one of the five elementary functions. 
The optimal solutions $\alpha$ are plotted in Figure\,\ref{cat2201}, which are very different between $N=1$ and $N=10$ cases for the same type of control. The scalar field of the optimal solution when $N=10$ does not differ much from that when $N=1$ of the same control type (see Figure\,\ref{cat2202} second and third rows). 
All of these facts indicate that under single control types investigated in this work, modulating the time segmentation is not very efficient in cost reduction. 

\subsection{Optimization by combined control types} 
\label{sec_combined_optimalmixing}

\begin{figure}[htbp]
\begin{center}
\includegraphics[scale=0.3]{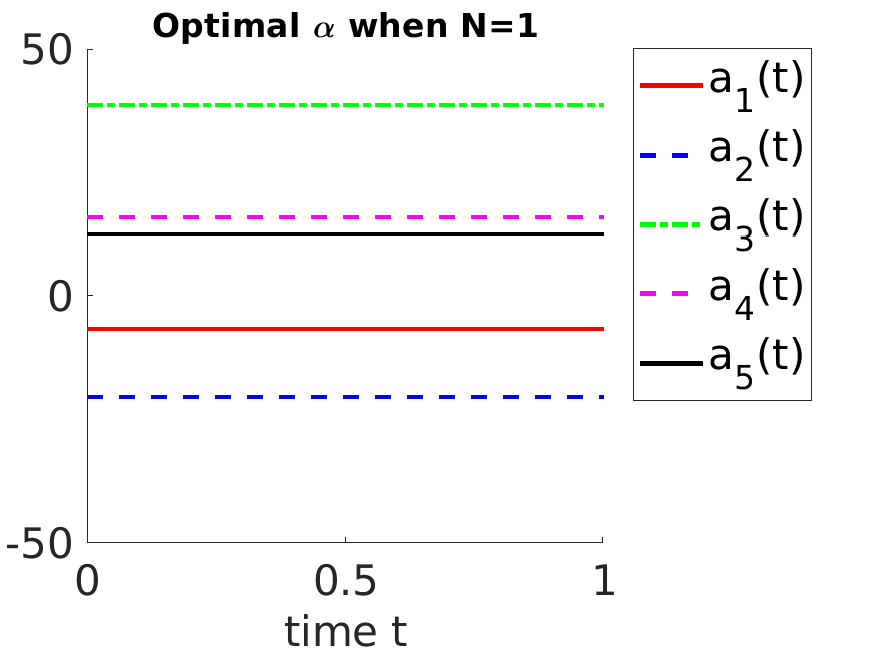}
\includegraphics[scale=0.3]{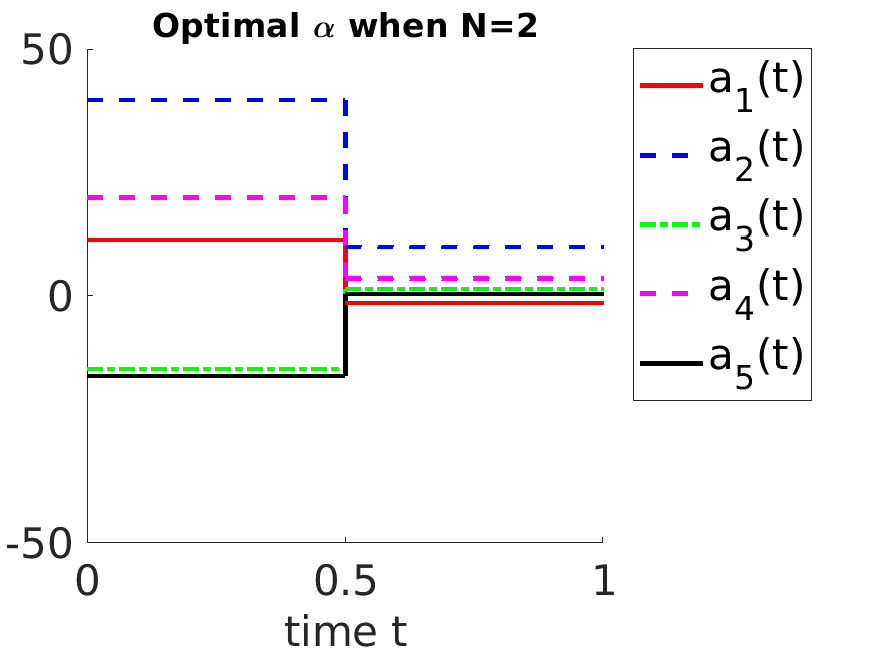}
\includegraphics[scale=0.3]{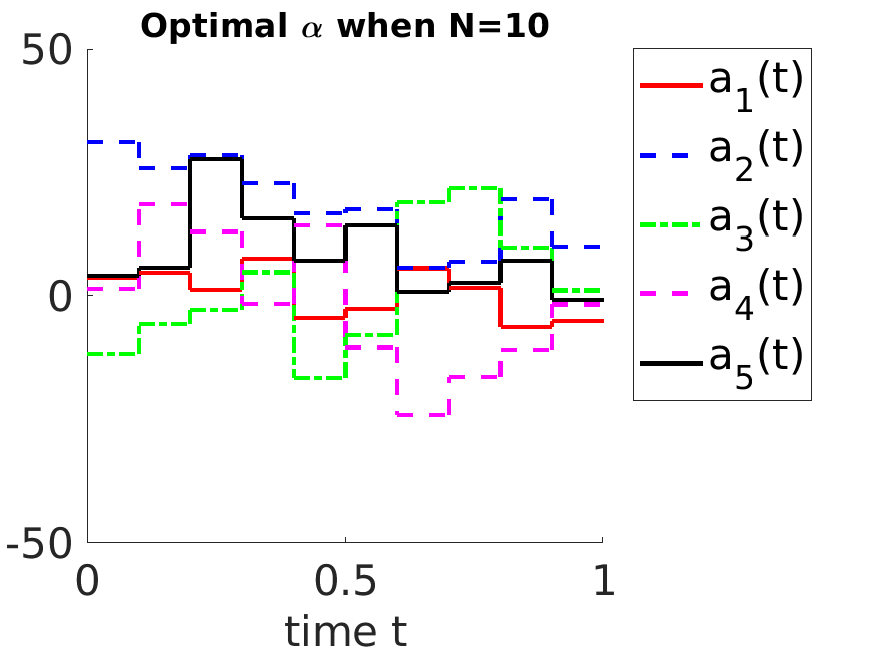}
\end{center}
\caption{Optimal solutions $\alpha=(\alpha_1(t), \cdots, \alpha_5(t))$ in $g=\sum_{i=1}^5 \alpha_i(t) g^b_i$, 
where $g^b_1=1$, $g^b_2=\cos(\omega)$, $g^b_3=\sin(\omega)$, $g^b_4=\cos(2\omega)$, and $g^b_5=\sin(2\omega)$.
\label{fig_alltypes_alpha} 
}
\end{figure}

\begin{figure}[htbp]
\begin{center}
\includegraphics[scale=0.07]{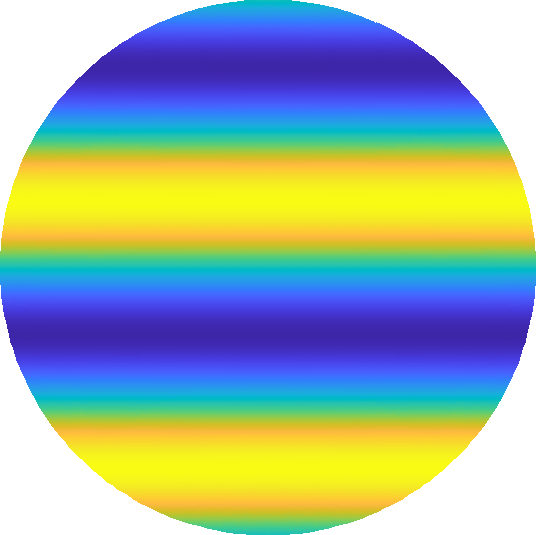}
\includegraphics[scale=0.07]{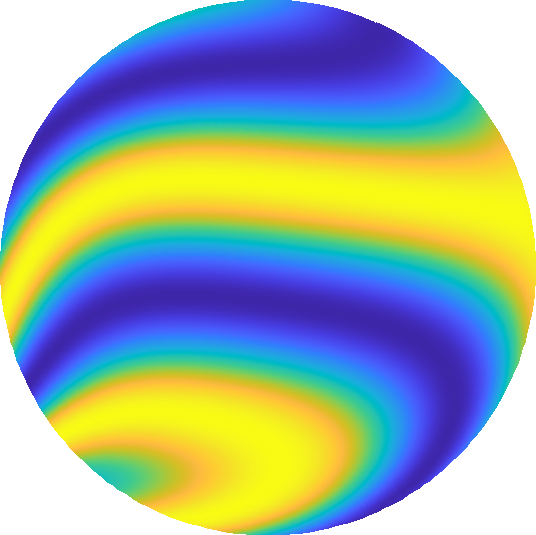}
\includegraphics[scale=0.07]{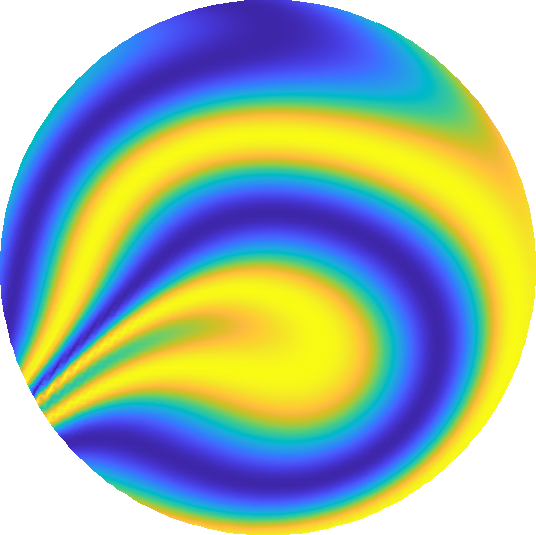}
\includegraphics[scale=0.07]{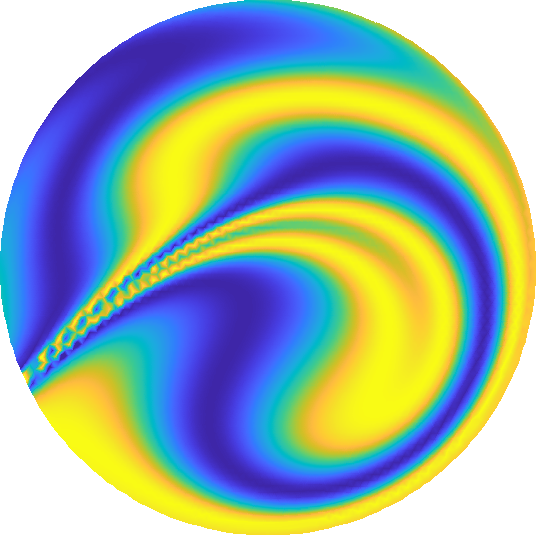}
\includegraphics[scale=0.07]{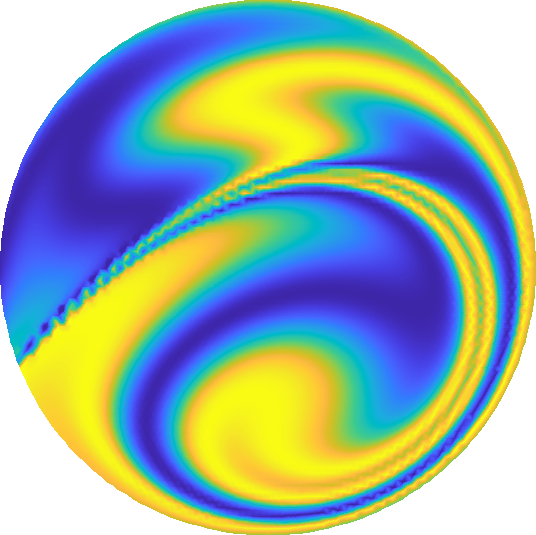}
\includegraphics[scale=0.07]{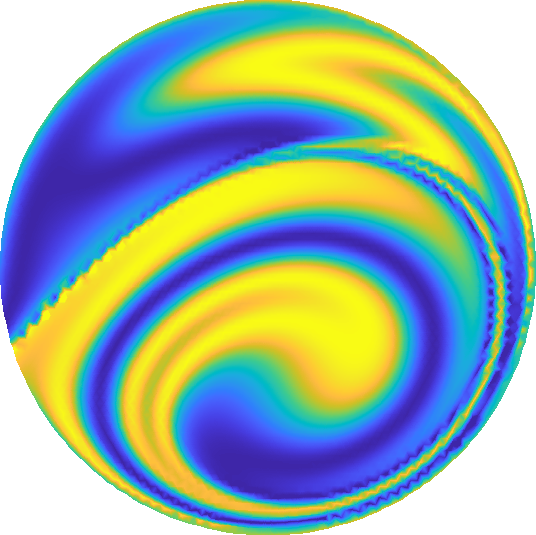}
\includegraphics[scale=0.07]{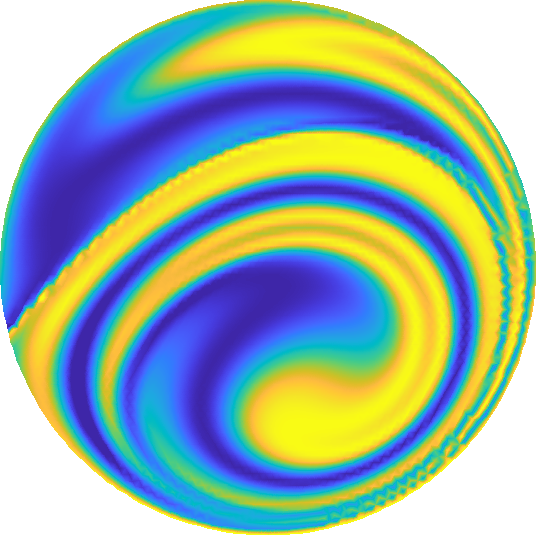}
\includegraphics[scale=0.07]{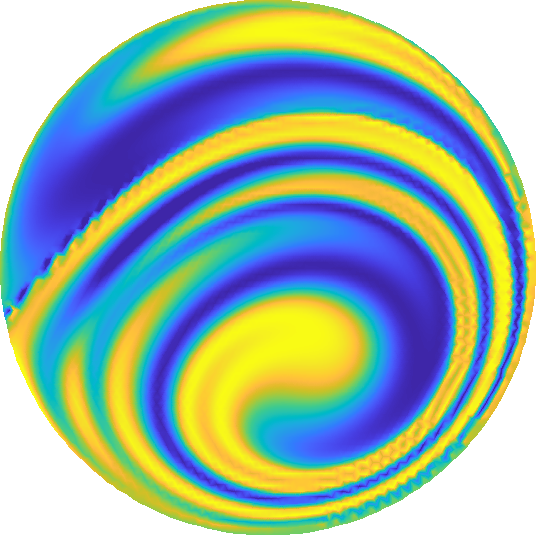}
\includegraphics[scale=0.07]{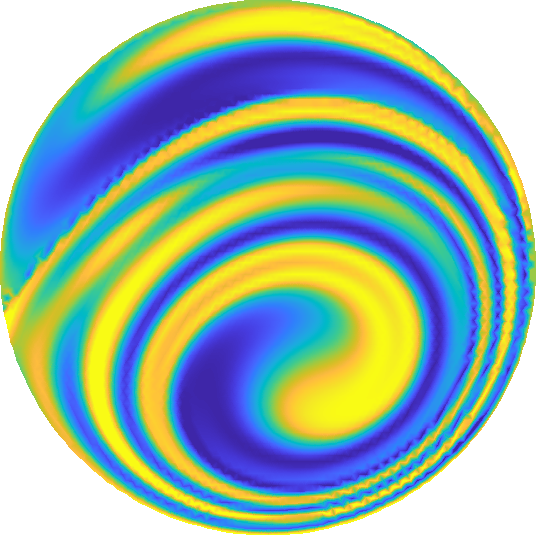}
\includegraphics[scale=0.07]{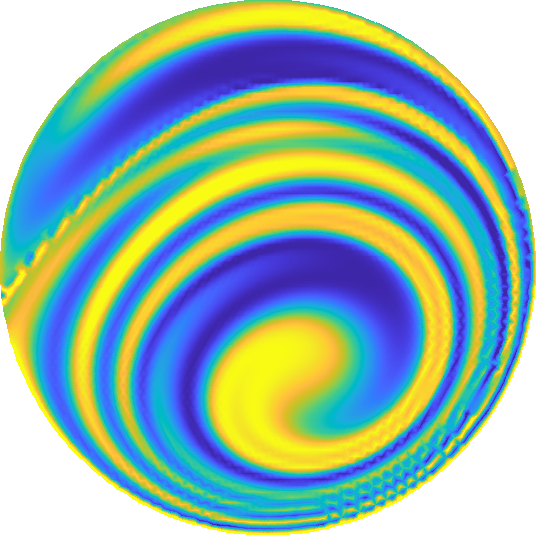}
\includegraphics[scale=0.07]{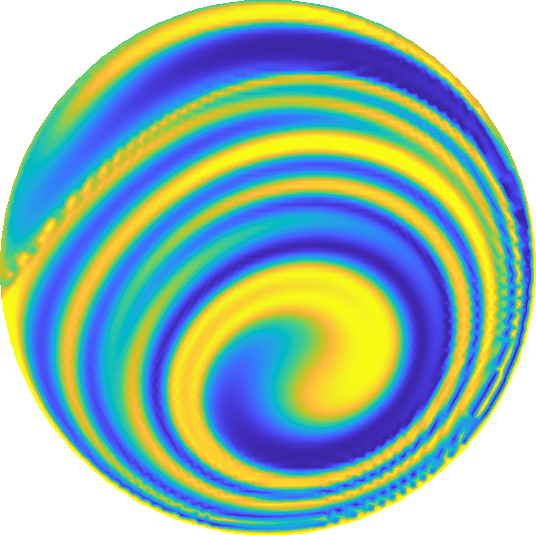}
\\
\includegraphics[scale=0.07]{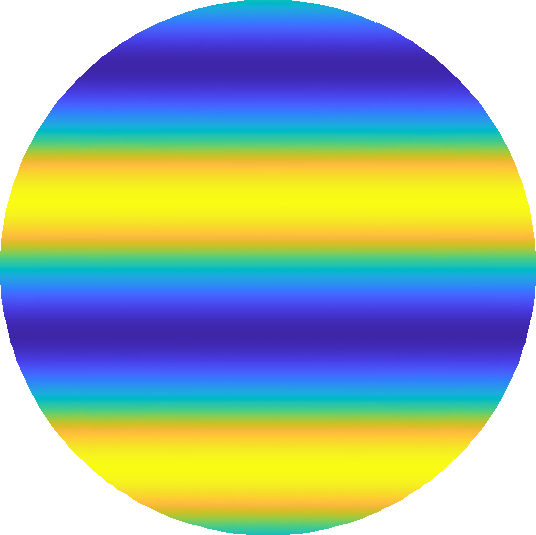}
\includegraphics[scale=0.07]{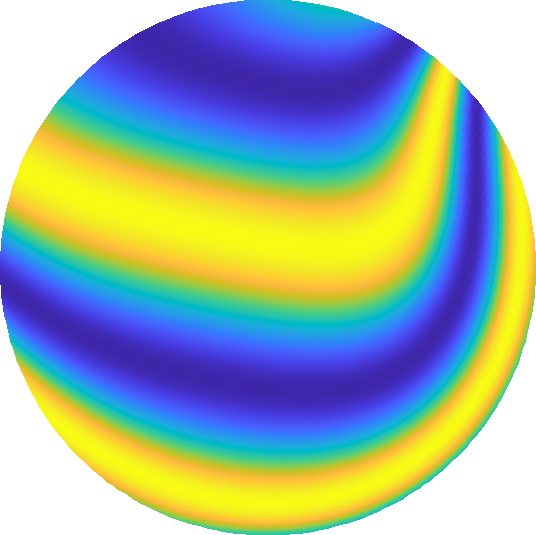}
\includegraphics[scale=0.07]{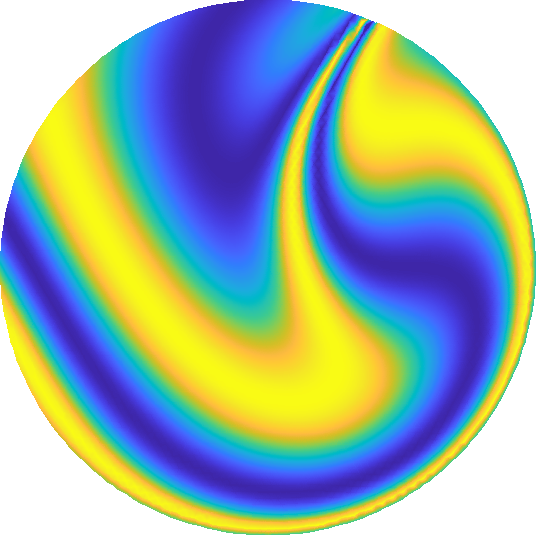}
\includegraphics[scale=0.07]{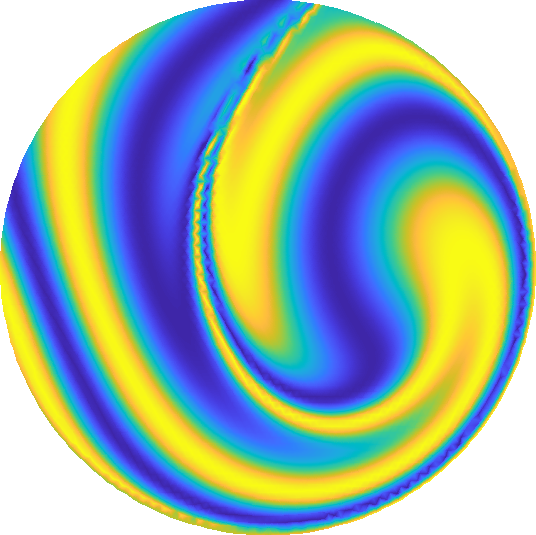}
\includegraphics[scale=0.07]{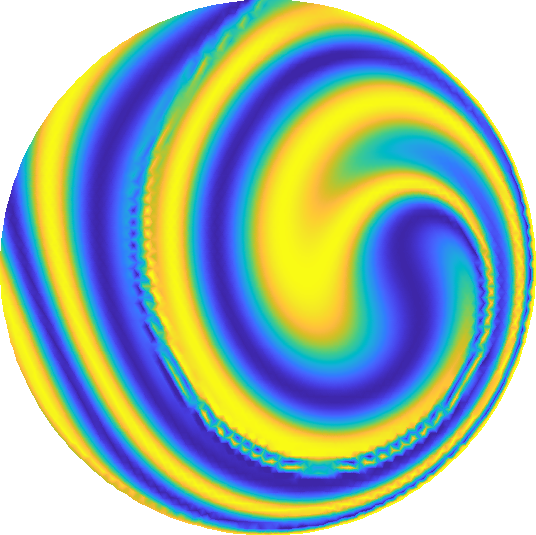}
\includegraphics[scale=0.07]{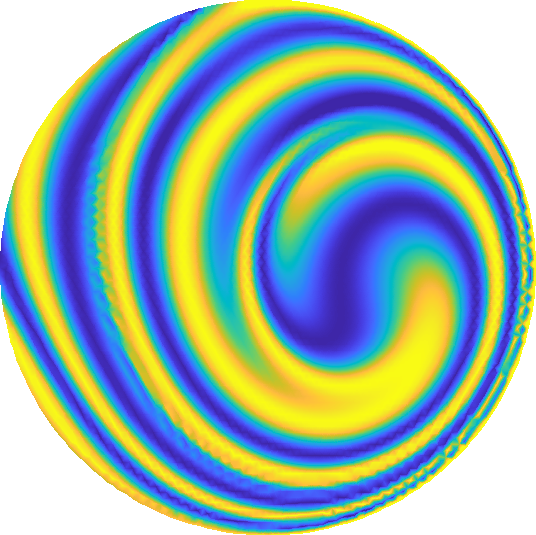}
\includegraphics[scale=0.07]{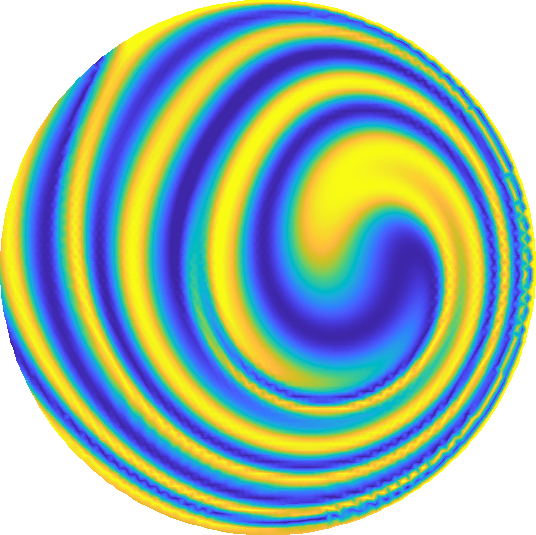}
\includegraphics[scale=0.07]{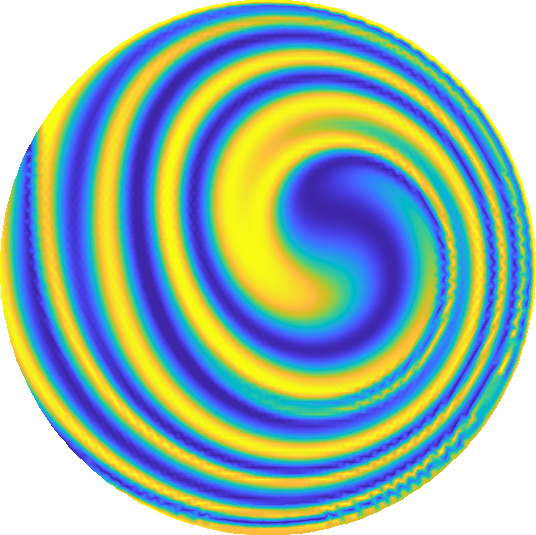}
\includegraphics[scale=0.07]{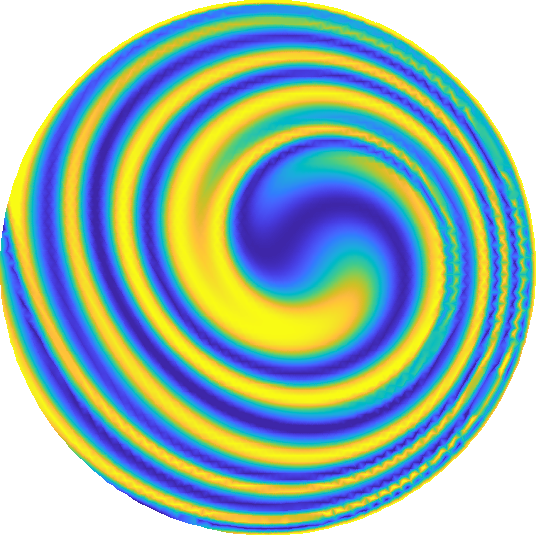}
\includegraphics[scale=0.07]{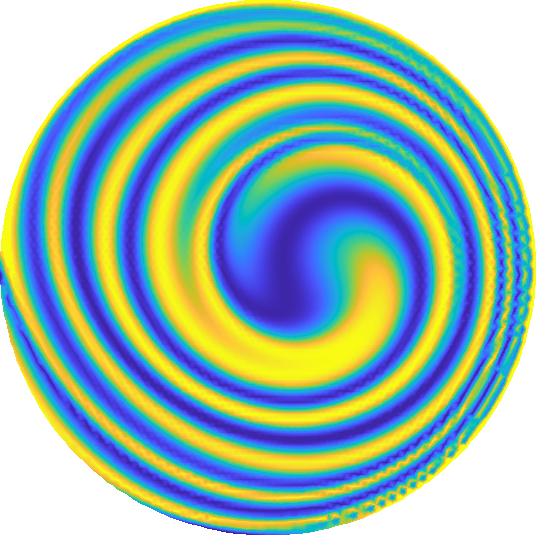}
\includegraphics[scale=0.07]{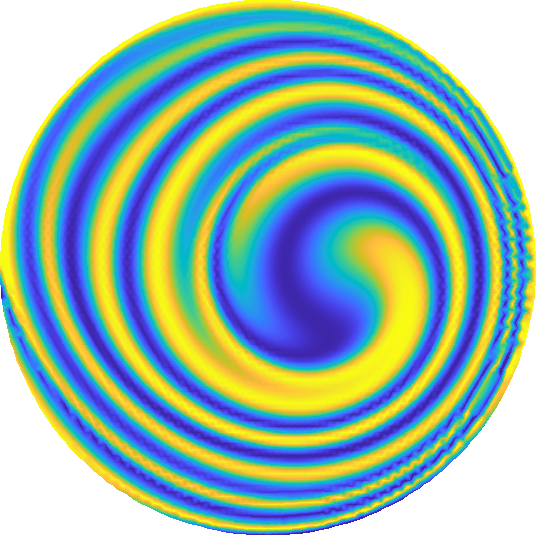}
\\
\includegraphics[scale=0.07]{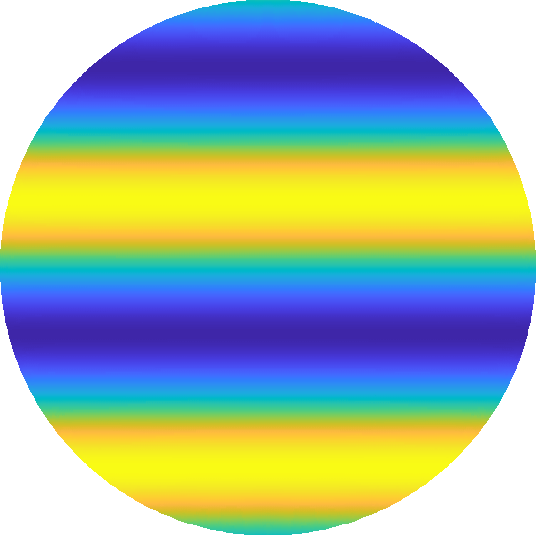}
\includegraphics[scale=0.07]{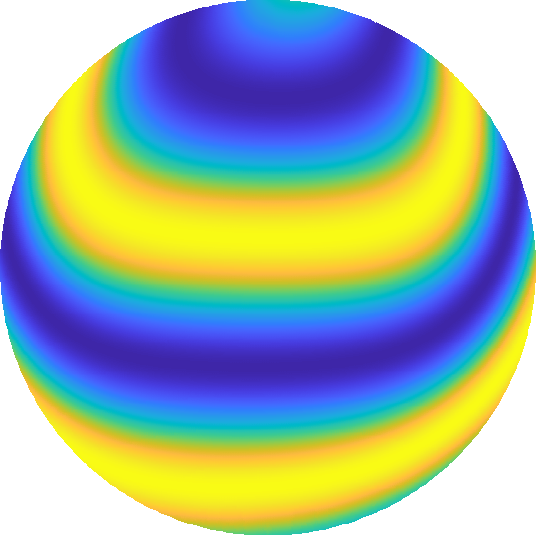}
\includegraphics[scale=0.07]{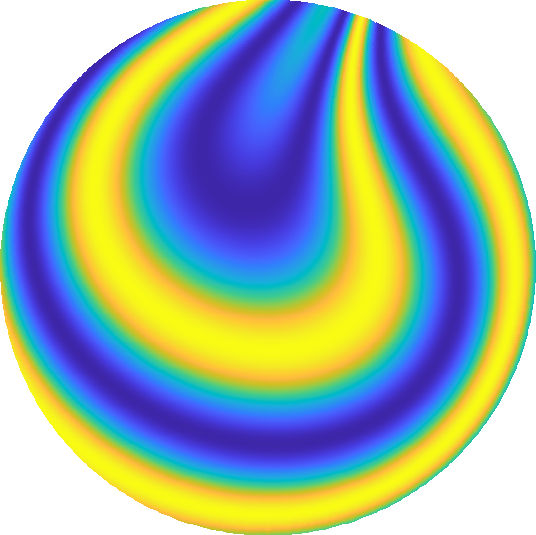}
\includegraphics[scale=0.07]{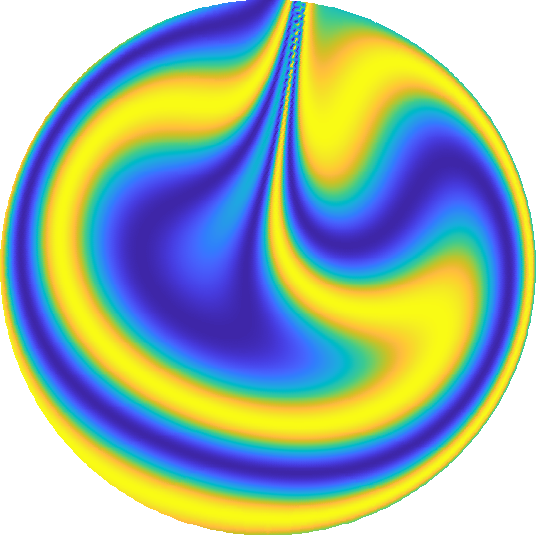}
\includegraphics[scale=0.07]{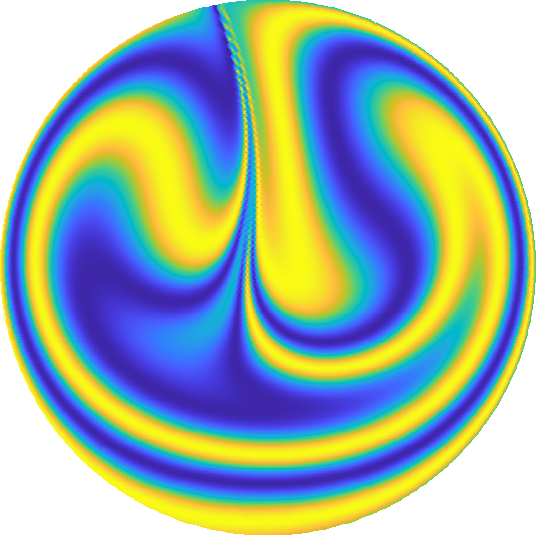}
\includegraphics[scale=0.07]{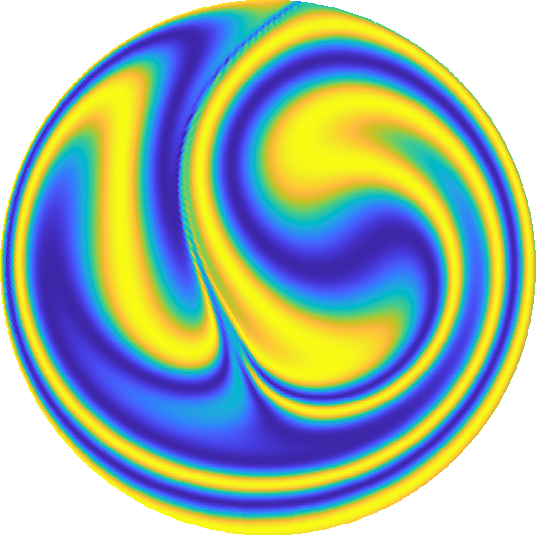}
\includegraphics[scale=0.07]{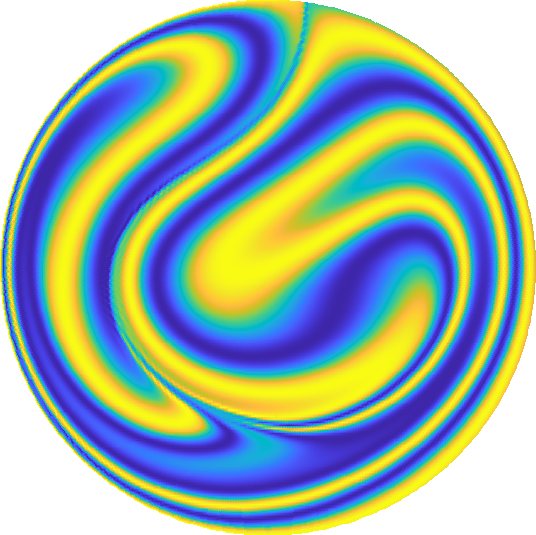}
\includegraphics[scale=0.07]{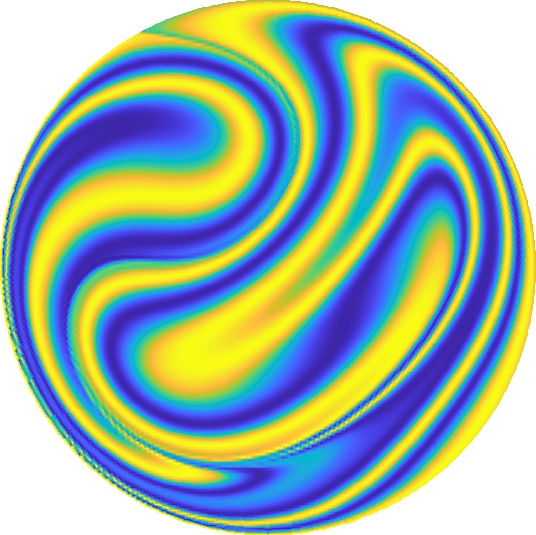}
\includegraphics[scale=0.07]{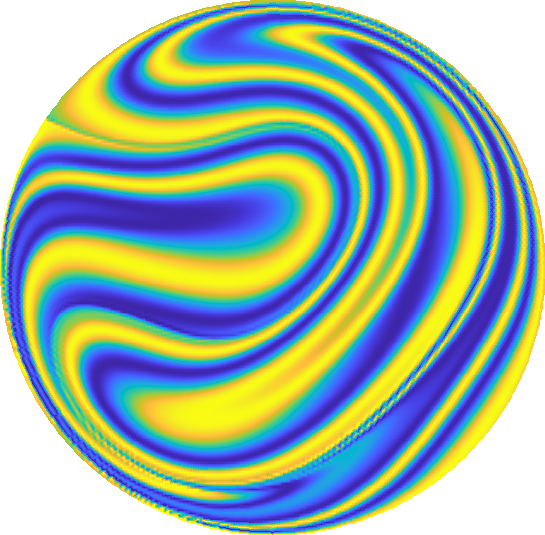}
\includegraphics[scale=0.07]{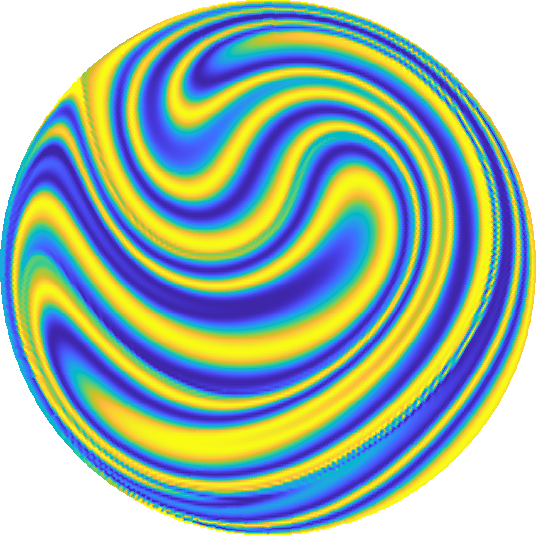}
\includegraphics[scale=0.07]{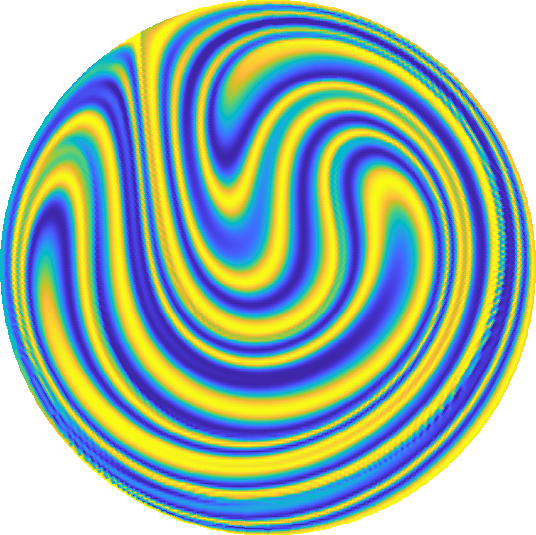}
\end{center}
\caption{Snapshots of $\theta$ at $t=0, 0.1, \cdots, 1$ of the optimal solutions  with five types of control. First row: $N=1$. Second row: $N=2$. Third row: $N=10$. 
\label{fig_alltypes} 
}
\end{figure}

\begin{figure}[htbp]
\begin{center}
\includegraphics[scale=0.3]{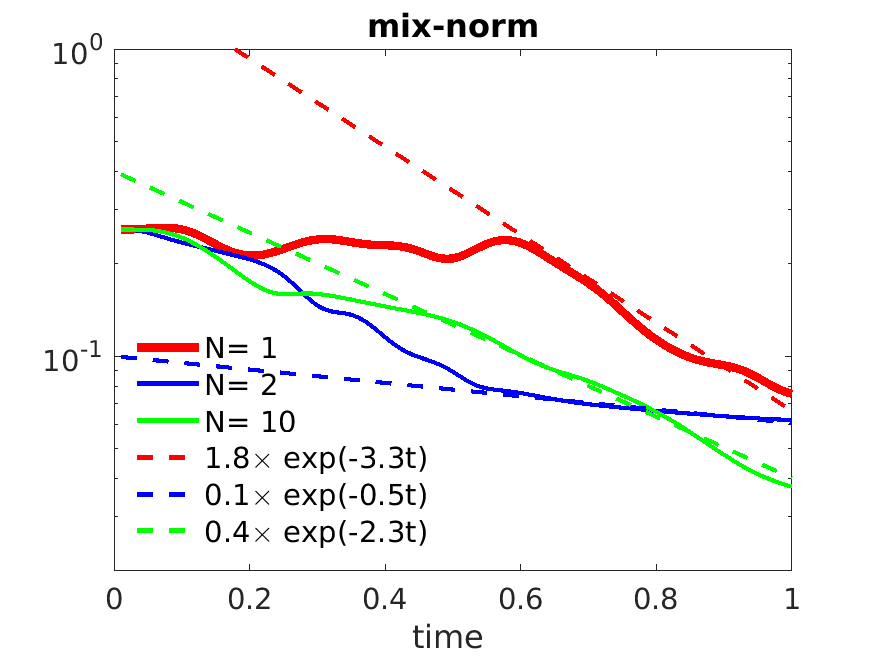}
\end{center}
\caption{ Mix-norm decay over time of three optimal solutions with the combined control types.
\label{fig_exp_decay} 
}
\end{figure}

\begin{figure}[htbp]
\begin{center}
\includegraphics[scale=0.3]{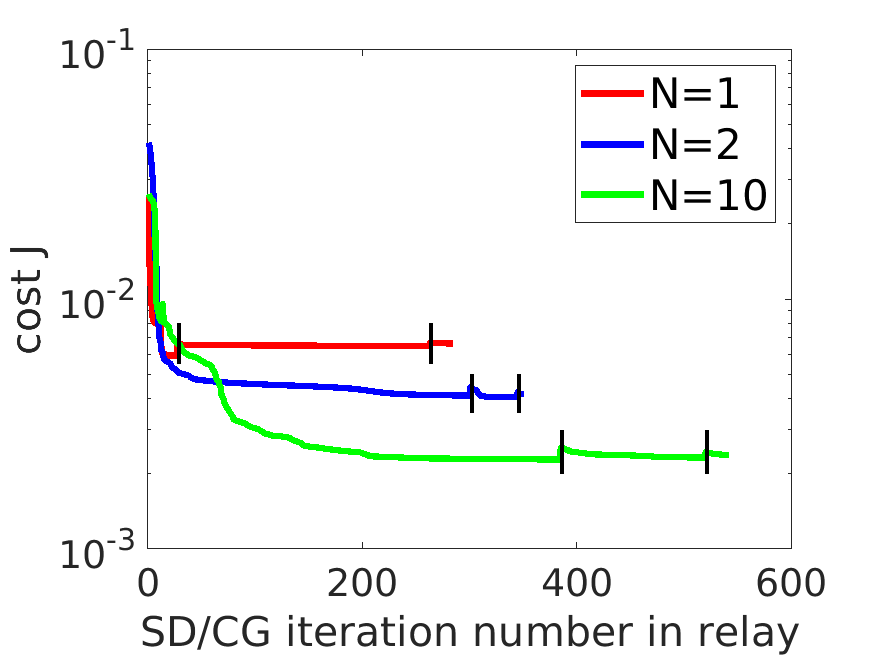}[a]
\includegraphics[scale=0.3]{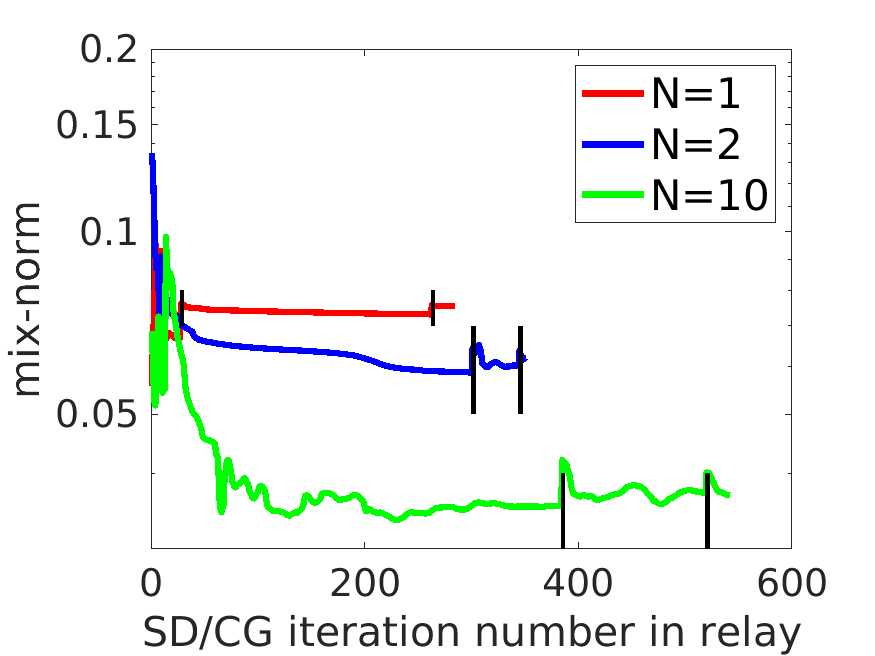}[b]
\includegraphics[scale=0.3]{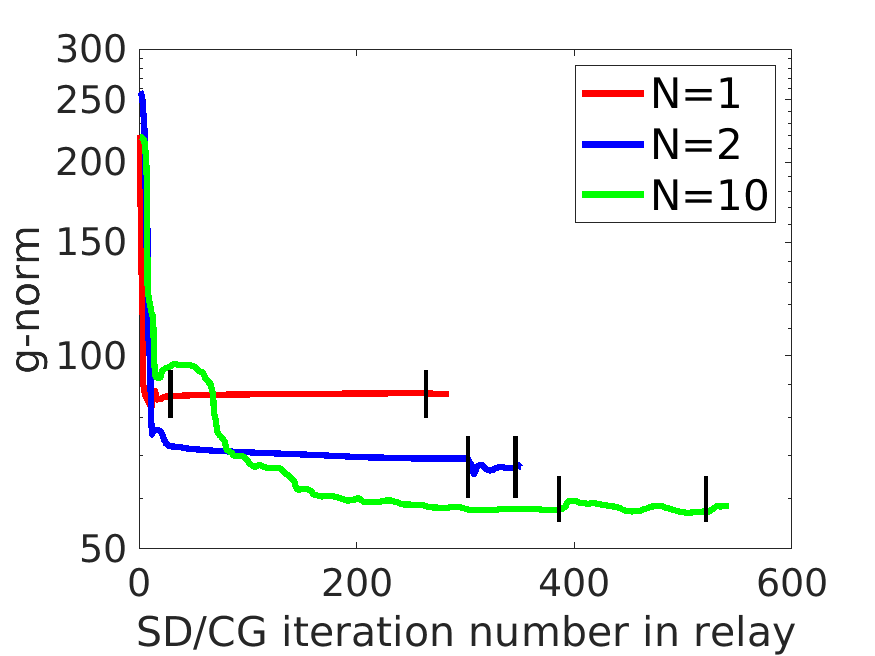}[c]
\end{center}
\caption{
Decays of the cost [a],  mix-norm [b], and  g-norm [c] with respect to iterations in the relay algorithm. The small vertical black bars represents the relay moments when a coarse mesh is replaced with a fine mesh.
\label{fig_alltypes_comp} 
}
\end{figure}

In this section, all the five types of controls used in the last section are combined together to steer mixing, where the time segmentation number $N$ is chosen as $N=1, 2, 10$. The optimal solutions of the control are shown in Figure\,\ref{fig_alltypes_alpha}.  The snapshots of time evolution of the density in the optimal mixing of each value of $N$ are illustrated in Figure\,\ref{fig_alltypes}, which show that the morphology is more complicated when $N$ is larger. Furthermore, when $N$ is larger, the mix-norm, g-norm, and the cost of the optimal solution become smaller (Table\,\ref{table_combined_properties}). 
The minimum cost of the combined control types is 2.37e-3, which is 28\% of the smallest cost 8.42e-3 of only one control type, corresponding to $g^b=\sin(2\omega)$ 
in Table\,\ref{Table0cat_1098}. 
This supports the usage of multiple control types to reduce the cost. 
The mix-norms of these optimal solutions demonstrate the exponential decay in the time window $[0.6, 1]$ (Figire\,\ref{fig_exp_decay}).
Figure\,\ref{fig_alltypes_comp} illustrates the details how the mix-norm, g-norm, and cost decrease with the iteration number in the relay algorithm.

\begin{table}[htbp]
\begin{center}
\caption{Information of optimal solutions when all five types of controls are combined
\label{table_combined_properties}
}
\begin{tabular}{|c|c|c|c|c|}
\hline
$N$ &  mix-norm & g-norm & cost  \\
\hline
1 & 7.55e-2 & 8.71e+1 & 6.64e-3  \\
\hline
2 & 6.20e-2 & 6.70e+1  & 4.17e-3  \\
\hline
10 & 3.68e-2 & 5.83e+1 & 2.37e-3 \\
\hline
\end{tabular}
\end{center}
\end{table}

\section{Conclusions}
\label{conclusions}
This work is the first numerical study of optimal mixing through tangential force exerted on the boundary in the unsteady Stokes flows. In the absence of diffusion, transport and mixing occur due to pure advection.  Built upon the theoretical foundation laid by Hu and Wu, an accurate and efficient optimization algorithm is proposed. The entire algorithm is sophisticated due to the nature of the problem and has many new techniques, which are summarized below. 
\begin{enumerate}
\item[(1)] The boundary control is focused on a finite number of basis functions with time segmentation.  Given the zero initial velocity field, the linear relation between the flow and the control allows the generation of the velocity basis before the optimization process, thus saving the simulation time. 

\item[(2)] The computation of the gradient of the cost functional  is crucial to the numerical accuracy, where a hybrid method is developed to treat different control basis functions with appropriate methods (finite difference or variational formula).

\item[(3)] The combination of several line search methods and descent direction choices are investigated. Specifically, the following two pairs work well: the backtracking with the steepest descent method, and  the exact line search  with the conjugate gradient method.  The simulations demonstrate that the latter performs slightly better than the former in most simulations, but not  significantly. 

\item[(4)] A relay process is placed on the top of this optimization algorithm by repeatedly refining the search from a coarser mesh to a finer mesh. Numerical tests in  Section\,\ref{sec_2Dtest} show that this process produces accuracy results while significantly saving the computational time. 

\end{enumerate}

The numerical simulations reveal the following physical features of mixing by the boundary control design. 
\begin{enumerate}
\item[(1)] The mixing efficacy of only one single type of control function may be limited by the fixed flow pattern, as shown in Section\,\ref{sec_mixingfeatures}. The different  control types derived from  $\cos(\omega)$, $\sin(\omega)$, $\cos(2\omega)$, $\sin(2\omega)$ have separatrices in the domain. But when these types are combined and added the Type 1 control, the separatrices are eliminated. This is consistent with the observation in \cite{thiffeault2011moving}, where the wall rotation removes the separatrices produced by the internal mixing.  Furthermore, the time segmentation of a control, similar to the chaotic mixing strategy, can furthermore increase the mixing efficacy.

\item [(2)] The result of the boundary control can be comparable to that of the internal control, which can be seen from the comparison of the mixed density in Section\,\ref{sec_combined_optimalmixing} with those in \cite{mathew2007optimal},  where the velocity field is generated by the internal stirring.  In addition, it is observed that the mix-norm of the scalar field under the optimal  boundary control reaches the exponential decay rate.

\end{enumerate}

Another unique feature of this work is the use of the dynamic control, where a force is modulated to steer mixing. In contrast, all the existing works from other researchers mentioned at the beginning of Section\,\ref{sec_objectives} have employed the kinematic control, that is, a velocity field is directly modulated. One intrinsic difference between these two types of controls is the inertia, i.e., the perseverance of the motion until it is changed by a force.  In the case of dynamic controls,  the velocity takes a certain time to accelerate from zero to a field with effective mixing when the force is started, and another time duration to decelerate to negligible magnitude after the force is withdrawn. This can be seen clearly in Figure\,\ref{fig_umax}. However, in the case of kinematic controls,  a prescribed velocity field is modulated in an arbitrary manner without consideration of any inertia effects. 
Therefore, the dynamic control would better represent the reality in the mixing problems where the inertia effect is significant.

This work focuses on the tangential boundary force control with the Navier slip boundary conditions, which can model the tangential cilia beating in the inner membrane of vertebrate organs. As described in Section\,\ref{sec_examples}, there are many examples of boundary driven mixing in nature and industry, including rotating wall driven mixing, mircomixers with acoustic waves, and artificial cilia mixing.  Therefore, there is a big potential to extend this work to these applications and beyond. Furthermore, it is interesting to study the effects of combining it with internal controls for optimal mixing problems.

\vspace{0.1in}
\noindent \textbf{Acknowledgments}\,\, 
W. Hu was partially supported by the NSF grant DMS-2111486.
J. Wu was partially supported by the National Science Foundation of USA under grant DMS 2104682 and the AT\&T Foundation at Oklahoma State University. 
This work was supported in part by computational resources
and services provided by HPCC of the Institute for Cyber-Enabled Research at Michigan State University through a collaboration program of Central Michigan University, USA.

\section{Appendix}

\subsection{Derivation of the Gâteaux  derivative\label{appendix_derivation_Gateaux}}

The rigorous derivation of the  first-order  optimality system  for $U_{ad}=L^2(0, T; L^2(\Gamma))$ has been addressed in \cite{hu2020approximating}, using an approximating  control approach. 
Here we provide a short and formal derivation by assuming that all the involved  functions are sufficiently smooth and all the operations are valid.

\begin{thm} \label{app_thm1}
With the governing equations \eqref{EQ01}--\eqref{ini}, the 
Gâteaux  derivative of $J$ with respect to $g$ in the direction $\varphi\in U_{ad}$ is given by
\begin{equation}
DJ(g; \varphi) = \int_0^T (\theta(g) \nabla\rho(g), L\varphi) \,dt
+ \gamma \int_0^T \langle g,\varphi\rangle_\Gamma \,dt,
\end{equation}
where  $\rho(g)$ is the adjoint state satisfying  \eqref{adj}--\eqref{adj_final} and $L\varphi$ is the velocity field governed by the Stokes system
\eqref{Stokes1}--\eqref{ini} with $v_0=0$ and   the tangential  boundary control $g$ replaced by $\varphi$.
\end{thm}
\begin{pf}
Recall from \eqref{cost} that   
$
J(g) = \frac{1}{2} (\Lambda^{-2} \theta(T), \theta(T)) + \frac{\gamma}{2}\int_0^T \langle g,g\rangle _\Gamma dt.
$
Taking the Gâteaux  derivative of $J$ at $g$ in the direction $\varphi$ gives 
\begin{equation}
DJ(g; \varphi) =  (\Lambda^{-2} \theta(T), D\theta(g;\varphi)(T)) + \gamma\int_0^T \langle g, \varphi\rangle _\Gamma dt,
\label{pp100}
\end{equation}
where $D\theta(g;\varphi)(T)$ is the Gâteaux  derivative of $\theta$ at $g$ in the direction $\varphi$ at time $T$. 
Let $z\triangleq D\theta(g;\varphi)$ and $w\triangleq Dv(g;\varphi)$. 
Then $z$ satisfies 
\begin{align}
&\frac{\partial z}{\partial t} 
+v\cdot \nabla z +w \cdot \nabla\theta=0, \label{pp101}\\
 &z( 0)=0. \label{pp102}
\end{align}
Using the notation $v(g)=L(g)$ and the linearity of $L$ when $v_0=0$, we have $w=DL(g;\varphi)=L(\varphi)$, which is also divergence free and $L(\varphi)\cdot n|_{\Gamma}=0$.
Next, taking the inner produce of   \eqref{pp101} with  $\rho$ and integrating with respect to $t$ over $[0, T]$, we get 
\begin{equation}
\int_0^T \left(\frac{\partial z}{\partial t},\rho\right) \,dt
+
\int_0^T (v\cdot \nabla z, \rho) \,dt
+ 
\int_0^T (L(\varphi) \cdot\nabla\theta, \rho) \,dt=0.
\end{equation}
Using $(v\cdot \nabla z, \rho)= -( v\cdot \nabla\rho, z)$ and \eqref{pp102}, the above equation becomes
\begin{equation}
(\rho(T), z(T)) - 
\int_0^T \left(z, \frac{\partial \rho}{\partial t}\right) \,dt
-
\int_0^T (v\cdot \nabla \rho, z) \,dt
+ 
\int_0^T (L(\varphi)\cdot\nabla\theta, \rho) \,dt=0. \label{EST_z}
\end{equation}
Since $\rho$ satisfies \eqref{adj} and \eqref{adj_final}, it follows  from \eqref{EST_z} that 
\begin{equation}
(\Lambda^{-2}\theta(T), D\theta(g,\varphi)(T))=(\rho(T), z(T))
= -\int_0^T (L(\varphi)\cdot\nabla\theta, \rho) dt.
\label{pp103}
\end{equation} 
Finally,  plugging \eqref{pp103} into \eqref{pp100} yields
\begin{align*}
DJ(g;\varphi) = -\int_0^T (L(\varphi)\cdot\nabla\theta, \rho) dt + \gamma \int_0^T \langle g,\varphi\rangle_\Gamma dt  
=  \int_0^T (\theta\nabla \rho, L(\varphi)) dt + \gamma \int_0^T \langle g,\varphi\rangle_\Gamma dt.
\end{align*}
\end{pf}
\begin{remark}
This theorem still holds when the initial velocity $v_0\ne 0$.  In this case, $v(g) = L(g) + \tilde{v}$ where $\tilde{v}$ is the velocity field generated by $v_0$ through the Stokes system \eqref{Stokes1}--\eqref{Stokes3} with $g=0$.  Since $\tilde{v}$ is independent of $g$,  $D\tilde{v}(g;\varphi)=0$.   Thus, $Dv(g;\varphi) = DL(\varphi)(g,\varphi) + D\tilde{v}(g,\varphi)=L(\varphi)$. Then the same proof holds.

\end{remark}

\subsection{Proof of Proposition \ref{lemma_invariance_innerprod} \label{proof_prop}}
\begin{pf}
For any $s\in[0,T]$, taking the  inner product of \eqref{EQ01} with $\rho$ over $ \Omega$ and integrating  in time from $s$ to $T$ gives
\begin{equation}
 \int_s^T (\theta_t, \rho)\,dt + 
 \int_s^T ( v\cdot \nabla\theta, \rho)\,dt =0. \label{1EST_theta}
\end{equation}
Integration by parts leads to
\begin{align*}
( v\cdot \nabla\theta, \rho)=
\langle v\cdot n,  \theta\rho \rangle_\Gamma
-(\nabla \cdot v, \theta\rho)
-(\theta, v\cdot \nabla\rho)
=-(\theta, v\cdot \nabla\rho),
\end{align*}
where the conditions $\nabla \cdot v=0$ and $v\cdot n|_{\Gamma}=0$ are used. 
Thus,  \eqref{1EST_theta} becomes
 \begin{equation*}
\int_s^T (\theta,\rho)_t  -(\theta, \rho_t)dt
- \int_s^T (\theta, v\cdot \nabla\rho) dt=0.
\label{EST_rho}
\end{equation*}
This turns to 
\begin{equation*}
(\rho(T), \theta(T))
-(\rho(s), \theta(s))
=\int^T_s (\theta, \rho_t + v\cdot\nabla\rho)\,dt.
\end{equation*}
Since $\rho_t + v\cdot\nabla\rho=0$, 
\begin{equation}
(\rho(T), \theta(T))
=(\rho(s), \theta(s)).
\end{equation} 
\end{pf}

\subsection{Unsteady Stokes equations: Iterative projection/BDF2/Taylor-Hood finite element method}
\label{section_stokes_solver}
The standard Taylor-Hood P2/P1 elements are employed to approximate the velocity and pressure in the Stokes equations \eqref{Stokes1}--\eqref{Stokes3}. That is, the velocity is approximated by the continuous piecewise quadratic functions and the pressure by the continuous piecewise linear functions. Denote the triangulated domain as $\Omega_h$ where all the elements are triangles.   The finite element spaces are defined  as
\begin{eqnarray}
V_h &=& \{  
w=(w_1,w_2)\in (C^0(\Omega))^2: w\cdot n|_{\Gamma}=0, w_i|_K \in P^2(K), i=1,2, \forall K\subset \Omega_h \}, \label{fem_Vh} \\
Q_h &=& \{ q\in C^0(\Omega): q|_K\in P^1(K), \forall K\subset \Omega_h  \}, \label{fem_Qh}
\end{eqnarray}
where  $n$ is the unit outward normal  on the boundary.

The basis functions of $V_h$ are chosen as follow. Denote the inner nodes of the mesh as $x_i$, $i=1, \cdots, N_I$ and the boundary nodes as $x^B_j$, $j=1,\cdots, N_B$.
Denote $\phi_i$ as the scalar basis function that is continuous in $\Omega$, piecewise quadratic in each element, taking value $1$ at node $i$ and zero on all other nodes. Let vectors $e_1=(1,0)^T$ and $e_2=(0,1)^T$. At an inner node $x_i$, there are two basis functions of velocity, which are $\phi_i e_1$ and $\phi_i e_2$.
At a boundary node $x^B_j$, 
there is only one basis function, $\phi_j e_\tau$, where $e_\tau$ is the unit tangential vector at $x^B_j$.

The weak form of equations of \eqref{Stokes1}--\eqref{ini}  is finding $v\in V_h$ and $p\in Q_h$ such that for all $w\in V_h$ and $q\in Q_h$,
\begin{eqnarray}
\int_\Omega \frac{\partial v}{\partial t} \cdot w 
+ 2\int_\Omega \mathbb{D}(v)\cdot \mathbb{D}(w) 
+ \int_{\Gamma} k (v\cdot \tau) (w\cdot\tau)
- \int_\Omega p \nabla\cdot w
&\!\!\!\!=\!\!\!\!& \int_{\Gamma} g  (w\cdot \tau),
\label{eqn_weak1}\\
\int_\Omega q \nabla\cdot v  &\!\!\!\!=\!\!\!\!& 0,
\label{eqn_weak2}
\end{eqnarray}
where $v = (v_1, v_2)^T$, $w = (w_1, w_2)^T$, 
$\mathbb{D}(v)\cdot \mathbb{D}(w) =\frac{1}{4} \sum_{i,j=1,2} (\partial_i v_j+\partial_j v_i) (\partial_i w_j+\partial_j w_i)$.

An iterative projection method with BDF2 time discretization is used to solve the velocity and pressure \cite{zheng2023iterative}. Denote the numerical solution at the time step $t^s$ as $(v^{s}, p^{s})$. To obtain $(v^{s+1}, p^{s+1})$, we use the following iterations with index  $l$.
For $l=0,1,2,\cdots$, let 
$p^{s+1,0}=(2p^s-p^{s-1})$, and 
\begin{eqnarray}
\int_\Omega \frac{1.5 \tilde v^{s+1, l+1}-2 v^s+ 0.5 v^{s-1}}{\Delta t} \cdot w 
&+& 2\int_\Omega \mathbb{D}(\tilde v^{s+1,l+1})\cdot\mathbb{D}(w)
+ \int_{\Gamma} k (\tilde v^{s+1,l+1} \cdot\tau) (w\cdot \tau) 
\nonumber \\
 &=&
 \int_\Omega p^{s+1,l} \nabla\cdot w 
 + \int_{\Gamma} g (w\cdot \tau) , 
\quad \forall w\in V_h,
\label{pp001} \\
\int_\Omega \nabla\phi^{l+1} \cdot \nabla q &=& -\frac{1}{\Delta t} \int_\Omega q 
(\nabla\cdot \tilde v^{s+1,l+1}), 
\quad \forall q\in Q_h,
\label{pp002}
\\
\int_\Omega p^{s+1,l+1} q &=& \int_\Omega (p^{s+1,l}   + 1.5 \phi^{l+1} -  
 \nabla\cdot \tilde v^{s+1,l+1}) q, \quad \forall q\in Q_h,
\label{pp003} \\
\int_\Omega v^{s+1,l+1} \cdot w &=& \int_\Omega \tilde v^{s+1,l+1} \cdot w - \Delta t \phi^{l+1}  (\nabla\cdot w), \quad \forall w\in V_h.
\label{pp004}
\end{eqnarray}
The stopping criterion  for the iterations is chosen as when 
$||p^{s+1,l+1}- p^{s+1,l}||_{L^2(\Omega)} < \varepsilon_s$.  When convergent, we let $(v^{s+1}, p^{s+1})=(v^{s+1,l+1}, p^{s+1,l+1})$ and have the estimate   
\begin{eqnarray}
\bigg| \int_\Omega q (\nabla\cdot v^{s+1}) \bigg|<\varepsilon_s, 
\forall q\in Q_h.
\label{iterative_div_free}
\end{eqnarray}
The threshold $\epsilon_s$ is set as $10^{-10}$ in this work. Therefore, although the divergence of the numerical velocity is  not pointwise zero, it is almost zero in the weak sense.

\subsection{Transport equations: Discontinuous Galerkin method}
\label{sec_advection}
A standard Runge-Kutta Discontinuous Galerkin (RKDG) scheme \cite{CockburnShu2001RKDGreview} is used to solve the scalar $\theta$ governed by the transport equation \eqref{EQ01}, and the adjoint quantity $\rho$ from \eqref{adj}.
Define the discontinuous finite element space 
\begin{equation}
W^{DG}_{h,M_{DG}}=\{ w_h\in P^{M_{DG}}(K), \forall K\subset \Omega_h  \},
\label{DG_space}
\end{equation}
where $P^{M_{DG}}(K)$ denotes the set of  $M_{DG}$-th degree polynomials in each triangle $K$ of the discrete domain $\Omega_h$. 
To ensure stability, a Courant-Fredrichs-Lewy  (CFL) condition is used to determine the time step size $\Delta t$,
\begin{equation}
||v||_{\max} \cdot \frac{\Delta t}{h} \le \text{CFL}_{L^2}
\label{CFL_condition}
\end{equation}
where the constant $\text{CFL}_{L^2}$ for  degree $M_{DG}$ of polynomials is given in Table 2.2 of  \cite{CockburnShu2001RKDGreview}.

To show the idea, a  first-order  temporarily discretized numerical scheme is given as follows. Given the numerical solution $\theta^s\in W^{DG}_{h,M_{DG}}$ at time step $t^s$, we obtain $\theta^{s+1}\in W^{DG}_{h,M_{DG}}$ from 
\begin{eqnarray}
\int_K \frac{\theta^{s+1} - \theta^s}{\Delta t} \phi
+ \int_{e\subseteq \partial K} (v\cdot\hat{n}) \theta^s \phi 
- \int_K \theta (v\cdot \nabla\phi) =   \int_K \theta (\nabla\cdot v) \phi, 
\quad \forall \phi\in P^{M_{DG}}(K),
\end{eqnarray}
where $\hat{n}$ is the unit outward normal on  edge $e$ of $K$ and $\hat\theta^s$ is the numerical flux. The Godunov flux (see \cite{CockburnShu2001RKDGreview} page 206) is used, i.e., 
\begin{equation}
\hat \theta^s|_{\partial K} = 
\left\{
\begin{array}{cc}
\theta^s|_{K} & v\cdot \hat{n} >0,\\
\theta^s|_{K+} & v\cdot \hat{n} <0, 
\end{array}
\right.
\end{equation}
where $K+$ is the neighbour triangle that $K$ bounds across the edge $e$. In practice, we use a second order TVD-RK scheme in time and quadratic DG approximations in space ($M_{DG}=2$ in \eqref{DG_space}), of which the  details can be found in \cite[page 190]{CockburnShu2001RKDGreview}.

\subsubsection{Choices of basis functions of $P^{M_{DG}}(K)$ and quadrature rules}
The basis functions of $P^{M_{DG}}(K)$, $M_{DG}\ge 0$, $K\subset \Omega_h$  are chosen as follows. Denote the center point of $K$ as $(x_0, y_0)$
and a generic basis function as $\phi_{i,j}=(x-x_0)^i (y-y_0)^j$, $i\ge 0,  j\ge 0$, $i+j\le M_{DG}$. There are $M_t=(M_{DG}+1)(M_{DG}+2)/2$ such basis functions, or $dim( P^{M_{DG}}(K))=M_t$. 
For any smooth function $\theta(x,y)$, its representation $\theta_h\in P^{M_{DG}}(k)$ has the expression 
$\theta_h=\sum_{i,j\ge 0}^{i+j\le M} \theta_{i,j} \phi_{i,j}$, 
where $\theta_{i,j} = \frac{1}{i! j!} \frac{\partial^{i+j} \theta(x_0,y_0)}{\partial x^i \partial y^j} $.
Re-order these bases as $\psi_s=\phi_{i,j}$ where $s=(i+j)(i+j+1)/2+(j+1)$, which is a one-to-one correspondence from the double-index set $\{(i,j): i\ge 0, j\ge 0, i+j\le M_{DG} \}$ to the single-index set $\{1,\cdots, M_t \}$.

The mass matrix $A_{M_t\times M_t}$ on each triangle $K$ is 
\begin{equation}\label{def_DG_matrix}
A_{i,j} = \int_K \psi_i(x,y) \psi_j(x,y) dxdy, 
\quad i,j=1,\cdots, M_t.
\end{equation}
Suppose the above integral is approximated by the following quadrature rule, 
\begin{equation}\label{def_quadrature}
\int_K f(x,y) \approx \sum_{l=1}^{G} w_l f(x_l,y_l),
\end{equation}
where all the weight $w_l >0$.
Denote the resulting matrix generated from the above quadrature rule as $A^{G}$. The next lemma provides a necessary condition to ensure the invertibility of $A^G$.
\begin{lemma}
For the matrix $A^G$ to be invertible, the number of quadrature points in  the triangular integral \eqref{def_quadrature} which approximates \eqref{def_DG_matrix} must be greater than or equal to the number of basis functions of $P^M(K)$, that is, $G\ge M_t$.
\end{lemma}
\begin{pf}
For any $c\in \mathbb{R}^{M_t}$, 
$c^T A^G c =  \sum_{l=1}^{G} \sum_{i,j=1}^{M_t} w_l c_i \psi_i(x_l,y_l) \psi_j(x_l,y_l) c_j $. Let $f_l= \sum_{i=1}^{M_t} c_i \psi_i(x_l,y_l)$. Then $c^T A^G c =  \sum_{l=1}^{G} w_l f_l^2\ge 0$ since $w_l>0$. It is clear that the matrix $A^G$ is symmetric and positive semi-definite. To be invertible, it requires that $A^G$ is positive definite or $c^T A^G c=0$ has only the zero solution $c=0$. Because $w_l>0$ for  $l=1, \cdots, G$, it leads to $f_l=0$ for all $l$, i.e., $\sum_{i=1}^{M_t} \psi_i(x_l,y_l) c_i=0$. This system has $G$ linear equations and $M_t$ variables ($c_i$). If $G<M_t$, then this system must have free variables and thus nonzero solutions. 
\end{pf}

Some choices of basis functions and quadrature rules are given in Table\,\ref{table_basis_quadrature}. In the implementations with $M_{DG}=3$ or  $4$, a 16-point Gaussian quadrature rule on a triangle from \cite{Zhanglinbo2009} is used, which is exact for 8-th degree polynomials. As for the line integral,  a 16-point quadrature rule in \cite{DavisRabinowitz1956} is used, which is exact for polynomials of degree $\le 31$. In the implementations with $M=$0, 1, or 2, a 7-point Gaussian quadrature rule on a triangle is used, which is exact for 5-th degree polynomials, and  a 3-point quadrature rule is used for the line integral, which is exact for polynomials of degree $\le 5$.
\begin{table}[htp]
\caption{Choices of basis functions and quadrature rule on a triangle for DG method
\label{table_basis_quadrature}}
\vspace{-0.7cm}
\begin{center}
\begin{tabular}{|c|c|c|c|c|c|c|c|}
\hline
$M_{DG}$, order of polynomial &  0 & 1 & 2 & 3 & 4 & 5\\
\hline 
$M_t=(M_{DG}+1)(M_{DG}+2)/2$, dimension of $P^{M_{DG}}(K)$ & 1 & 3 & 6 & 10 & 15 & 21 \\
\hline
$G$, minimum number of quadrature points & 1 & 3 & 6 & 10 & 15 & 21\\ 
\hline
\end{tabular}
\end{center}
\end{table}

\subsection{A simple check of the numerical code for the solution of $v$, $\theta$ and $\rho$}
\label{sec_simple_check}
\begin{figure}[!htbp]
\begin{center}
\includegraphics[scale=0.4]{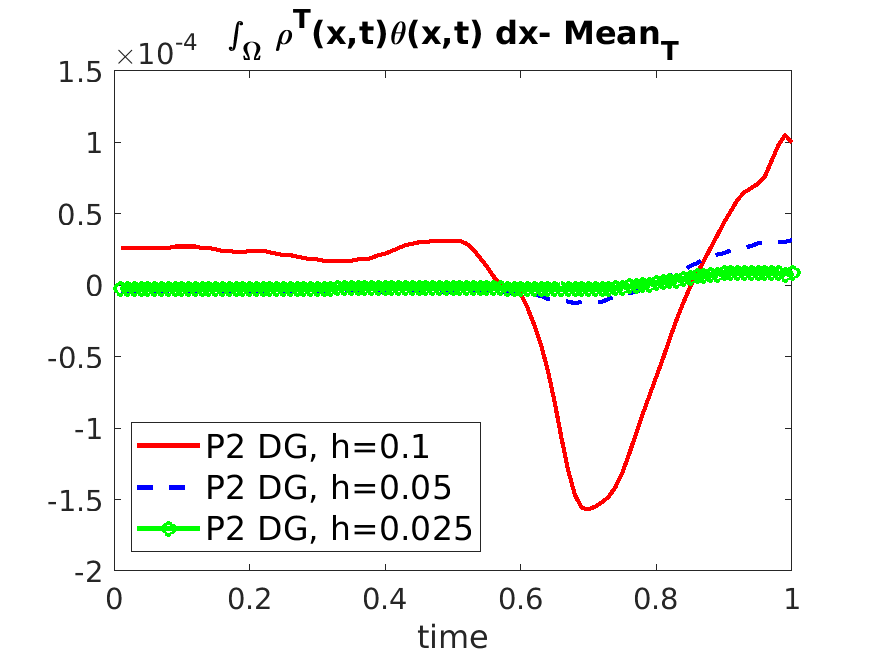}
\caption{
\label{testing_innerprod} 
A test for Proposition\,\ref{lemma_invariance_innerprod}: $(\int_\Omega \rho_T (x,t) \theta_T(x,t) dx - Mean_T)$ over time.  
Initial value $\theta_0=\sin(2\pi y)$, $g=10\cos(2\omega)$ if $t\in [0,0.5]$ and $g=20\sin(2\omega)$ if $t\in [0.5,1]$. $Mean_T$ is the mean value in time defined in \eqref{def_meanvalue}.
}
\end{center}
\end{figure}

We make use of Proposition\,\ref{lemma_invariance_innerprod}  to check the code  that  solves the velocity $v$ from the Stokes equations from given controls, evolves $\theta$ with $v$ from $t=0$ to $t=T$, computes $\rho(T)=\Lambda^{-2} \theta(T)$,  and transports $\rho(t)$ backward with $v$ from $t=T$ to $t=0$. We set  $\theta_0=\sin(2\pi y)$ and choose  control $g=10\cos(2\omega)$ when $t\in [0,0.5]$ and $g=20\sin(2\omega)$ when $t\in [0.5,1]$.
The velocity $v$ is computed  using the iterative projection scheme in Section\,\ref{section_stokes_solver} and  $\theta$ and $\rho$ are solved  by DGP2 ($M_{DG}=2$) method in Section\,\ref{sec_advection}. 
The test results are shown in Figure\,\ref{testing_innerprod},  where
\begin{equation}
Mean_T = \frac{1}{T} \int_0^T \int_\Omega \rho(x,t) \theta(x,t) dx dt.
\label{def_meanvalue}
\end{equation}
In this test, $T=1$. The maximum error of $(\int_0^T \int_\Omega \rho^T(x,t) \theta(x,t) \,dx - Mean_T)$ over $t\in [0,1]$ 
is 1.05e-4 when $h=0.1$, 3.15e-5 when $h=0.05$, and 8.70e-6 when $h=0.025$, which shows roughly second order convergence to zero when the mesh is refined. This partially verifies the code.

\subsection{Radially symmetric steady flow in the unit disk when $g=1$}
\label{sec_radialsymmetry}
In polar coordinates $(r,\varphi)$, denote the velocity as $v=v_r \hat{e}_r + v_\varphi\hat{e}_\varphi$, where ${e}_r$ and ${e}_\varphi$ are unit vectors in the direction $r$ and $\varphi$. 
The divergence free condition is $\nabla\cdot v = \frac{1}{r}\frac{ \partial (r v_r)}{\partial r}  + \frac{1}{r} \frac{\partial v_\varphi}{\partial\varphi}=0$. 
Under the radial symmetry assumption, $v_r=v_r(r)$, $v_\varphi=v_\varphi(r)$, $p=p(r)$, and $v_r(0)=v_\varphi(0)=0$. Thus, the divergence free condition becomes $\frac{1}{r}\frac{ \partial (r v_r)}{\partial r}  =0$, which gives $v_r(r)=0$ in the disk.

In general, 
$\nabla v + (\nabla v)^T = 
\left(\begin{array}{cc}
2\frac{\partial v_r}{\partial r} & \frac{1}{r} \frac{\partial v_r}{\partial\varphi} + \frac{\partial v_\varphi}{\partial r} - \frac{v_\varphi}{r} \\
\frac{1}{r} \frac{\partial v_r}{\partial\varphi} + \frac{\partial v_\varphi}{\partial r} - \frac{v_\varphi}{r} &
\frac{2}{r} \frac{\partial v_\varphi}{\partial\varphi} + \frac{v_r}{r}
\end{array}
\right)
$.
With radial symmetry and $v_r=0$, the steady state  momentum equations become
$\frac{\partial p}{\partial r}=0$ and 
$-\frac{\partial}{\partial r} \left(\frac{1}{r}
\frac{\partial (r v_\varphi)}{\partial r}\right) =0$
when $0<r<1$. 
The Navier-slip boundary condition on the unit circle becomes
$\frac{\partial v_\varphi}{\partial r} + (k-1) v_\varphi =g$.
These three equations admit a unique solution: $v_\varphi=\frac{g}{k}r$ and  $p$ is a constant.


\bibliographystyle{abbrv}

\bibliography{ref}

\begin{thebibliography}{10}

\bibitem{Ahmed2009}
D.~Ahmed, X.~Mao, B.~Juluri, and et~al.
\newblock A fast microfluidic mixer based on acoustically driven
  sidewall-trapped microbubbles.
\newblock {\em Microfluid Nanofluid}, 7:727, 2009.

\bibitem{alberti2019exponential}
{Alberti, G.}, {Crippa, G.}, and {Mazzucato, A.}
\newblock Exponential self-similar mixing by incompressible flows.
\newblock {\em Journal of the American Mathematical Society}, 32(2):445--490,
  2019.

\bibitem{Burden2016}
R.~Burden, J.~Faires, and A.~Burden.
\newblock {\em Numerical Analysis}.
\newblock Cengage Learning, 10th edition, 2016.

\bibitem{chakravarthy1996mixing}
{Chakravarthy, V. S.} and {Ottino, J. M.}
\newblock Mixing of two viscous fluids in a rectangular cavity.
\newblock {\em Chemical Engineering Science}, 51(14):3613--3622, 1996.

\bibitem{Chateau2018}
S.~Chateau, U.~d’Ortona, S.~Poncet, and J.~Favier.
\newblock Transport and mixing induced by beating cilia in human airways.
\newblock {\em Front. Physiol.}, page 161, 2018.

\bibitem{CockburnShu2001RKDGreview}
{Cockburn, B.} and {Shu, C.-W.}
\newblock Runge-kutta discontinuous galerkin methods for convection-dominated
  problems.
\newblock {\em Journal of Scientific Computing}, 16(3):173--261, 2001.

\bibitem{crippa2019polynomial}
{Crippa, G.}, {Luc{\`a}, R.}, and {Schulze, C.}
\newblock Polynomial mixing under a certain stationary euler flow.
\newblock {\em Physica D: Nonlinear Phenomena}, 394:44--55, 2019.

\bibitem{DavisRabinowitz1956}
{Davis, P.} and {Rabinowitz, P.}
\newblock Abscissas and weights for gaussian quadratures of high order.
\newblock {\em Journal of Research of the National Bureau of Standards},
  56:35--37, 1956.

\bibitem{Ding2014}
Y.~Ding, J.~Nawroth, M.~McFall-Ngai, and E.~Kanso.
\newblock Mixing and transport by ciliary carpets: A numerical study.
\newblock {\em Journal of Fluid Mechanics}, 743:124--140, 2014.

\bibitem{elgindi2019universal}
{Elgindi, T.M.} and {Zlato{\v{s}}, A.}
\newblock Universal mixers in all dimensions.
\newblock {\em Advances in Mathematics}, 356:106807, 2019.

\bibitem{Glowinski2022}
R.~Glowinski, Y.~Song, X.~Yuan, and H.~Yue.
\newblock Bilinear optimal control of an advection-reaction-diffusion system.
\newblock {\em SIAM Review}, 64(2):392--421, 2022.

\bibitem{gouillart2008slow}
{Gouillart, E.}, {Dauchot, O.}, { Dubrulle, B.}, {Roux, S.}, and {Thiffeault,
  J.-L.}
\newblock Slow decay of concentration variance due to no-slip walls in chaotic
  mixing.
\newblock {\em Physical Review E}, 78(2):026211, 2008.

\bibitem{gouillart2007walls}
{Gouillart, E.}, {Kuncio, N.}, {Dauchot, O.}, {Dubrulle, B.}, {Roux, S.}, and {
  Thiffeault, J.-L.}
\newblock Walls inhibit chaotic mixing.
\newblock {\em Physical review letters}, 99(11):114501, 2007.

\bibitem{gouillart2010rotation}
{Gouillart, E.}, {Thiffeault, J.-L.}, and { Dauchot, O.}
\newblock Rotation shields chaotic mixing regions from no-slip walls.
\newblock {\em Physical review letters}, 104(20):204502, 2010.

\bibitem{Griva2009}
{Griva, I.}, {Nash, S.G.}, and {Sofer, A.}
\newblock {\em Linear and nonlinear Optimization}.
\newblock SIAM, 2009.

\bibitem{gubanov2010towards}
{Gubanov, O.} and {Cortelezzi, L.}
\newblock Towards the design of an optimal mixer.
\newblock {\em Journal of Fluid Mechanics}, 651:27--53, 2010.

\bibitem{Guo2020}
H.~Guo, H.~Zhu, and S.~Veerapaneni.
\newblock Simulating cilia-driven mixing and transport in complex geometries.
\newblock {\em Phys. Rev. Fluids}, 5:053103, May 2020.

\bibitem{hinze2008optimization}
{Hinze, M.}, {Pinnau, R.}, {Ulbrich, M.}, and {Ulbrich, S.}
\newblock {\em Optimization with PDE constraints}, volume~23.
\newblock Springer Science \& Business Media, 2008.

\bibitem{hu2020approximating}
W.~Hu.
\newblock An approximating control design for optimal mixing by {S}tokes flows.
\newblock {\em Applied Mathematics \& Optimization}, 82:471--498, 2020.

\bibitem{hu2018boundarycontrol}
{Hu, W.}
\newblock Boundary control for optimal mixing by {S}tokes flows.
\newblock {\em Applied Mathematics \& Optimization}, 78(1):201--217, 2018.

\bibitem{hu2018boundary}
{Hu, W.} and {Wu, J.}
\newblock Boundary control for optimal mixing via {N}avier--{S}tokes flows.
\newblock {\em SIAM Journal on Control and Optimization}, 56(4):2768--2801,
  2018.

\bibitem{hu2019approximating}
{Hu, W.} and {Wu, J.}
\newblock An approximating approach for boundary control of optimal mixing via
  {N}avier--{S}tokes flows.
\newblock {\em Journal of Differential Equations}, 267(10):5809--5850, 2019.

\bibitem{Hui2018}
K.~Hui, R.~Ching, S.~Chan, J.~Nicholls, N.~Sachs, H.~Clevers, J.~Peiris, and
  M.~Chan.
\newblock Tropism, replication competence, and innate immune responses of
  influenza virus: an analysis of human airway organoids and ex-vivo bronchus
  cultures.
\newblock {\em Lancet Respir Med.}, 11:846--854, 2018.

\bibitem{iyer2014lower}
{Iyer, G.}, {Kiselev, A.}, and {Xu, X.}
\newblock Lower bounds on the mix norm of passive scalars advected by
  incompressible enstrophy-constrained flows.
\newblock {\em Nonlinearity}, 27(5):973, 2014.

\bibitem{kelliher2006navier}
{Kelliher, J. P.}
\newblock {N}avier--{S}tokes equations with {N}avier boundary conditions for a
  bounded domain in the plane.
\newblock {\em SIAM journal on mathematical analysis}, 38(1):210--232, 2006.

\bibitem{D0LC01173H}
Y.~Li, X.~Liu, Q.~Huang, A.~T. Ohta, and T.~Arai.
\newblock Bubbles in microfluidics: an all-purpose tool for micromanipulation.
\newblock {\em Lab Chip}, 21:1016--1035, 2021.

\bibitem{lin2011optimal}
{Lin, Z.}, {Thiffeault, J.-L.}, and {Doering, C. R.}
\newblock Optimal stirring strategies for passive scalar mixing.
\newblock {\em Journal of Fluid Mechanics}, 675:465--476, 2011.

\bibitem{lions1971optimal}
{Lions, J. L.}
\newblock Optimal control of systems governed by partial differential
  equations.
\newblock 1971.

\bibitem{liu2008mixing}
{Liu, W.}
\newblock Mixing enhancement by optimal flow advection.
\newblock {\em SIAM Journal on Control and Optimization}, 47(2):624--638, 2008.

\bibitem{Lukens2010}
S.~Lukens, X.~Yang, and L.~Fauci.
\newblock Using lagrangian coherent structures to analyze fluid mixing by
  cilia.
\newblock {\em Chaos}, 20:017511, 2010.

\bibitem{lunasin2012optimal}
{Lunasin, E.}, {Lin, Z.}, {Novikov, A.}, {Mazzucato, A.}, and {Doering, C. R.}
\newblock Optimal mixing and optimal stirring for fixed energy, fixed power, or
  fixed palenstrophy flows.
\newblock {\em Journal of Mathematical Physics}, 53(11):115611, 2012.

\bibitem{mathew2007optimal}
{Mathew, G.}, {Mezi{\'c}, I.}, {Grivopoulos, S.}, {Vaidya, U.}, and {Petzold,
  L.}
\newblock Optimal control of mixing in {S}tokes fluid flows.
\newblock {\em Journal of Fluid Mechanics}, 580:261--281, 2007.

\bibitem{mathew2005multiscale}
{Mathew, G.}, {Mezi{\'c}, I.}, and {Petzold, L.}
\newblock A multiscale measure for mixing.
\newblock {\em Physica D: Nonlinear Phenomena}, 211(1):23--46, 2005.

\bibitem{Nakamura2020}
R.~Nakamura, T.~Katsuno, Y.~Kishimoto, and et~al.
\newblock A novel method for live imaging of human airway cilia using wheat
  germ agglutinin.
\newblock {\em Sci Rep.}, 10:14417, 2020.

\bibitem{navier1823memoire}
{Navier, C.-L.}
\newblock M{\'e}moire sur les lois du mouvement des fluides.
\newblock {\em M{\'e}moires de {\'l}Acad{\'e}mie Royale des Sciences de
  {\'l}Institut de France}, 6:389--440, 1823.

\bibitem{Nawroth2017}
J.~Nawroth, H.~G. nad E.~Koch, E.~Heath-Heckman, J.~Hermanson, E.~Ruby,
  J.~Dabiri, E.~Kanso, and M.~McFall-Ngai.
\newblock Motile cilia create fluid-mechanical microhabitats for the active
  recruitment of the host microbiome.
\newblock {\em Proc. Natl. Acad. Sci.}, 114:9510, 2017.

\bibitem{seis2013maximal}
{Seis, C.}
\newblock Maximal mixing by incompressible fluid flows.
\newblock {\em Nonlinearity}, 26(12):3279, 2013.

\bibitem{Supatto2008}
W.~Supatto, S.~E. Fraser, and J.~Vermot.
\newblock An all-optical approach for probing microscopic flows in living
  embryos.
\newblock {\em Biophys. J.}, 95:L29, 2008.

\bibitem{thiffeault2012using}
{Thiffeault, J.-L.}
\newblock Using multiscale norms to quantify mixing and transport.
\newblock {\em Nonlinearity}, 25(2):R1, 2012.

\bibitem{thiffeault2011moving}
{Thiffeault, J.-L.}, {Gouillart, E.}, and {Dauchot, O.}
\newblock Moving walls accelerate mixing.
\newblock {\em Physical Review E}, 84(3):036313, 2011.

\bibitem{Islam2022}
T.~{ul Islam}, Y.~Wang, I.~Aggarwal, Z.~Cui, H.~{Eslami Amirabadi}, H.~Garg,
  R.~Kooi, B.~Venkataramanachar, T.~Wang, S.~Zhang, P.~Onck, and J.~{den
  Toonder}.
\newblock Microscopic artificial cilia - a review.
\newblock {\em Lab on a Chip}, XX(X), Apr. 2022.

\bibitem{vikhansky2002enhancement}
{Vikhansky, A.}
\newblock Enhancement of laminar mixing by optimal control methods.
\newblock {\em Chemical Engineering Science}, 57(14):2719--2725, 2002.

\bibitem{yao2014mixing}
{Yao, Y.} and {Zlatos, A.}
\newblock Mixing and un-mixing by incompressible flows.
\newblock {\em arXiv preprint arXiv:1407.4163}, 2014.

\bibitem{Zhanglinbo2009}
{Zhang, L.}, {Cui, T.}, and {Liu, H.}
\newblock A set of symmetric quadrature rules on triangles and tetrahedra.
\newblock {\em Journal of Computational Mathematics}, 27:89--96, 2009.

\bibitem{zheng2023iterative}
X.~Zheng, K.~Zhao, J.~Wu, W.~Hu, and D.~Du.
\newblock {Iterative projection method for unsteady Navier-Stokes equations
  with high Reynolds numbers}, 2023.
\newblock arXiv 2304.07963.

\end{thebibliography}

\end{document}